\documentclass[prl,twocolumn,aps,showpacs,amsmath,amssymb,superscriptaddress,floatfix,longbibliography]{revtex4-1}

\usepackage{color}
\usepackage[dvipsnames]{xcolor}
\usepackage{bm}
\usepackage{graphicx}
\usepackage{braket}
\usepackage[caption=false]{subfig}

\usepackage[pdftex]{hyperref}
\hypersetup{colorlinks = true, urlcolor = blue, linkcolor = blue, citecolor = blue}

\begin{document}
\title{Critical properties of the measurement-induced transition in random quantum circuits}
\author{Aidan Zabalo}
\affiliation{Department of Physics and Astronomy, Center for Materials Theory, Rutgers University, Piscataway, NJ 08854 USA}
\author{Michael J. Gullans}
\affiliation{Physics Department, Princeton University, Princeton, New Jersey 08544, USA}
\author{Justin H. Wilson}
\affiliation{Department of Physics and Astronomy, Center for Materials Theory, Rutgers University, Piscataway, NJ 08854 USA}
\author{Sarang Gopalakrishnan}
\affiliation{Department of Engineering Science and Physics, CUNY College of Staten Island, Staten Island, New York 10314, USA and Initiative for the Theoretical Sciences, CUNY Graduate Center, New York, New York 10016 USA}
\author{David A. Huse}
\affiliation{Physics Department, Princeton University, Princeton, New Jersey 08544, USA}
\affiliation{Institute for Advanced Study, Princeton, New Jersey 08540, USA}
\author{J. H. Pixley}
\affiliation{Department of Physics and Astronomy, Center for Materials Theory, Rutgers University, Piscataway, NJ 08854 USA}

\date{\today}

\begin{abstract}
We numerically study the measurement-driven quantum phase transition of Haar-random quantum circuits in $1+1$ dimensions.  By analyzing the tripartite mutual information we are able to make a precise estimate of the critical measurement rate $p_c = 0.17(1)$.  We extract estimates for the associated bulk critical exponents that are consistent with the values for percolation, as well as those for stabilizer circuits, but differ from previous estimates for the Haar-random case. 
Our estimates of the surface order parameter exponent appear different from that for stabilizer circuits or percolation, but we cannot definitively rule out the scenario where all exponents in the three cases match.
Moreover, in the Haar case the prefactor for the entanglement entropies $S_n$ depends strongly on the R\'enyi index $n$; for stabilizer circuits and percolation this dependence is absent. Results on stabilizer circuits are used to guide our study and identify measures with weak finite-size effects. We discuss how our numerical estimates constrain theories of the transition.
\end{abstract}

\maketitle

\begin{figure}[tb]
\centering
           \includegraphics[width=\columnwidth]{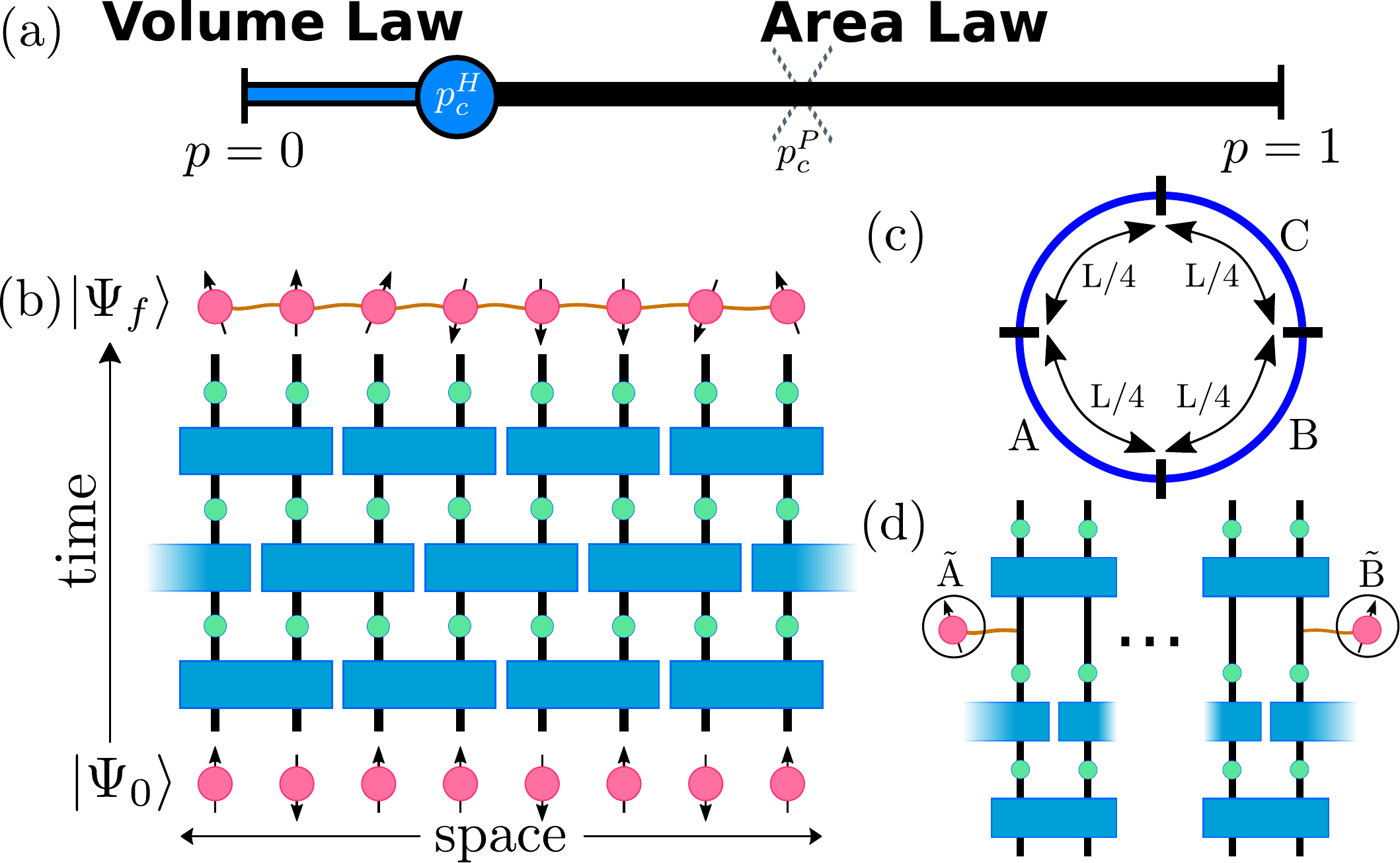}
   \caption{(a) Phase diagram with $p_c^H$ and $p_c^P$ marking the separation between volume law and area law entanglement in the R\'enyi entropies $n \ge 1$ and $n=0$, respectively. (b) 
Depiction of the model:   Blue rectangles represent two-site entangling gates and the green circles denote local projective measurements that are performed with a probability $p$. (c) The geometry used to compute the TMI, partitioning the system with periodic boundary conditions into four equal-length segments. (d) The set-up to probe the order parameter correlation function through entangling the local system qubits at time $t=t_0=2L$ with two ancilla qubits seperated by a distance $r-r'$ and computing their mutual information at later times.}
  \label{fig:model}
\end{figure}

Characterizing phase transitions in the dynamics of nonequilibrium quantum systems is a key open question in quantum statistical physics. So far, nonequilibrium phase transitions have been studied primarily for isolated quantum systems~\cite{polkovnikov_review, nhreview} and for steady states of dissipative systems \cite{opensystemsbook,Sieberer16}.  One much-studied case is the many-body localization transition~\cite{nhreview}, which can be seen either~(i) as a dynamical transition at which thermalization slows down and stops as a parameter (e.g., the disorder strength in a spin chain) is tuned or~(ii) as an \emph{entanglement} transition at which the many-body eigenstates of the system change from volume-law to area-law entangled.  Recently, a different type of entanglement transition was discovered~\cite{Vasseur2018,Skinner2019, LCF2018} in the steady-state entanglement of the states produced by individual quantum trajectories~\cite{Zoller87,mcd,CarmichaelBook,Plenio98} of a repeatedly-measured quantum many-body system.  As the system is measured at an increasing rate, this single-trajectory entanglement goes from volume-law to area-law (see Fig. \ref{fig:model}(a))~\cite{Skinner2019, LCF2018, ChoiBaoQiAltman2019,Gullans2019,BaoChoiAltman2019,Szyniszewski2019,LCF2019,Ludwig2019,ChanNandkishore2019}. 

A measurement-driven transition is expected for quantum chaotic dynamics whether temporally random~\cite{Skinner2019, LCF2018} or Hamiltonian~\cite{Tang2019}.  Current studies have focused on quantum circuits acting on an array of qudits (of local Hilbert space dimension $q$); these are believed to be generic models of chaotic quantum dynamics~\cite{nrvh,rpv,Keyserlingk2017,Khemani2017,Rakovszky2018,Chan-2018,Chang-2018}.  
In specific cases, analytic results (or large-scale simulations) exist. For the Hartley entropy (i.e., rank of the reduced density matrix) and in the $q \rightarrow \infty$ limit, mappings to percolation have been found~\cite{Skinner2019, Vasseur2018,Ludwig2019,BaoChoiAltman2019}.
%
For stabilizer circuits, efficient classical simulations~\cite{LCF2018, LCF2019,Gullans2019,Gullans2019b} have been implemented in one-dimensional systems with $q=2$.  All three methods agree (within numerical precision for the stabilizer circuits) on the order-parameter and correlation-length critical exponents (respectively $\eta$ and $\nu$) at the transition; all of them, likewise, predict that the steady-state R\'enyi entanglement entropies, $S_n^{(A)} = (1 - n)^{-1} \log_2 \mathrm{Tr} \rho_A^n$ [where $A$ denotes a contiguous subsystem of length $L$ in a one-dimensional system, and $\rho_A$ its reduced density matrix]  should scale as $S_n \sim \alpha_n \ln L$. For stabilizer circuits and in the large-$q$ limit, $\alpha$ is independent of $n$. However, the value of $\alpha$ seems to be different in each of these solvable cases, suggesting that in \emph{some} respects these are distinct critical phenomena.

The present work analyzes the physically relevant, but analytically intractable, limit of Haar-random circuits with $q = 2$. Some numerical results exist for this case~\cite{Skinner2019,BaoChoiAltman2019,LCF2019} but are inconclusive because the values of the critical exponents are sensitive to the estimate for the critical point and choice of scaling ansatz. We circumvent these issues by studying the tripartite mutual information (TMI), which is found to have minimal finite-size drifts and allows us to reliably locate the critical point with minimal scaling assumptions on small system sizes.  The TMI is finite at the critical point, vanishes in the area-law phase, and diverges in the volume-law phase; thus, curves for different sizes cross at the critical point, allowing one to locate it reliably~\cite{Gullans2019, Kitaev2006, LevinWen2006}.  Having located the critical point, we estimate critical exponents; the correlation length exponent $\nu$ and the bulk anomalous dimension $\eta$~\cite{Gullans2019b} (described below) for the Haar case are close to or equal to those for percolation.  The surface critical exponent, however, appears to differ from both stabilizer circuits and percolation, suggesting that the Haar model lies in a separate universality class~\cite{Cardy84}.  The R\'enyi entropies $S_n$ (for $n \geq 1$) appear to be logarithmic at the critical point, but with a strongly $n$-dependent prefactor: the entanglement spectrum has a nontrivial critical structure. To guide our study of Haar-random circuits, we analyze small stabilizer circuits using the same methods: our results for small sizes reliably predict the exponents found from much larger sizes, showing that our observables have weak finite-size effects in stabilizer circuits, and thus seem likely to also be well-behaved for Haar-random circuits.

\emph{Models}.---We focus on two different models of random circuits in a ``brick-layer'' geometry with local projective measurements, as shown in Fig.~\ref{fig:model}(b). We start from a trivial product state $|\Psi_0 \rangle$ then time evolve in the presence of measurements. In the following, we consider two circuit models specified by the distribution of the gates. The local two-qubit gates $U_{i,i+1}$ (depicted as blue rectangles) are drawn from a Haar-random distribution for the Haar circuit model and for the stabilizer circuit model they are sampled uniformly from the Clifford group.  We expect the behavior of the Haar model to capture the generic behavior of systems undergoing chaotic unitary dynamics interspersed with projective measurements.
At each space-time ``site'' $(j)$ [shown as a green circle in Fig.~\ref{fig:model}(b)] with probability $p$ we make a measurement of the $z$-component of the spin $S^z_j$, project onto the measured value of $S^z$, and normalize the state.  For the Haar simulation, we exactly time evolve the state, while for the Clifford gates we initialize the system in a stabilizer state and dynamically update a generating set for the stabilizer group of this state \cite{Aaronson04}. To reduce finite size effects, we use periodic boundary conditions for a system size $L$, unless otherwise specified. We define one time step as one layer of gates and one layer of measurements.

\emph{Locating the critical point}.---Natural diagnostics of the transition are the bipartite R\'enyi entropies $S_n$, 
which saturate to a steady state on times $t \sim L$.  However, these entropies diverge logarithmically with $L$ at the critical point. Locating the critical point via $S_n$ requires one to account for the logarithm, which makes the finite-size scaling behavior less constrained. To circumvent this issue, we focus on the TMI between regions $A$, $B$, and $C$ as depicted in Fig.~\ref{fig:model}(c) 
\begin{eqnarray}\label{i3def}
&\mathcal{I}_{3,n}&(A,B,C) \equiv S_n(A)+S_n(B)+S_n(C) - S_n(A\cup B)
\nonumber
\\
&-&S_n(A\cup C)-S_n(B\cup C)+S_n(A\cup B\cup C).
\end{eqnarray}
We run the circuit out to time $t=4L$  so that the data is solely dependent on system size~\cite{suppmat}. In the area law phase, $\mathcal{I}_3$ is asymptotically zero for large $L$ because all the contributions to it come from boundary terms, and the boundary terms cancel out exactly in Eq.~\eqref{i3def}. In the volume law phase, it is negative and proportional to $L$, as the ``bulk'' contributions from regions $A$, $B$, and $C$ get subtracted out twice.  We find that $\mathcal{I}_3$ is finite and negative at the critical point. Within the minimal cut picture 
(which does not strictly apply to $n>0$, but appears to qualitatively capture some aspects of the transition) 
one can understand the behavior of $\mathcal{I}_3$ analytically~\cite{suppmat}. We remark that within the minimal cut picture, the mutual information $\mathcal{I}_{2,n}(A, C) \equiv S_n(A) + S_n(C) - S_n(A \cup C)$ should also be a constant at the critical point. Empirically, however, $\mathcal{I}_2$ has large finite-size drifts at small sizes~\cite{suppmat}.

\begin{figure}[tb]
\centering
           \includegraphics[width=0.49\columnwidth]{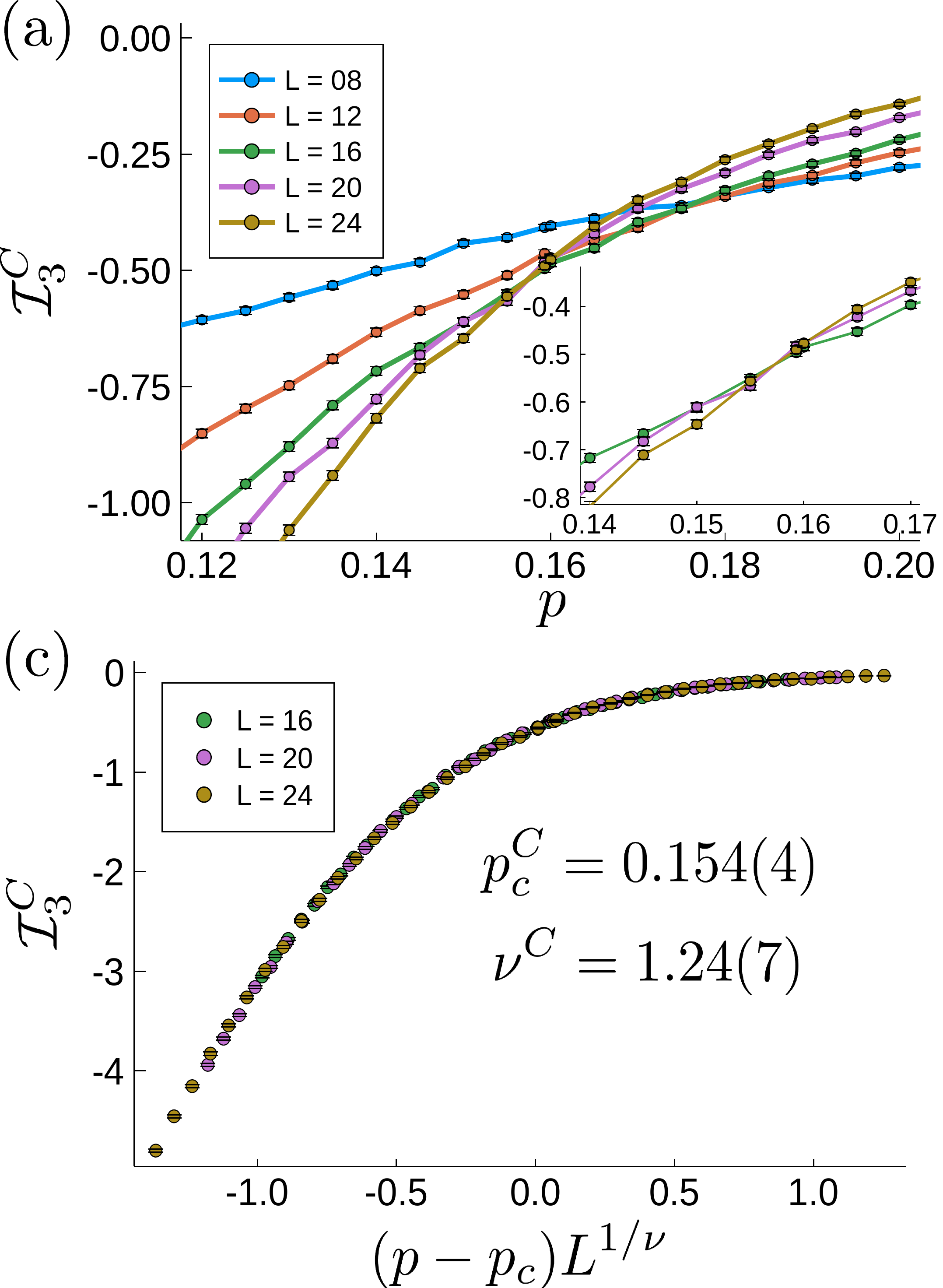}
           \includegraphics[width=0.49\columnwidth]{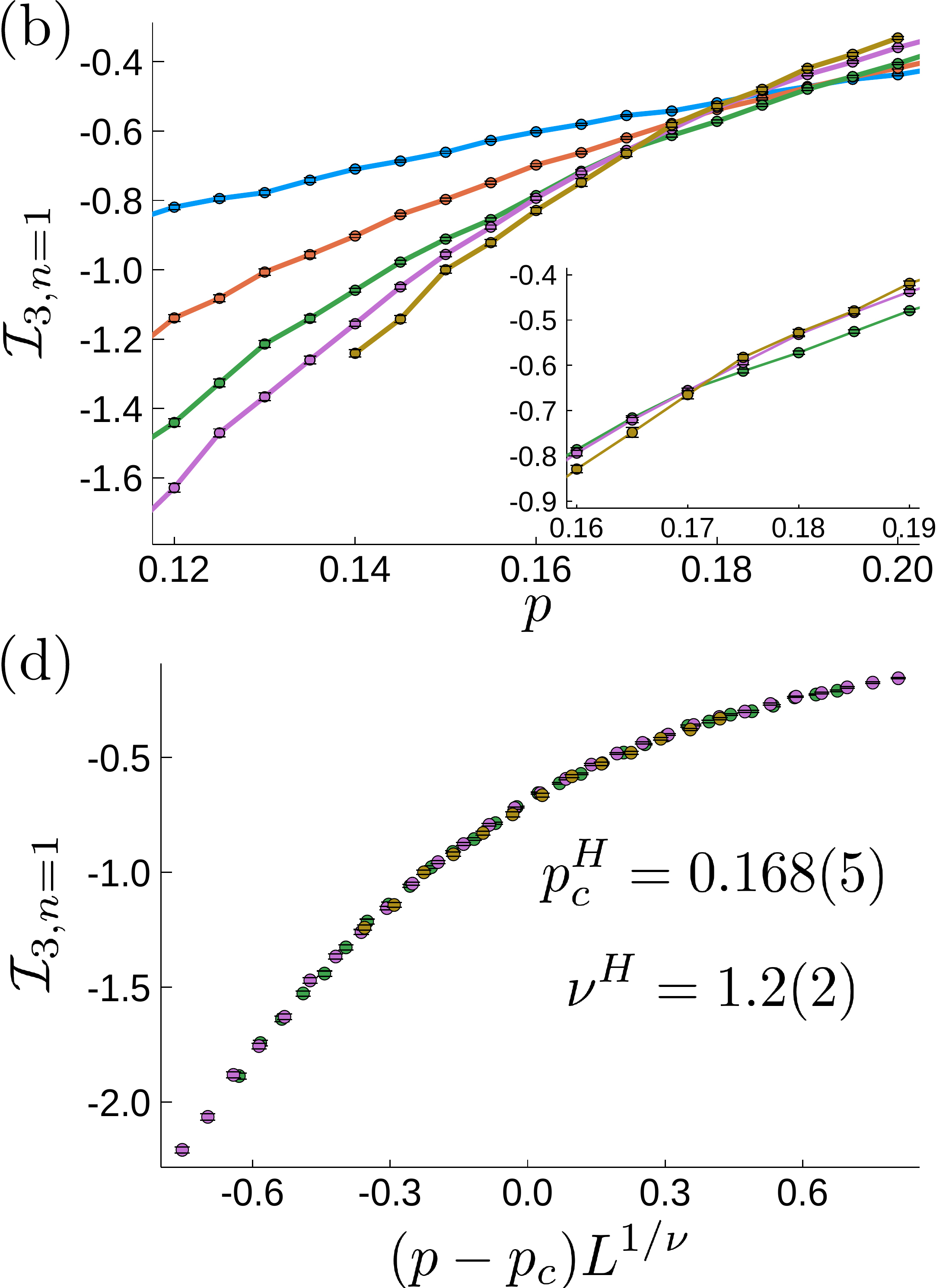}
   \caption{\emph{Tripartite mutual information (TMI) near the transition.} TMI near the transition for a circuit with (a) Clifford  and (b) Haar gates. Scaling collapse of the data for (c) Clifford and (d) Haar gates.}
  \label{fig:I3}
\end{figure}

We now turn to our numerical results on general $\mathcal{I}_{3,n}$. We find that $\mathcal{I}_{3,n}$ is an O(1) system size independent number at criticality for all values of $n$ we have considered~\cite{suppmat}. Thus, consistent with the minimal-cut argument as well as previous results on stabilizer circuits~\cite{Gullans2019}. 
Our results for $\mathcal{I}_{3,n=1}$ are shown in Fig.~\ref{fig:I3}(a) for Clifford gates and (b) for Haar gates at late times ($t=4L$) and similar system sizes. The TMI is negative for all $p$ and the data for different system sizes has a clear crossing for system sizes $L=16,20,24$. For stabilizer circuits this crossing yields an estimate of $p_c^C=0.154(4)$ 
, close to the critical value obtained up to sizes $L=512$.  For the Haar case, we estimate $p_c^H = 0.168(5)$. The value of $\mathcal{I}_{3,n}(p_c)$ is $L$-independent; for Haar gates we find $\mathcal{I}_{3,n=1}(p_c^H) \approx -0.66(8)$ and for stabilizer circuits $\mathcal{I}^C_{3}(p_c^C) \approx -0.56(9)$.  
The location of the crossings of the TMI for $n > 1$~\cite{suppmat} give estimates of $p_c$ that agree to within error bars with the result for $n = 1$. 

\emph{Correlation-length exponent}.---For $\mathcal{I}_{3,n=1}$ at late times ($t=4L$), we apply the scaling hypothesis
\begin{equation}\label{i3scalingform}
\mathcal{I}_{3,n=1}(p,L) \sim f(L^{1/\nu}(p-p_c))
\end{equation}
where $f(x)$ is a scaling function and $\nu$ is the correlation length exponent.  As shown in Fig.~\ref{fig:I3}(c) and (d) we find excellent data collapse that yields $\nu^C = 1.24 (7)$ and $\nu^H = 1.2(2)$, respectively. Our results for stabilizer circuits on small sizes agree with results on much larger system sizes up to $L=512$. We also obtain $\nu^H$ for various other R\'enyi indices and find that $\nu^H$ varies across $1.2(2)$ to $1.4(1)$ from $n=0.7$ to $\infty$ (see Table~\ref{tab:table1} and \cite{suppmat}), suggesting that $\nu^H$ is constant for all $n\ge 1$.

For $n > 1$ one can see that $p_c$ and $\nu$ are $n$-independent. All $S_n$ ($n > 1$) are upper- and lower-bounded by $S_\infty$: $S_\infty \leq S_n \leq n/(n-1) S_\infty$. Thus if any $S_n$ with $n > 1$ scales as a volume law, so must the others. Assuming single-parameter scaling as in Eq.~\eqref{i3scalingform}, $\nu^H$ is also independent of $n$: In the volume-law phase, by assumption, $S_n(L) \sim f_n(L/\xi_n)$ where $f_n(x) \sim \alpha_n \ln x$ for $x \ll 1$ and $f_n(x) \sim \alpha'_n x$ for $x \gg 1$ as well as $\xi_n \sim |p-p_c|^{-\nu_n^H}$. Then at very large length-scales, $S_n/S_\infty \sim \alpha'_n/\alpha'_\infty (\xi_\infty/\xi_n)$. This quantity cannot get parametrically large without violating the bounds on $S_n$, so $\xi_\infty/\xi_n$ must approach a constant, so $\nu^H$ is $n$-independent for $n > 1$.   For $n = 1$ these bounds do not apply. 
One can argue that $p_c$ remains $n$-independent for $0 < n < 1$ \emph{assuming} the entanglement transition coincides with the purification transition for an ancilla qubit (see below). For $p > p_c$, the ancilla purifies exponentially at a rate $t \sim L$; its smaller Schmidt coefficient decreases exponentially at this rate, so all its nonzero R\'enyi entropies vanish on timescales $\sim L$, yielding the same $p_c$ for all $n > 0$. 


%


The values of $\nu^H$ and $\nu^C$ are similar; indeed, within our numerical uncertainty both are consistent with the percolation exponent $\nu^P=4/3$.  For a more thorough comparison between stabilizer and Haar circuits we now turn to order-parameter correlations.

\emph{Order parameter}.---A \emph{local} bulk order parameter for the volume-law phase can be defined as follows \cite{Gullans2019b}. We run the circuit out to a steady state, then place one of the system spins into a Bell state with a reference qubit (an ancilla) $R$ at time $t = t_0$. We continue running the unitary-projective dynamics on the system. At $t_0$ the state of the system and $R$ is $|\psi_0\rangle = \frac{1}{\sqrt{2}} |A \uparrow\rangle - |B \downarrow \rangle$, where $|A\rangle, |B\rangle$ are orthogonal states of the system that are locally distinguishable at $t_0$. The order parameter is then $S_1(\rho_R)$, where $\rho_R$ is the density matrix of $R$ at a time $(t-t_0)\gg L$. This behaves differently in the two phases: In the area-law phase, measurements collapse the local state of the system that is coupled to $R$, thus disentangling $R$ and driving the order parameter to zero. In the volume-law phase the states $A$ and $B$ become indistinguishable under local measurements, and thus remain linearly independent under the dynamics, so the reference qubit stays entangled with the system and the order parameter remains nonzero.   Analogous surface order parameters can be defined by entangling $R$ with the initial state at $t_0=0$, or by using open boundary conditions and entangling it with an end spin.  
The estimate of $p_c$ obtained from order parameter dynamics agrees with that obtained from the TMI~\cite{suppmat}.

At $p_c$ the bulk order parameter decays very slowly; to get a cleaner numerical signal we study its two-point correlation function, which we access by introducing two ancilla qubits, $\tilde A$ and $\tilde B$, and entangling them with circuit qubits at spacetime points $(r,t_0)$ and $(r',t_0)$ [see Fig.~\ref{fig:model}(d)]. We define the connected order-parameter correlation as the mutual information between these ancillas. We fix $r-r'$ and determine the time dependence of the mutual information between $\tilde A$ and $\tilde B$, denoted $C(t - t_0)$. We consider two separate geometries: (i)~periodic boundary conditions, with ancillas connected to antipodal sites ($r-r' = L/2$), and (ii)~open boundary conditions, with ancillas connected to spins at each edge ($r-r'=L-1$). In both cases we start from a product state and run the circuit out to a time $t_0=2L$, introduce the ancilla qubits, maximally entangle them, and track their mutual information. 
Through a conformal transformation, the scaling dimension of case~(ii) can be related to the surface exponent $\eta_\parallel$ \cite{suppmat}.

\begin{figure}[tb]
\centering
           \includegraphics[width=0.49\columnwidth]{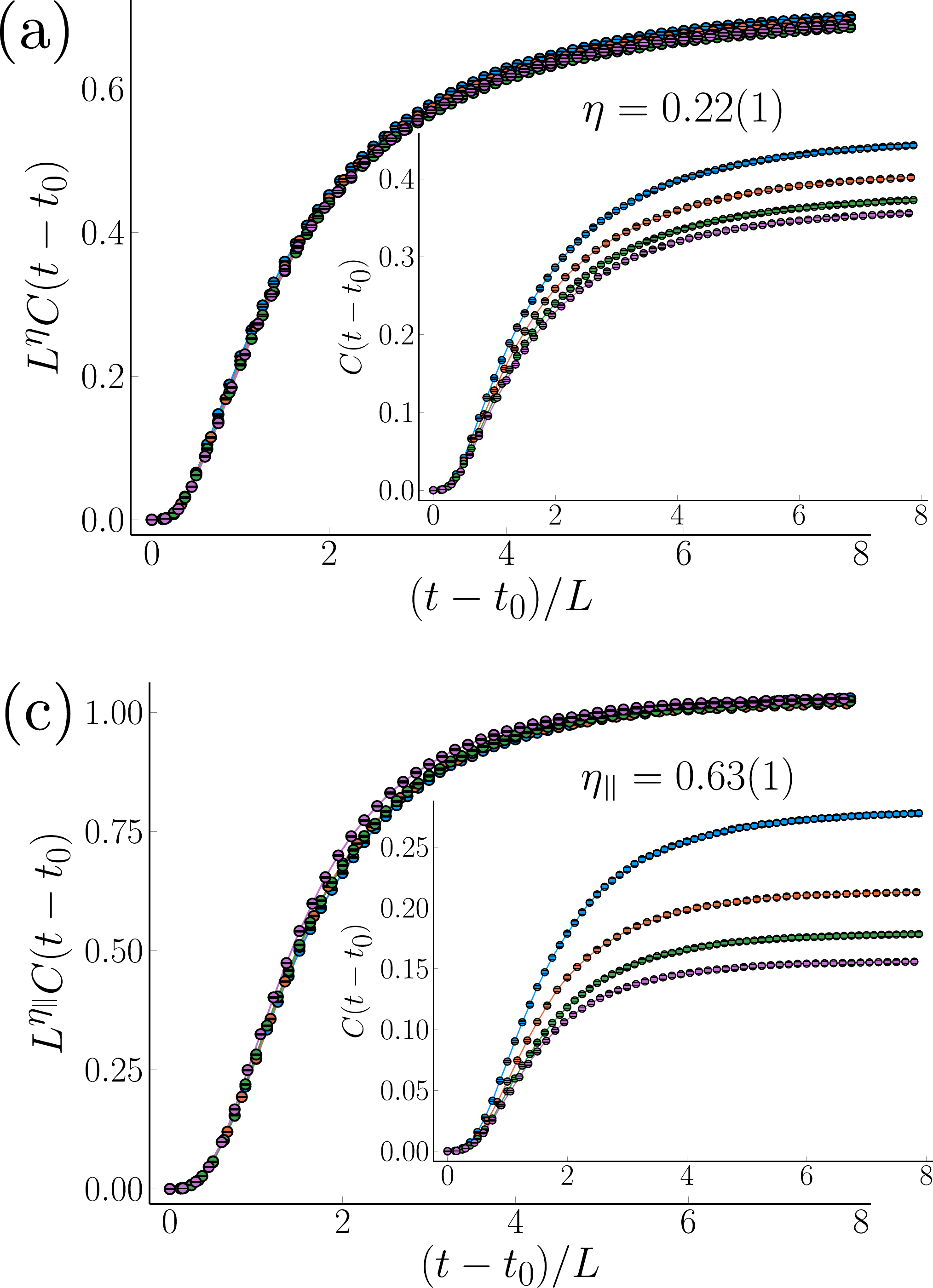}
           \includegraphics[width=0.49\columnwidth]{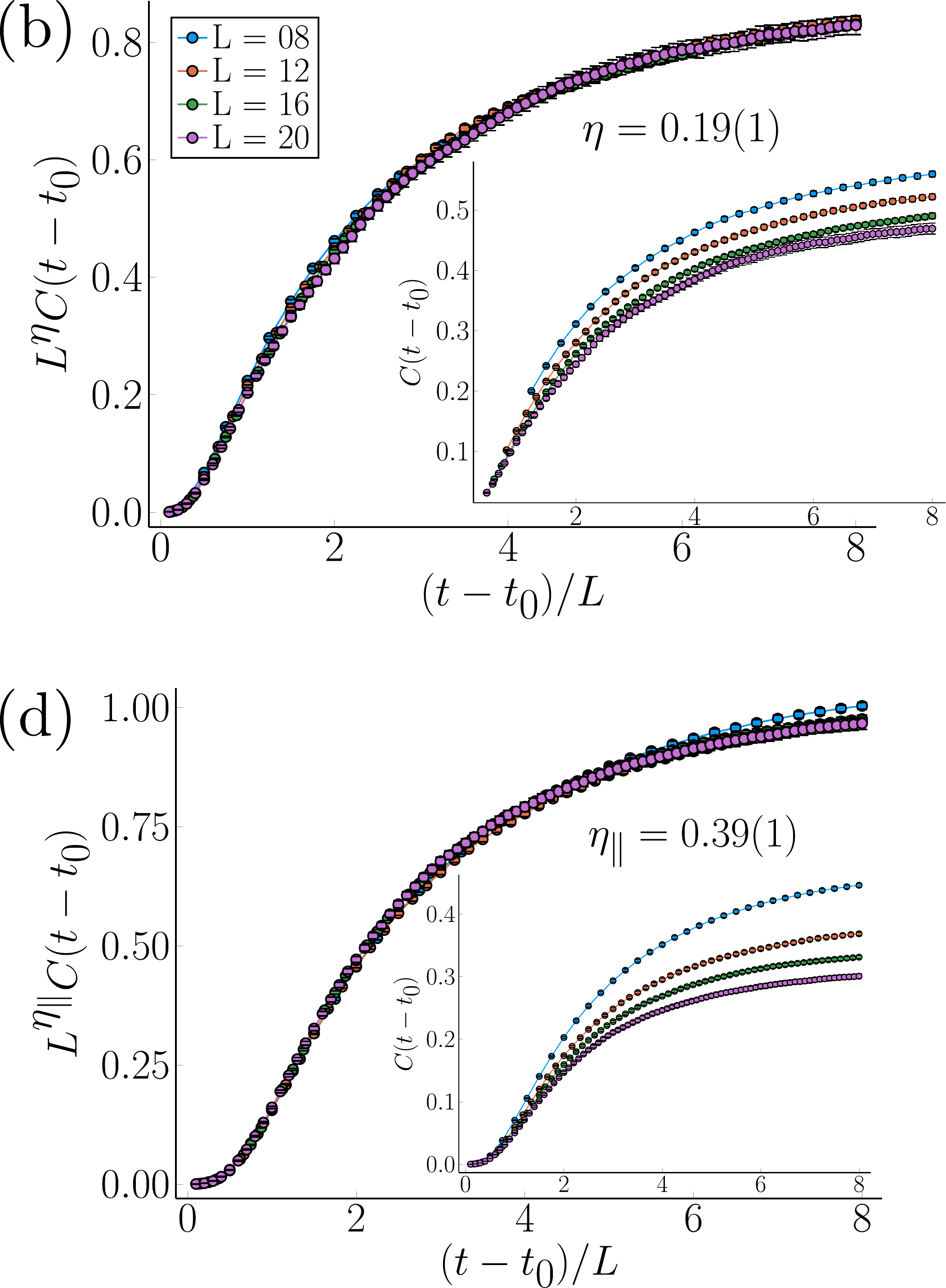}
   \caption{\emph{Scaling collapse of mutual information between two ancilla qubits (inset: unscaled data).} Mutual information between ancilla entangled at $r-r' = L/2$ for (a) Clifford and (b) Haar gates with periodic boundary conditions. Mutual information between ancilla entangled at $r=1$, $r' = L$ for (c) Clifford and (d) Haar gates with open boundary conditions.}
  \label{fig:eta}
\end{figure}

Our results for $C(t-t_0)$ are shown in Fig.~\ref{fig:eta}. In both the volume law phase, $p < p_c$, and the area law phase, $p>p_c$, $C(t-t_0) \sim \exp(-L/\xi)$ for $L\gg\xi$, where $\xi$ here is a finite correlation length away from $p_c$.  At criticality, we numerically estimate the dynamic exponent $z = 1.06(4)$ \cite{suppmat}, which is consistent  with previous work 
that finds $z=1$~\cite{Vasseur2018,Skinner2019,LCF2018,LCF2019}. Using this, the single-parameter scaling hypothesis implies the form
\begin{equation}
C(t-t_0,L) \sim L^{-\eta} g((t-t_0)/L)
\label{eqn:cfc}
\end{equation}
where $\eta$ and $g(x)$ depend on the boundary conditions used.  As demonstrated in Fig.~\ref{fig:eta}, we find excellent data collapse for system sizes $L=12,16,20$ for both Haar and Clifford gates, and a summary of the exponents are given in Table~\ref{tab:table1}. For periodic boundary conditions 
we 
find $\eta^C= 0.22(1)$ for Clifford gates and $\eta^H = 0.19(1)$ for Haar gates. Again, this result for Clifford gates agrees with a similar analysis at much larger $L$~\cite{Gullans2019b}.  These bulk exponents are both within uncertainties of the percolation value $\eta^P=5/24$.  To estimate the surface critical exponent we consider open boundary conditions 
and find $\eta_\parallel^H = 0.39(1)$ and $\eta_\parallel^C = 0.63(1)$, the latter of which is consistent with results obtained in different geometries on sizes up to $L = 1024$ \cite{Gullans2019b} and close to the percolation value of $2/3$. In stabilizer circuits, $\eta_\parallel$ extracted from the geometry used in Fig.~\ref{fig:eta}(c) has the smallest finite size effects, and the large discrepancy between these values suggests that the Haar and stabilizer circuits are in different universality classes \cite{Cardy84}. 
We have also checked whether this discrepancy persists in other geometries that have larger finite size effects for the stabilizer circuits~\cite{suppmat}. 
The statistical error in the collapse for these other quantities is not high, but certain exact relations based on the scaling hypothesis are not satisfied,  so there are potentially large systematic uncertainties in our estimate of $\eta_{\parallel}$. Although our results \emph{suggest} a different exponent, we cannot rule out a scenario in which the surface exponent is also consistent with percolation.

Finally, the order parameter dynamics for Haar and stabilizer circuits is qualitatively different. In stabilizer circuits, the ancilla jumps from fully mixed to fully pure in a single timestep; by contrast, in the Haar case, individual realizations purify gradually \cite{suppmat}.


\begin{table}[h!]
  \begin{center}
    \begin{tabular}{ccccccc}
 \hline
\hline
     $n$ & 1 & 2 & 5
      & $\infty$ & C & P\\
    \hline
    $p_c$ & 0.168(5) & 0.162(3) & 0.168(4) 
    & 0.170(4) & 0.154(4) & 0.5\\
    \hline
  $\nu$ & 1.2(2) & 1.3(1) & 1.4(1) 
  & 1.4(1) & 1.24(7) & 1.33\\
  \hline
  $\eta$ & 0.19(1) & 0.25(1) & 0.26(1)
   & 0.26(1) & 0.22(1) & 0.21\\
    \hline
  $\eta_\parallel$ & 0.39(1) & 0.49(1) & 0.49(2) 
  & 0.49(2) & 0.63(1) & 0.67\\
  \hline
    $\eta_\perp$ & 0.23(2) & 0.31(2) & 0.34(1) 
    & 0.34(1) & 0.43(2) & 0.44\\
  \hline
  $\alpha(n)$ & 1.7(2) & 1.2(2) & 0.9(1) & 0.7(1) & 1.61(3) & 0.55
  \\
       \hline
       \hline
    \end{tabular}
  \end{center}
  \caption{Table listing critical properties as a function of R\'enyi index $n$. The column $C$ corresponds to the $n$ independent results for the stabilizer circuit at small $L$ and P to the exact results from percolation provided to two digits of accuracy~\cite{perc}.}
  \label{tab:table1}
\end{table}

\begin{figure}[tb]
\centering
           \includegraphics[width=0.99\columnwidth]{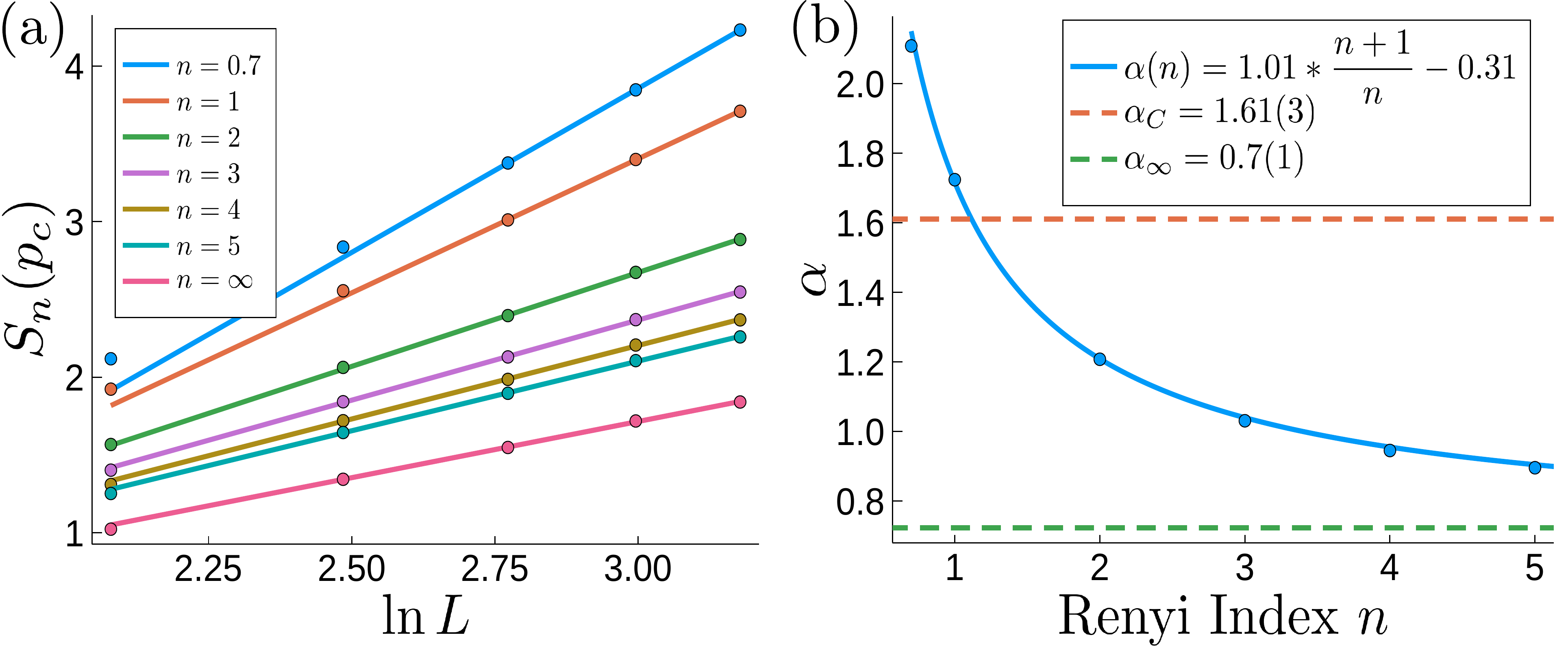}
   \caption{\emph{Properties of the R\'enyi entropies at criticality.} (a) The R\'enyi entropies show a $\ln L$ dependence near the critical point estimated by $\mathcal{I}_{3,n}$ for $n = 0.7,1,2,3,4,5,\infty$, with fits shown as solid lines. (b) The coefficient of the $\ln L$ term has a strong R\'enyi index dependence that is well described by a functional form $a(1 + 1/n) + b$.}
  \label{fig:Sn}
\end{figure}

\emph{R\'enyi Entropies}.---We now turn to the behavior of the R\'enyi entropies, which provides a clear distinction between Haar and stabilizer circuits. For stabilizer circuits at criticality we find $S_n(p_c,L) \sim \alpha_C \ln L$ for all $n$ with $\alpha_C = 1.61(3)$ on system sizes up to $L=24$, which agrees well with a similar fit out to much larger $L$ [yielding 1.63(3)]. In contrast, our data for Haar random circuits has a clear dependence on $n$, as shown in Fig.~\ref{fig:Sn}: we find
\begin{equation}
S_n(p_c,L) \sim \alpha(n) \ln L, \,  \, \alpha(n) = 0.7(1) + 1.0(1)/n.
\end{equation}
This fit is consistent with our direct estimate of $\alpha(n =\infty) \approx 0.7$. Interestingly, $\alpha(\infty)$ is close to the percolation value (=$\sqrt{3}/\pi \approx 0.55$ for periodic boundary conditions~\cite{Cardy00}), while $\alpha(1)$ is not far from the stabilizer value.

\emph{Discussion}.---
The critical properties obtained here are summarized in Table~\ref{tab:table1}. 
With our improved estimate of $p_c$, any differences in the bulk critical exponents between percolation and the Haar and stabilizer circuit transitions are within our uncertainties.  Haar and stabilizer circuits apparently differ in the surface critical exponent $\eta_\parallel$ and clearly differ in the coefficients $\alpha(n)$ of the log-divergence in the R\'enyi entropies at criticality.
Constraints imposed by conformal invariance imply that a different value of the surface critical exponent from percolation would imply that the Haar-random model is in a separate universality class \cite{Cardy84}.  The R\'enyi-dependence fits to a form $b + a(1 + 1/n)$, which is reminiscent of the scaling form for unitary conformal field theories, $a(1+1/n)$~\cite{lefevre}; however, in the present case one needs an offset to fit the data, so the critical wavefunctions at the measurement-induced transition differ from critical ground states.  We stress that these results are beyond any current analytic estimates, and come from being in the fully quantum regime: the R\'enyi-dependence is trivial in all the solvable limits.

The overall picture that emerges from our results is that the distinctions between the three known classes (percolation, stabilizer circuits, Haar-random circuits) of measurement-induced criticality are rather subtle:  The correlation length and bulk order-parameter exponents are consistent in all three cases to within our present error estimates.  However, the entanglement entropies at the critical points are clearly different, and the surface-order parameter exponents appear to differ.  Understanding why these superficially distinct critical phenomena look so similar is an important challenge for future work.

\begin{acknowledgments}
We thank Ehud Altman, Soonwon Choi, Vedika Khemani, Andreas Ludwig, Adam Nahum, Xiaoliang Qi, Jonathan Ruhman, Brian Skinner, and Romain Vasseur for helpful discussions. A.\ Z.\ is partially supported through a Fellowship from the Rutgers Discovery Informatics Institute. M.G.\ and D.H.\ are supported in part by the DARPA DRINQS program.  S.G.\ acknowledges support from NSF Grant No. DMR-1653271.   D.H.\ is also supported in part by a Simons Fellowship. J.H.P.\ is partially supported by Grant No. 2018058 from the United States-Israel Binational Science Foundation (BSF), Jerusalem, Israel.
S.G.\ and J.H.P.\ initiated this work at the Kavli Institute for Theoretical Physics, which is supported by NSF Grant No. PHY-1748958 and
J.H.W, S.G. and J.H.P.\  performed part of this work at the Aspen Center for Physics, which is supported by NSF Grant No. PHY- 1607611.
The authors acknowledge the Beowulf cluster at the Department of Physics and Astronomy of Rutgers University, the Office of Advanced Research Computing (OARC) at Rutgers, The State University of New Jersey (http://oarc.rutgers.edu) for providing access to the Amarel cluster, the Rutgers Discovery Informatics Institute for providing access to the Caliburn~\cite{Caliburn} cluster supported by Rutgers and the State of New Jersey, and associated research computing resources that have contributed to the results reported here.
\end{acknowledgments}

\bibliographystyle{apsrev4-1}
\bibliography{references}

\pagebreak
\widetext
\begin{center}

\end{center}
\textbf{\large Supplemental Material: Critical properties of the measurement-induced transition in random quantum circuits}

\setcounter{equation}{0}
\setcounter{figure}{0}
\setcounter{table}{0}
\setcounter{page}{1}
\renewcommand{\theequation}{S\arabic{equation}}
\setcounter{figure}{0}
\renewcommand{\thefigure}{S\arabic{figure}}
\renewcommand{\thepage}{S\arabic{page}}
\renewcommand{\thesection}{S\arabic{section}}
\renewcommand{\thetable}{S\arabic{table}}
\makeatletter

\section{Numerical Methods}
When calculating quantities such as the bipartite entanglement entropy, bipartite mutual information, or tripartite mutual information, there is an even-odd effect that occurs depending on whether or not there is a gate across the location of the cut. To deal with these effects we use the following prescription when calculating a quantity $Q$:
\begin{enumerate}
\item Perform a layer of gates on sites [1,2], [3,4], ...
\item Calculate quantity $Q^{(1)}$
\item Perform a layer of measurements
\item Perform a layer of gates on sites [2,3], [4,5], ...
\item Calculate quantity $Q^{(2)}$
\item Average results: $Q = \frac{1}{2}\left(Q^{(1)} + Q^{(2)}\right)$
\end{enumerate}

\section{R\'enyi Index n = 0}
\begin{figure}[htbp]
\centering
\subfloat[]{\includegraphics[width=.30\columnwidth]{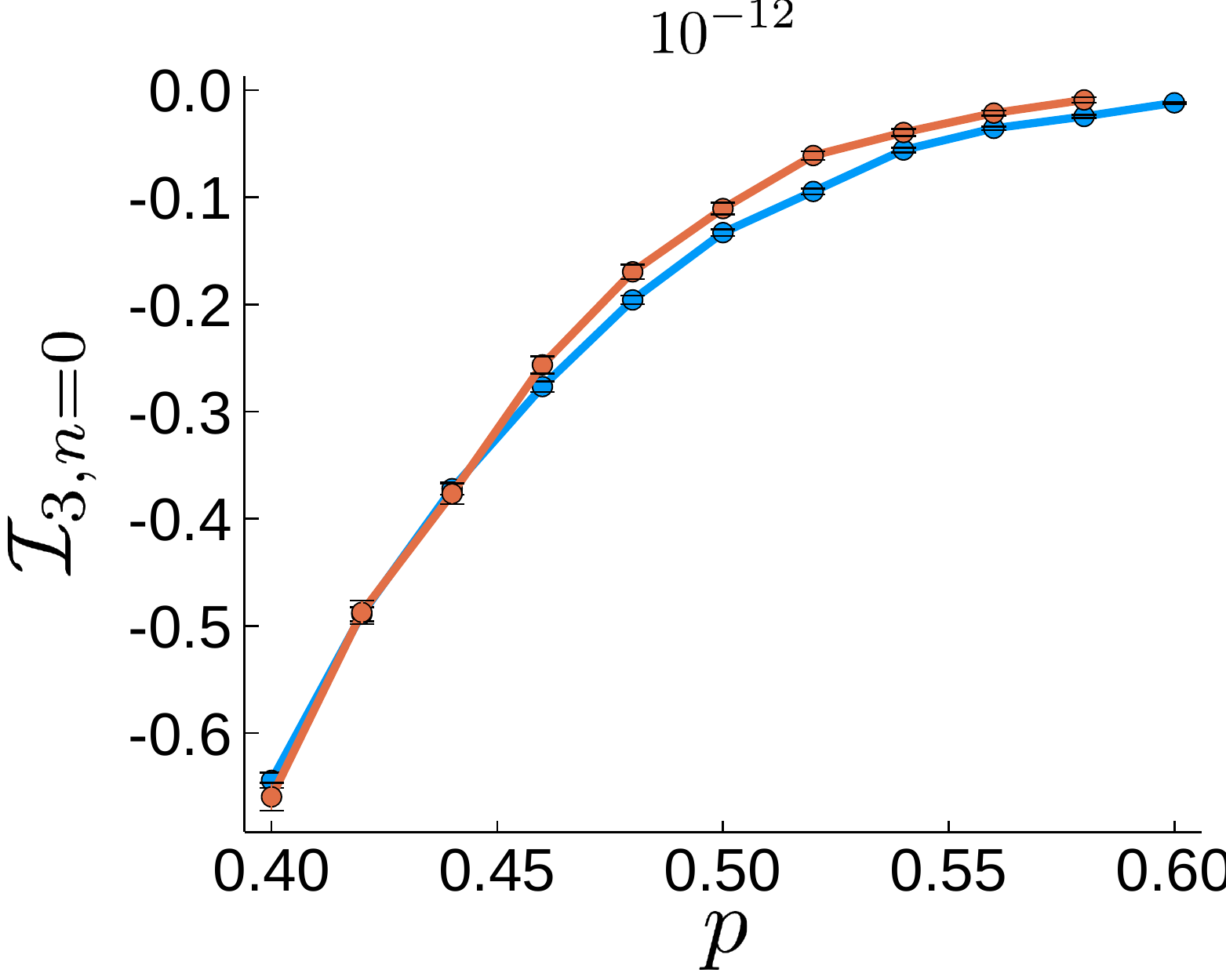}}
\subfloat[]{\includegraphics[width=.30\columnwidth]{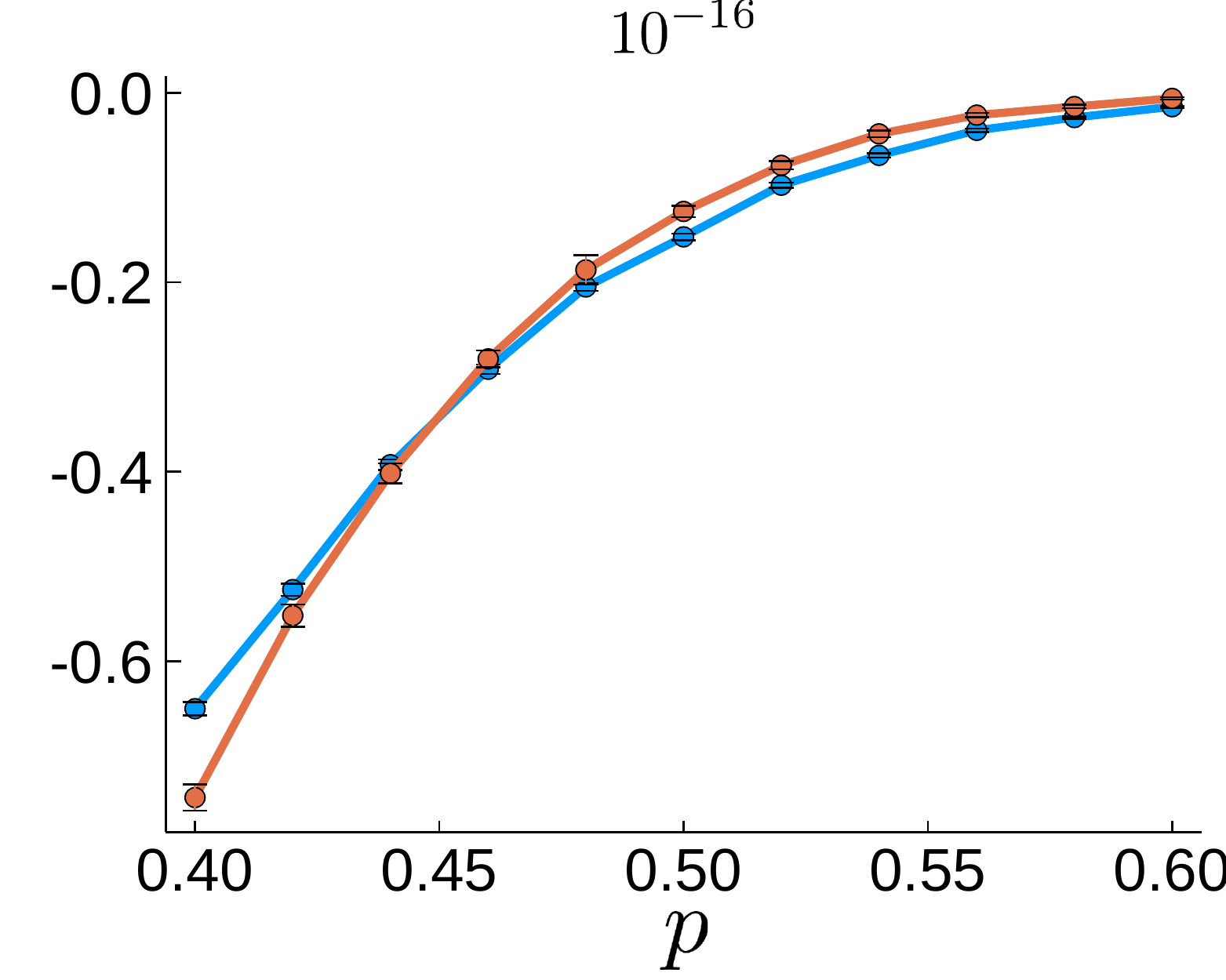}}
\subfloat[]{\includegraphics[width=.30\columnwidth]{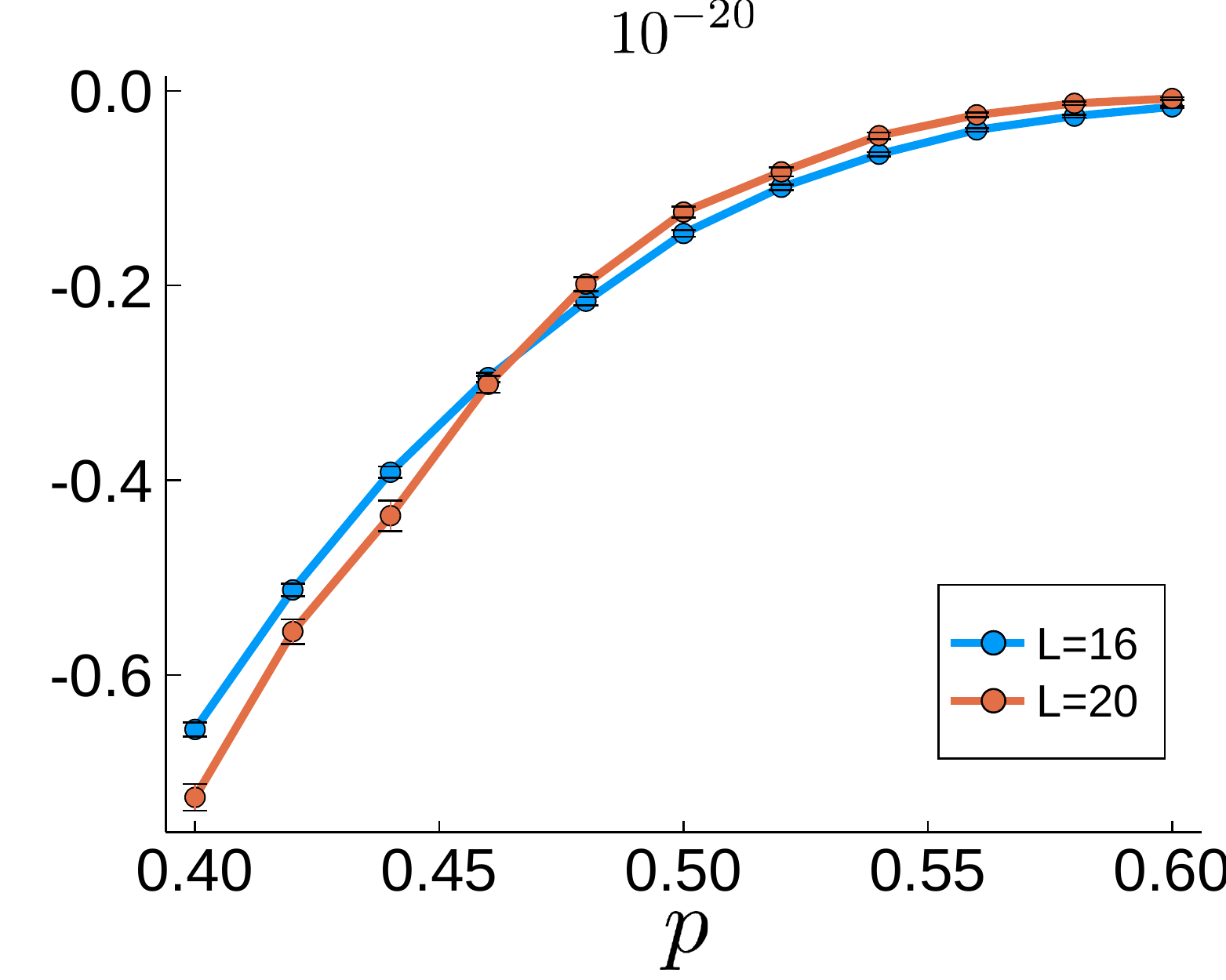}}
\caption{\emph{Sensitivity to cutoff.} The R\'enyi index $n=0$ raises all eigenvalues to the zeroth power and it is necessary to introduce a cutoff to prevent arbitrarily small eigenvalues from contributing to the sum. For a fixed set of system sizes, the crossing in $\mathcal{I}_{3,n=0}$ drifts towards a larger $p_c$ as the cutoff is increased. This leads to large errors when estimating $p_c$ and $\nu$.}
\label{fig:I3cutoff}
\end{figure}

\begin{figure}[htbp]
\centering
\subfloat[]{\includegraphics[width=.35\columnwidth]{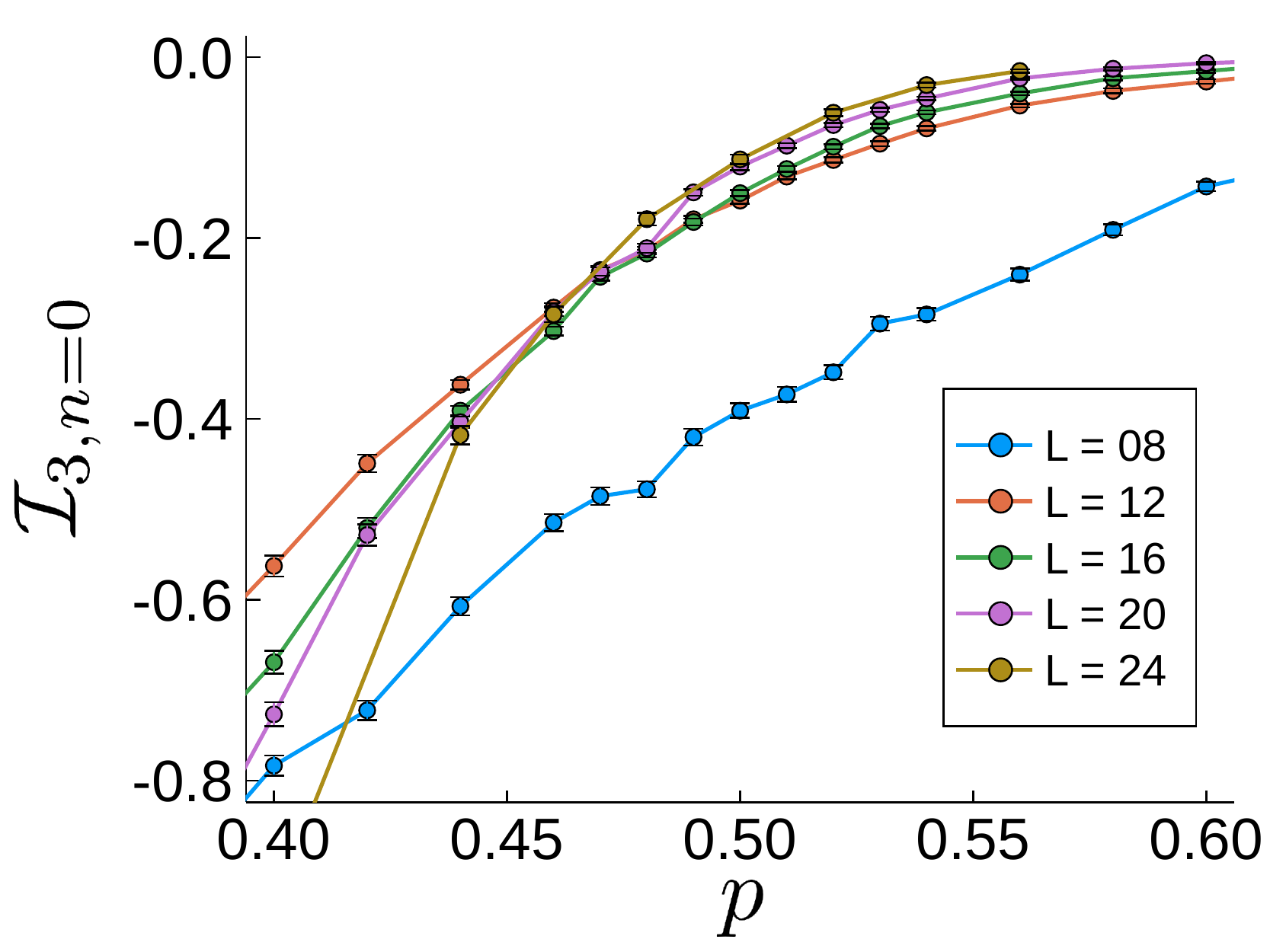}}
\subfloat[]{\includegraphics[width=.35\columnwidth]{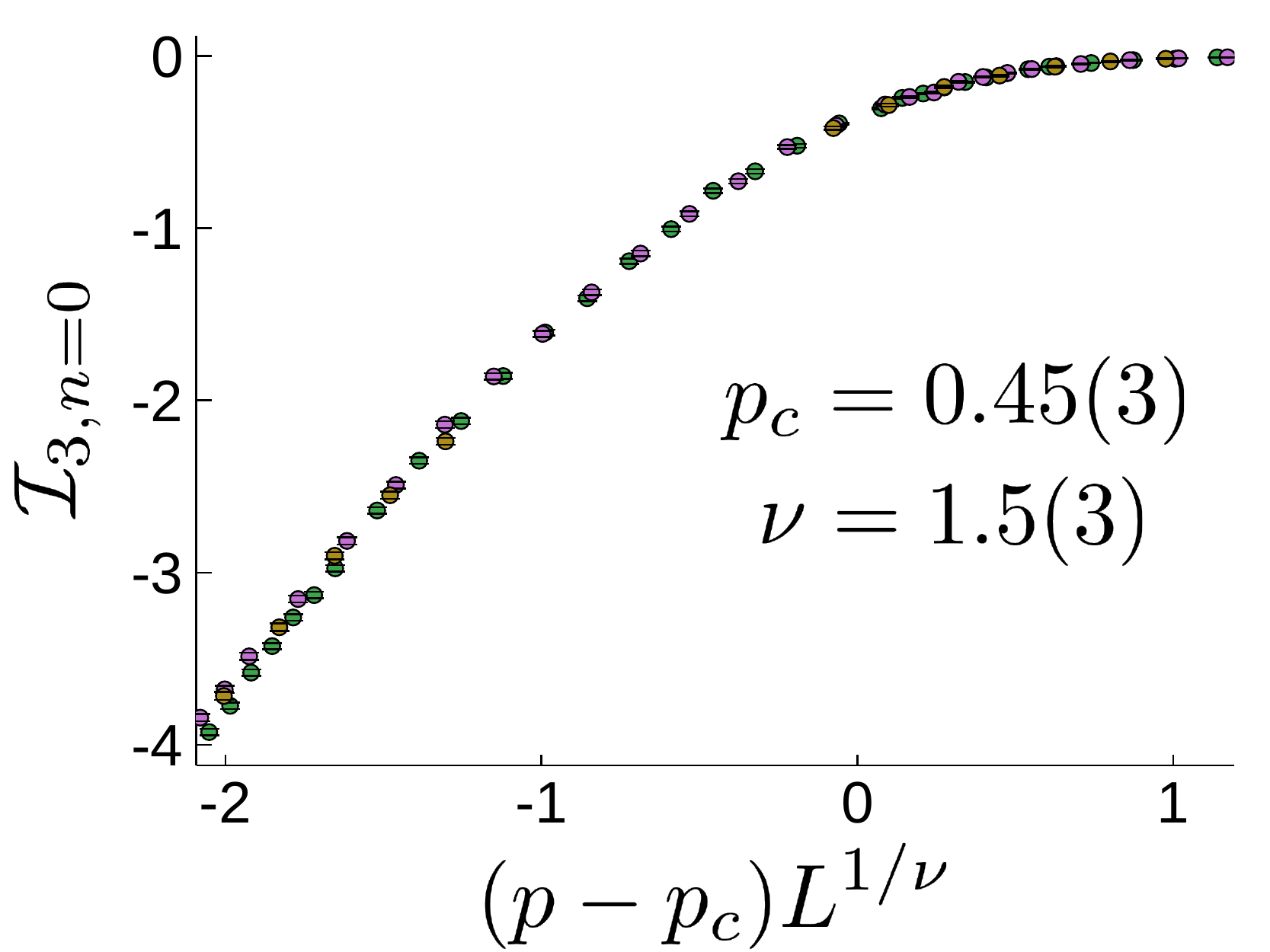}}
\caption{\emph{R\'enyi index $n=0$ tripartite mutual information.} The (a) tripartite mutual information and (b) data collapse for a cutoff given by machine epsilon for the binary64 number format. The estimated values of $p_c$ and $\nu$ are slightly different than the exact value given by the 2-D percolation mapping.}
\label{fig:I30}
\end{figure}

The quantities with R\'enyi index $n=0$ are difficult to calculate numerically since all eigenvalues must be raised to the zeroth power in the sum. One is then forced to introduce a cutoff to prevent arbitrarily small eigenvalues from contributing to the sum. This poses significant difficulties for estimating $p_c$ from the tripartite mutual information, $\mathcal{I}_{3,n=0}$, as shown by its sensitivity to the cutoff in Fig. \ref{fig:I3cutoff}. Choosing the cutoff to be machine epsilon for the binary64 number format, $\approx 2.22 \cdot 10^{-16}$, the critical values are estimated to be $p_c = 0.45(3)$ and $\nu = 1.5(3)$ as shown shown in Fig. \ref{fig:I30}. The estimated values are different than the exact values, $p_c = 0.5$ and $\nu = 4/3$, given by the 2-D percolation mapping.

\section{R\'enyi Index n $\geq$ 1}
\begin{figure}[htbp]
\centering
\subfloat[]{\includegraphics[width=.40\columnwidth]{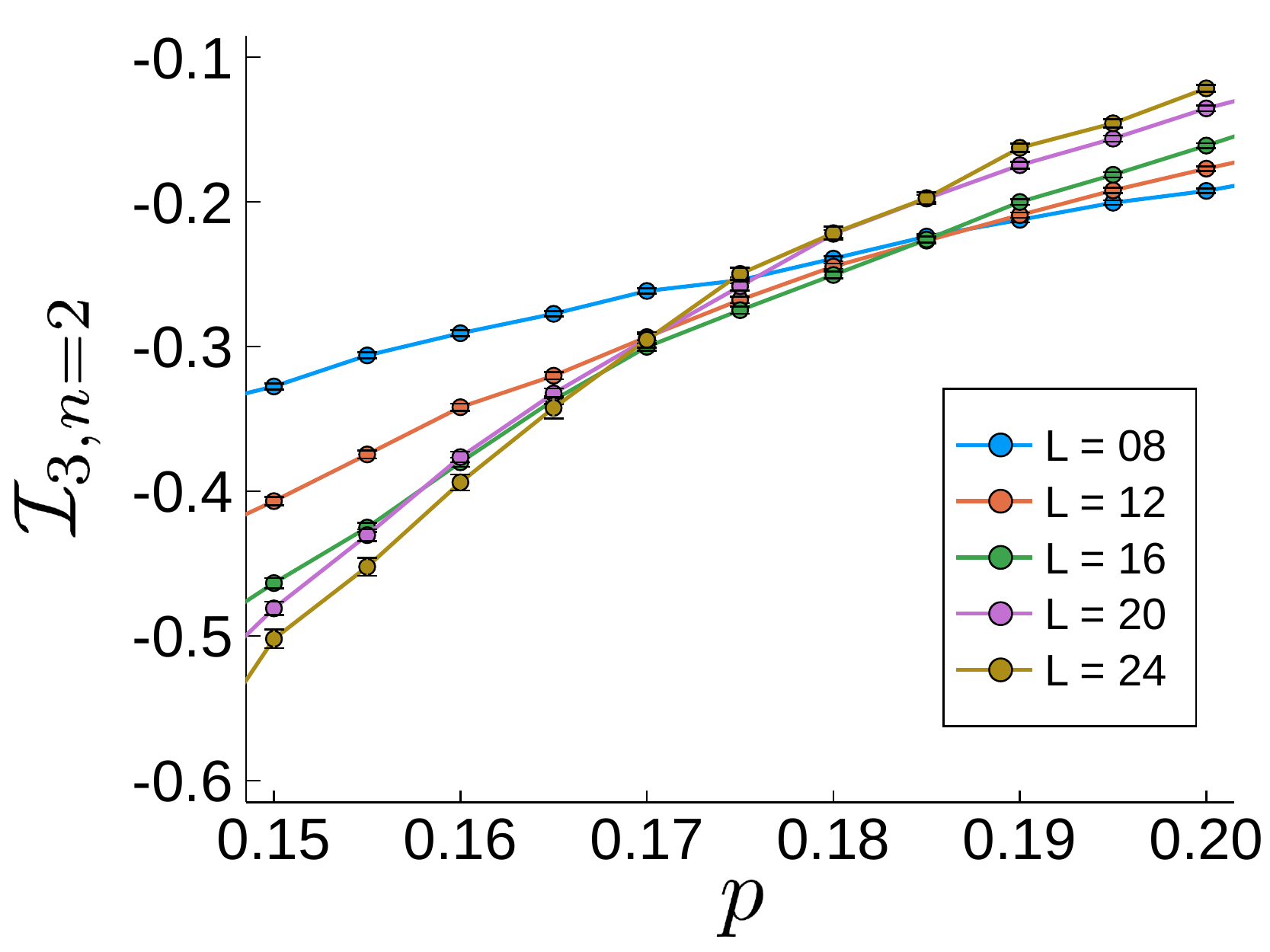}}
\subfloat[]{\includegraphics[width=.40\columnwidth]{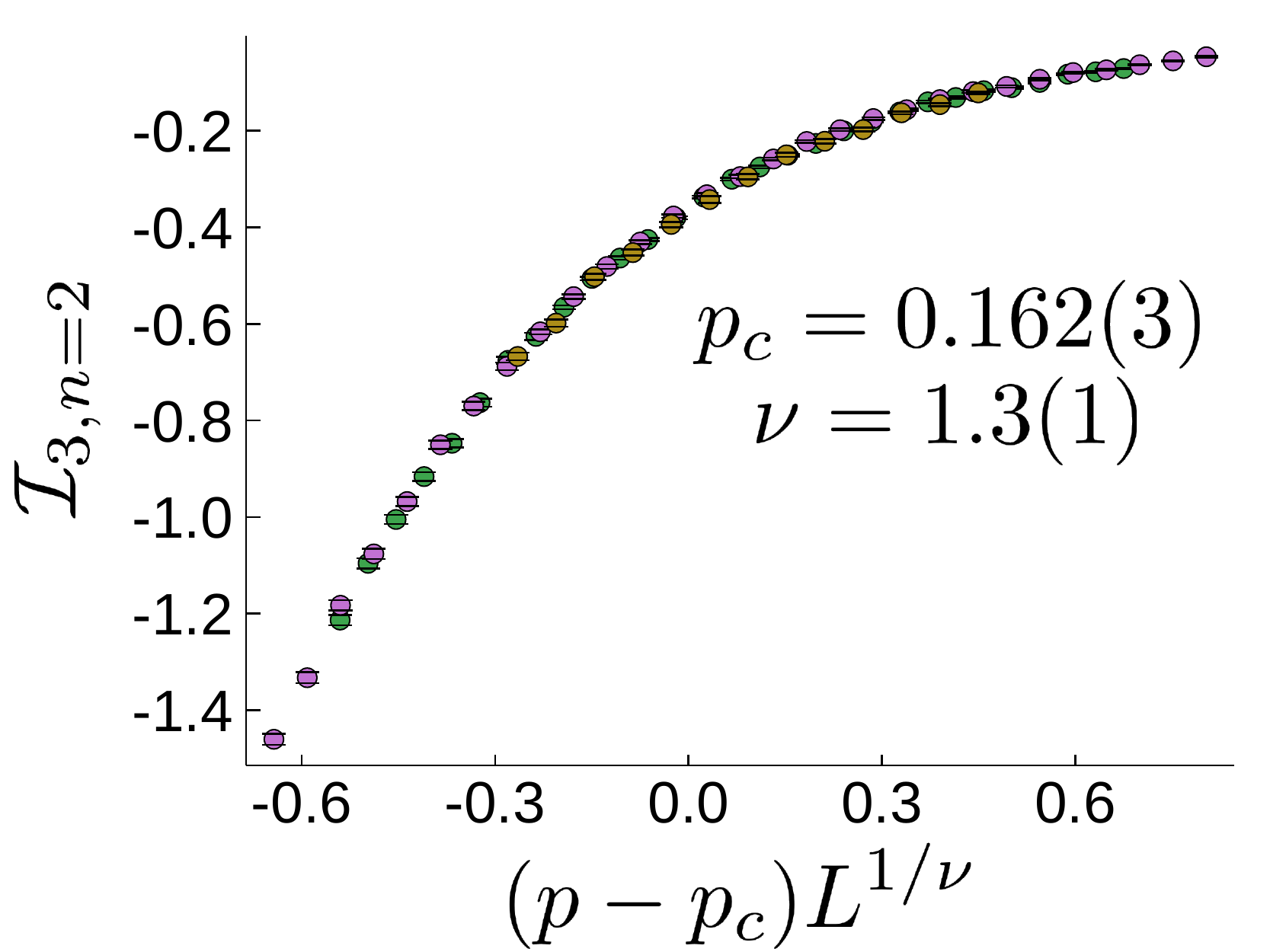}} \\
\subfloat[]{\includegraphics[width=.40\columnwidth]{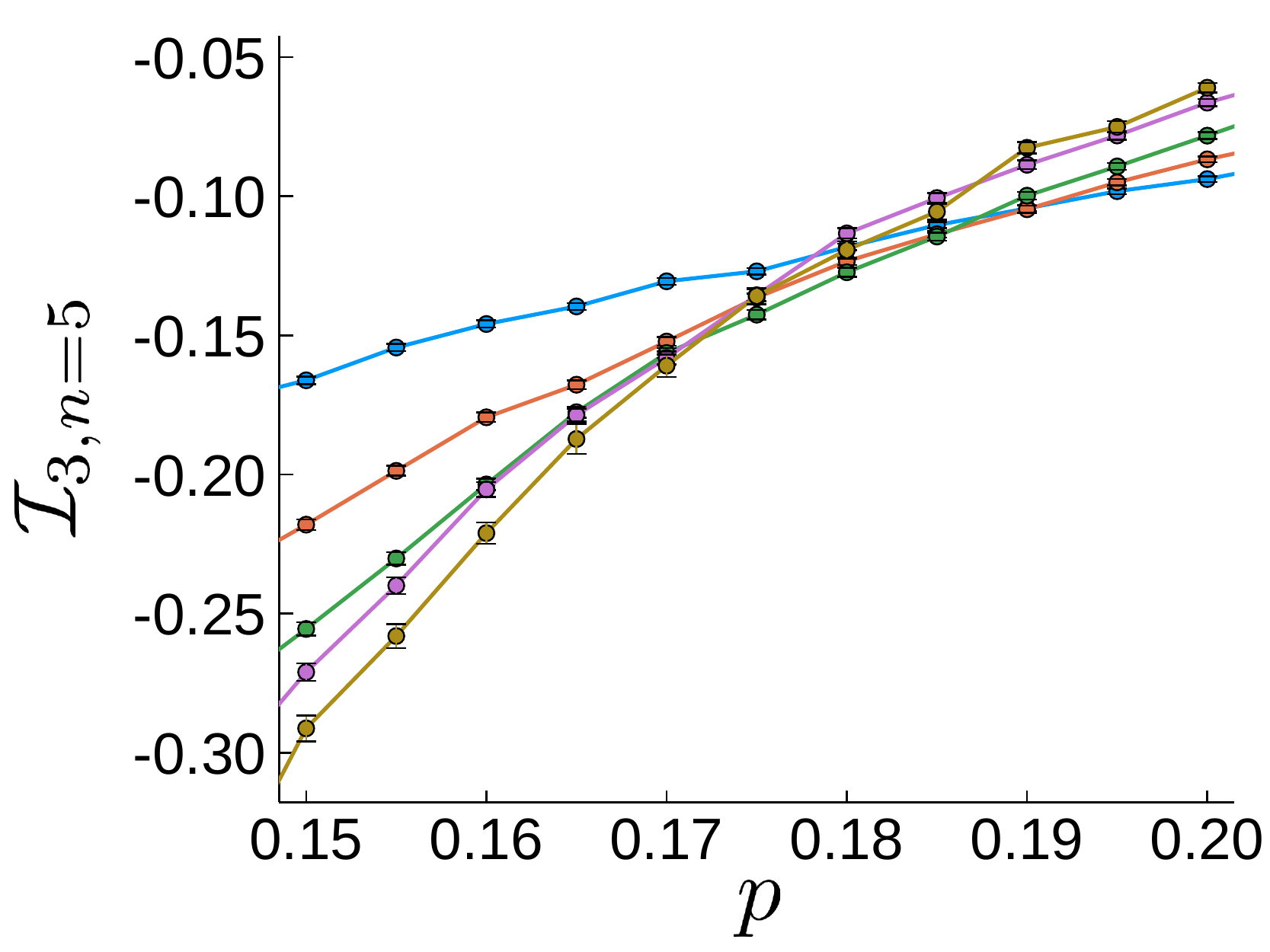}}
\subfloat[]{\includegraphics[width=.40\columnwidth]{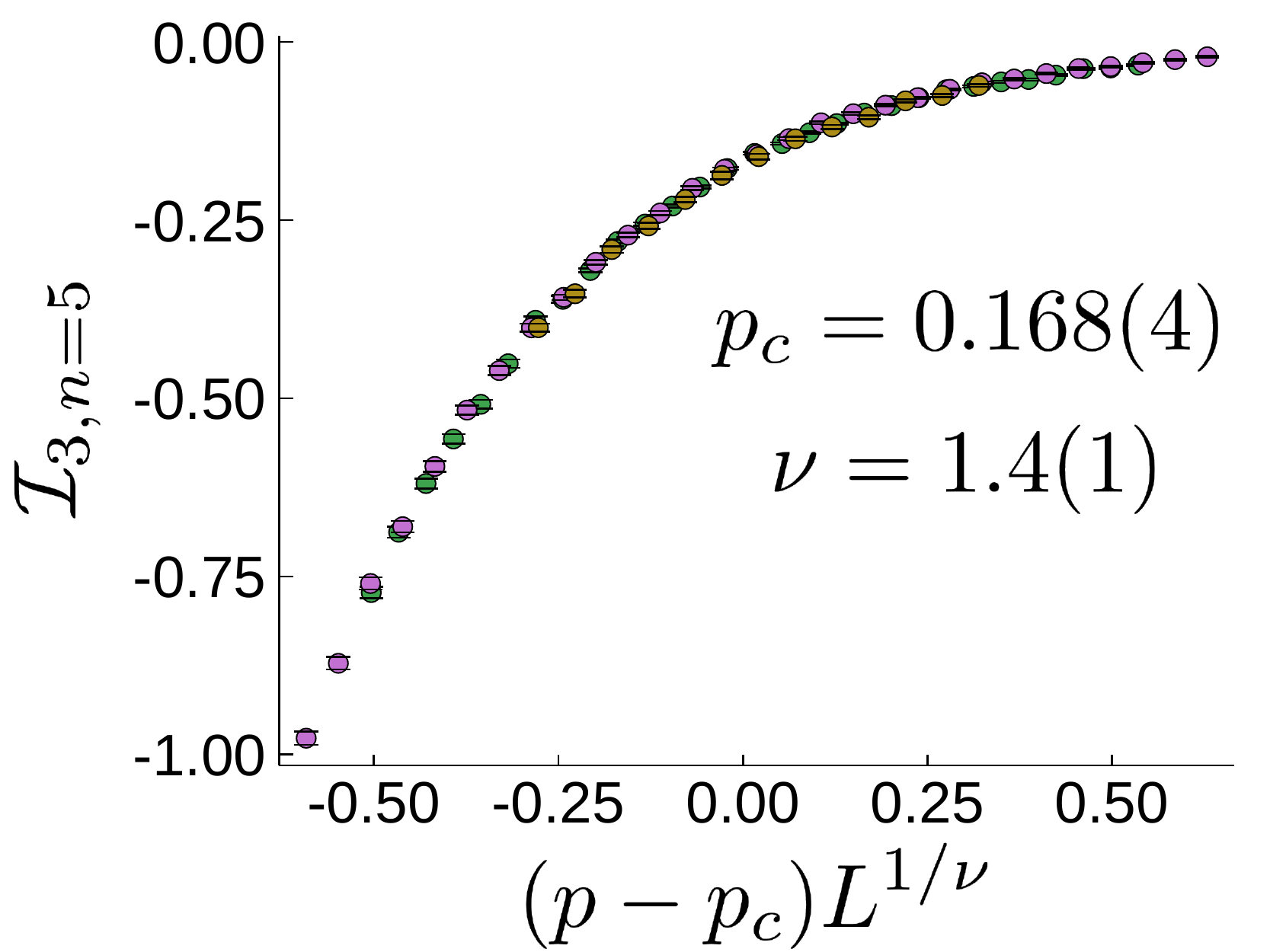}} \\
\subfloat[]{\includegraphics[width=.40\columnwidth]{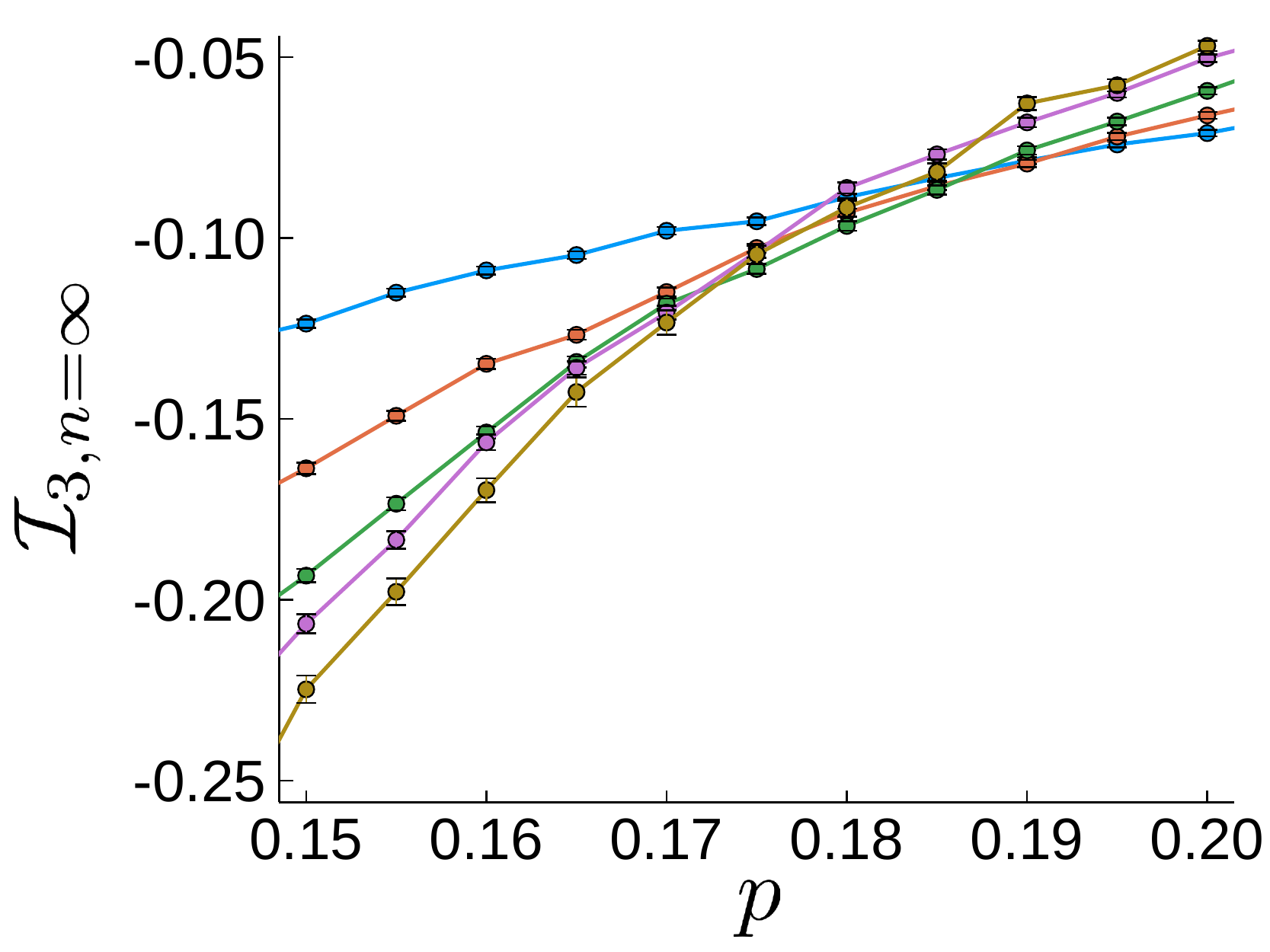}}
\subfloat[]{\includegraphics[width=.40\columnwidth]{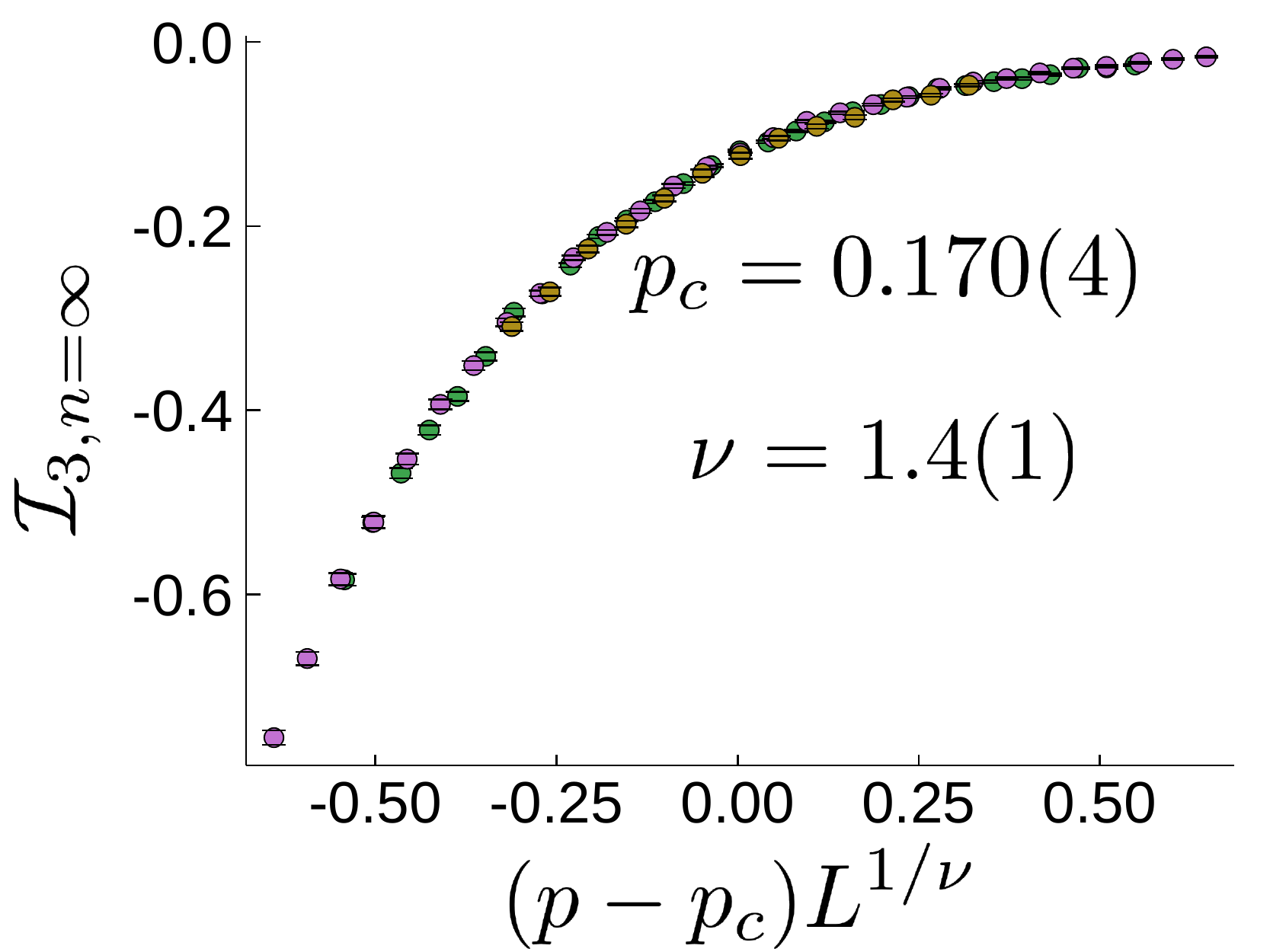}}
\caption{\emph{Tripartite mutual information near the transition.} The crossing of $\mathcal{I}_3$ for R\'enyi indices $n = 2, 5, \mbox{and} \infty$ is shown in (a), (c), and (e), respectively. The corresponding data collapse using the critical values determined from the finite size scaling analysis is shown in (b), (d), and (f).}
\label{fig:I3s}
\end{figure}

The data for the R\'enyi indices $n \gtrsim 0.1$ do not suffer from the numerical accuracy issues described in the previous section. This allows for the determination of an accurate value for the critical measurement rate, $p_c$. The main paper contains the results of the TMI and the data collapse from a finite size scaling analysis for $n=1$. In Fig. \ref{fig:I3s}, the other R\'enyi's ($n = 2,5,\infty$) are shown to have similar quality of data and collapse.

\section{Data Collapse}
In order to determine the critical measurement rate, $p_c$, and the critical exponent, $\nu$, we perform a finite size scaling analysis based on the procedure outlined by Kawashima and Ito \cite{Ito1993}. In summary, if we assume $\mathcal{I}_{3,n} = F\left[(p-p_c)L^{1/\nu}\right]$ for some arbitrary scaling function $F$, then we expect the data for different system sizes to collapse on each other for the appropriate choices of $p_c$ and $\nu$ when plotting $y = \mathcal{I}_{3,n}$ vs $x = (p-p_c)L^{1/\nu}$.

To judge the quality of collapse we study the objective function $O(p_c,\nu)$ defined as,
\begin{equation}
O(p_c,\nu) \equiv \frac{1}{n-2}\sum_{i=2}^{n-1}w(x_i,y_i,d_i\mid x_{i-1},y_{i-1},d_{i-1},x_{i+1},y_{i+1},d_{i+1}),
\label{eq:obj}
\end{equation}
where $x_i = (p_i - p_c)L_i^{1/\nu}$, $y_i = \mathcal{I}_{3,n,i}$, and $d_i = \sigma_{y_i}$ are the scaled data sorted such that $x_1 < x_2 < ... < x_n$. The quantity $w(x,y,d \mid x',y',d',x'',y'',d'')$ is defined as
\begin{align}
w &\equiv \left(\frac{y - \bar{y}}{\Delta(y-\bar{y})}\right)^2 \\
\bar{y} &\equiv \frac{(x''- x)y' - (x'-x)y''}{x''-x'} \\
\left[\Delta(y-\bar{y})\right]^2 &\equiv d^2 + \left( \frac{x''-x}{x''-x'}d'\right)^2 + \left( \frac{x'-x}{x''-x'}d''\right)^2.
\end{align}
Minimizing Eq. \ref{eq:obj} corresponds to minimizing the deviation of each point $(x_i,y_i)$ from the line determined by its adjacent points $(x_{i-1},y_{i-1})$ and $(x_{i+1},y_{i+1})$. For an ideal collapse, $O(p_c,\nu)$ attains its minimum value $=1$. 

\begin{figure}[htbp]
\centering
\includegraphics[width=.40\columnwidth]{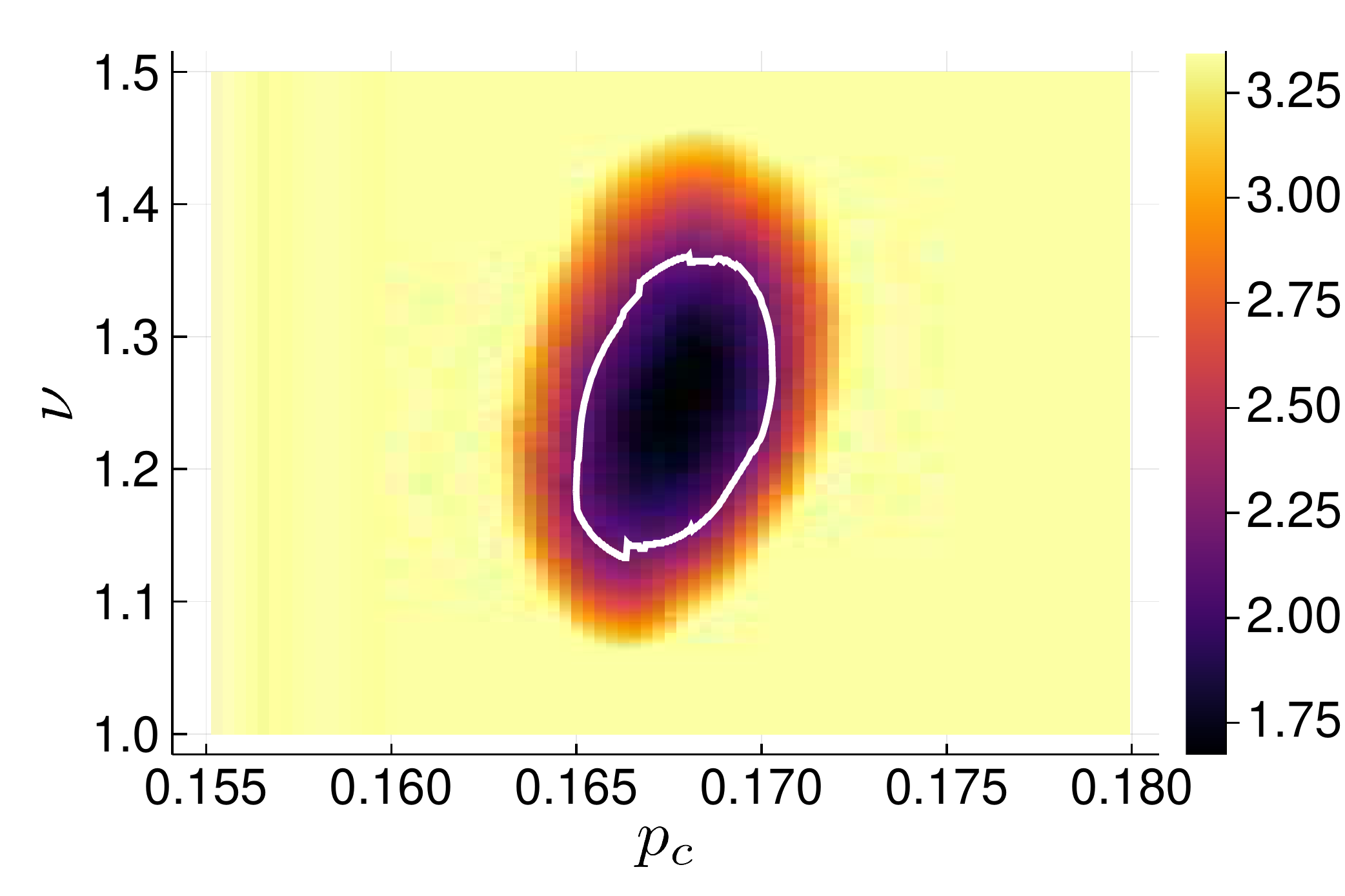}
\caption{\emph{Color plot of the objective function $O(p_c,\nu)$.} The critical values are determined from the global minimum of the objective function $O(p_c,\nu)$. To estimate the error bars in the critical values we take a region (outlined in white) around the minimum value of the objective function such that $O(p_c,\nu) < 1.3\cdot O(p_{c}^*,\nu^*)$.}
	\label{fig:obj}
\end{figure}

\begin{figure}[htbp]
\centering
\subfloat[]{\includegraphics[width=.40\columnwidth]{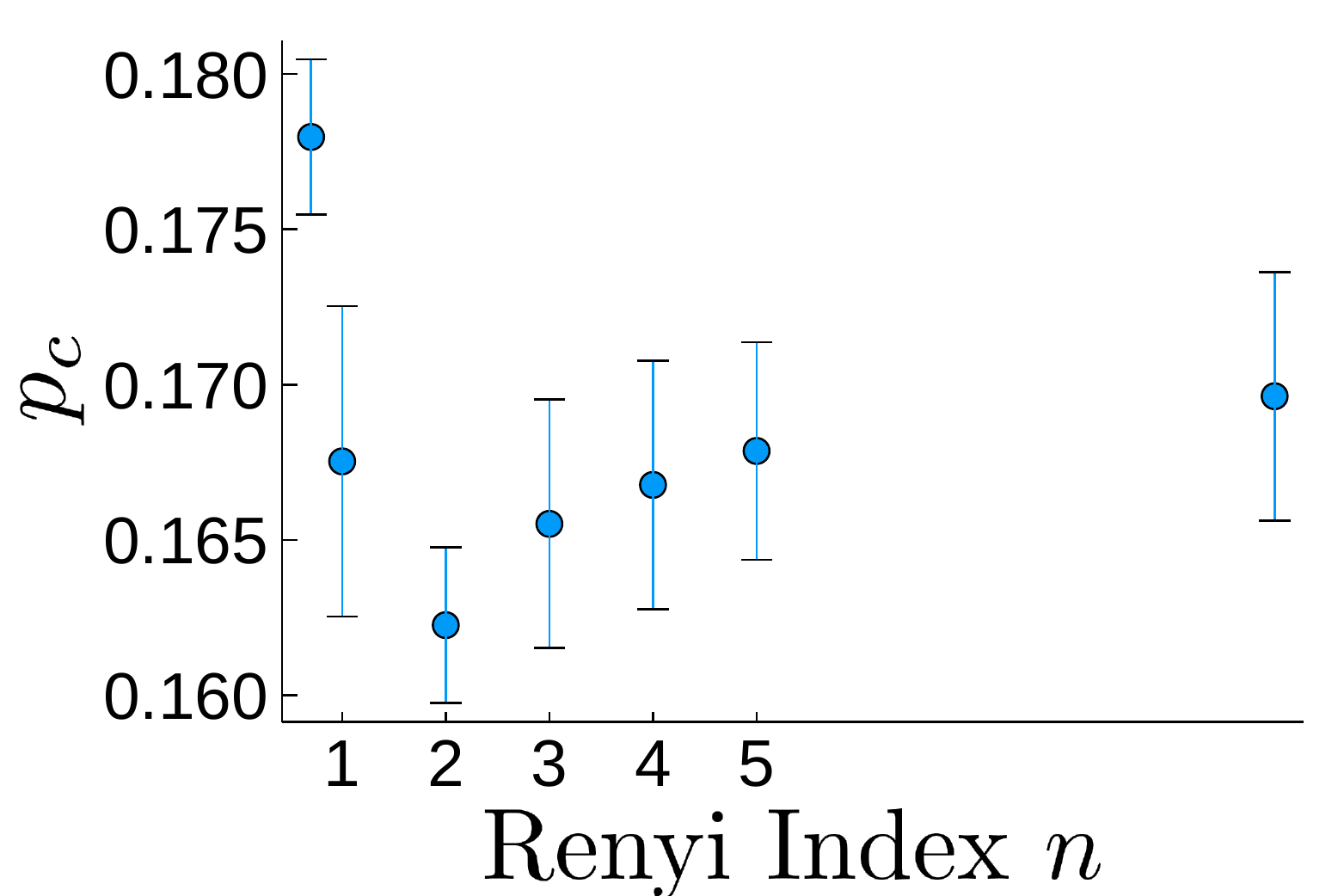}}
\subfloat[]{\includegraphics[width=.40\columnwidth]{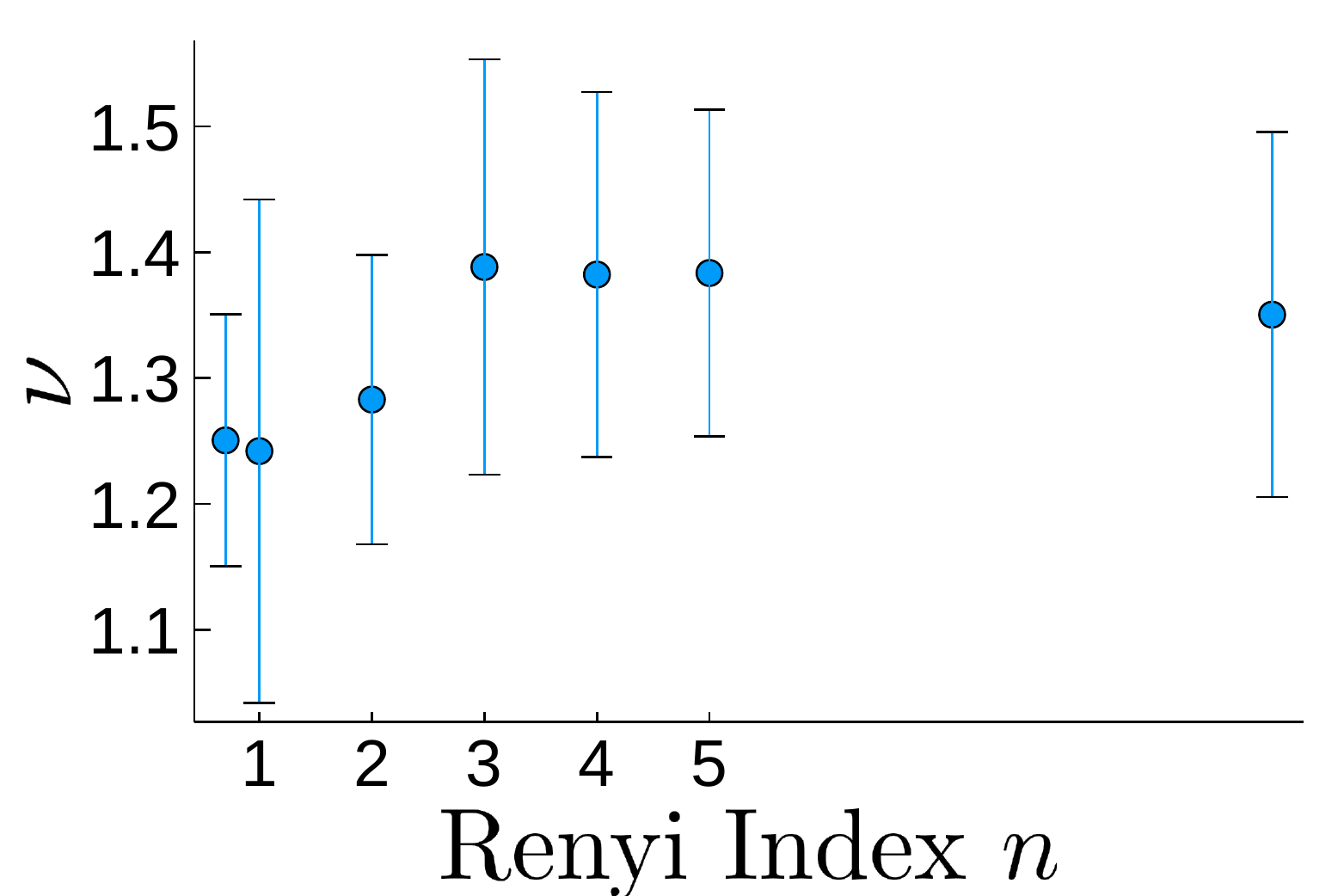}}
\caption{\emph{R\'enyi index dependence of the critical values.} The critical values $p_c$ and $\nu$ obtained from the global minimum of the objective function $O(p_c,\nu)$. The unlabeled data point corresponds to $n \rightarrow \infty$.}
\label{fig:crit}
\end{figure}

For a given dataset, the value of Eq. \ref{eq:obj} can be plotted for different values of $p_c$ and $\nu$ as shown in Fig. \ref{fig:obj}. The estimate of the critical values is given by the location of the global minimum: $(p_c^*,\nu^*)$. To estimate the error in the values  $p_c^*$ and $\nu^*$, a region around the minimum value is taken such that $O(p_c,\nu) \leq 1.3\cdot O(p_c^*,\nu^*)$, shown as the white contour. Repeating the procedure for the different R\'enyi indices results in the critical values shown in Fig. \ref{fig:crit}.

\section{Bipartite Mutual Information}
\begin{figure}[htbp]
\subfloat[]{\includegraphics[width=.30\columnwidth]{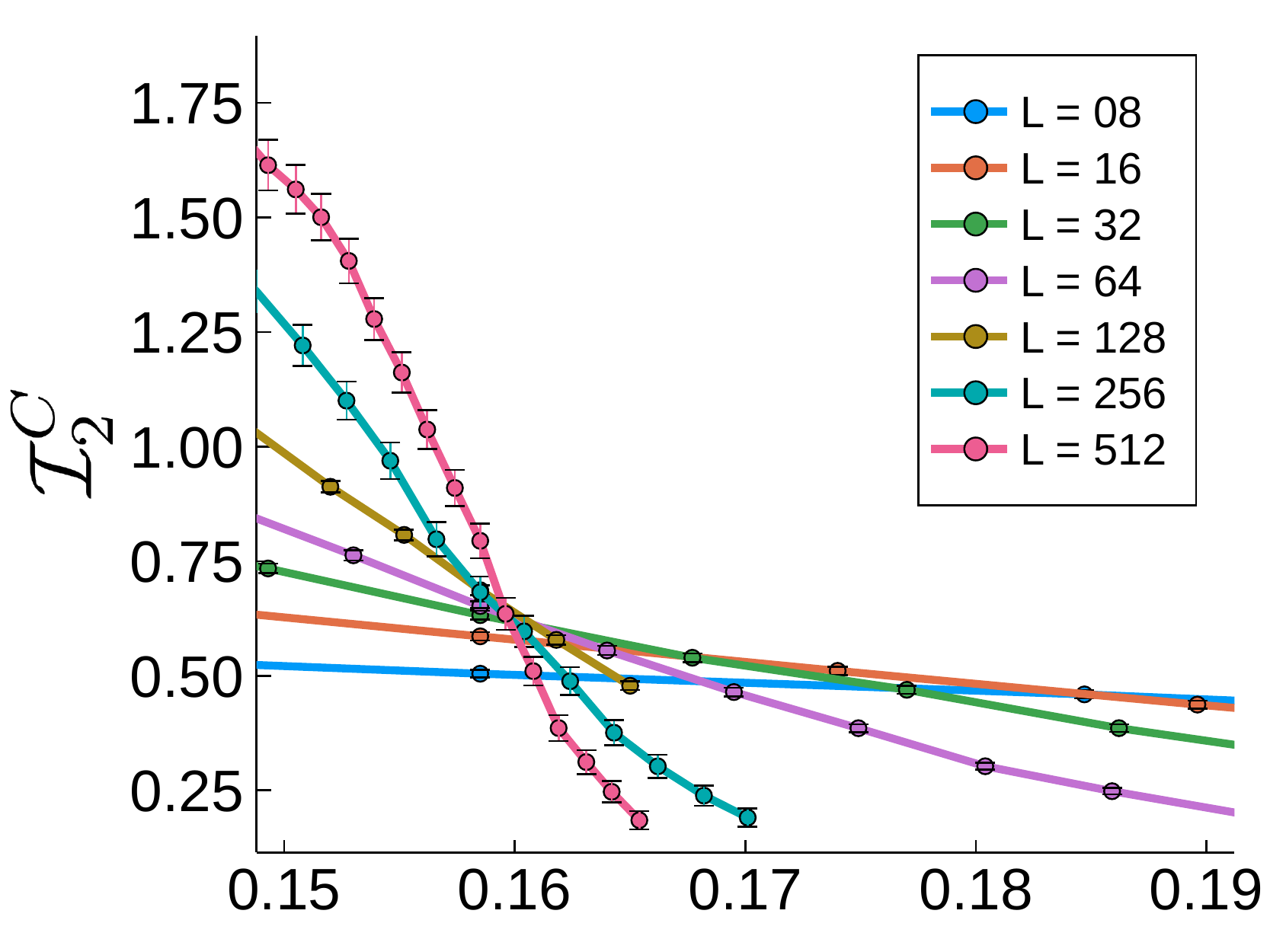}}
\subfloat[]{\includegraphics[width=.30\columnwidth]{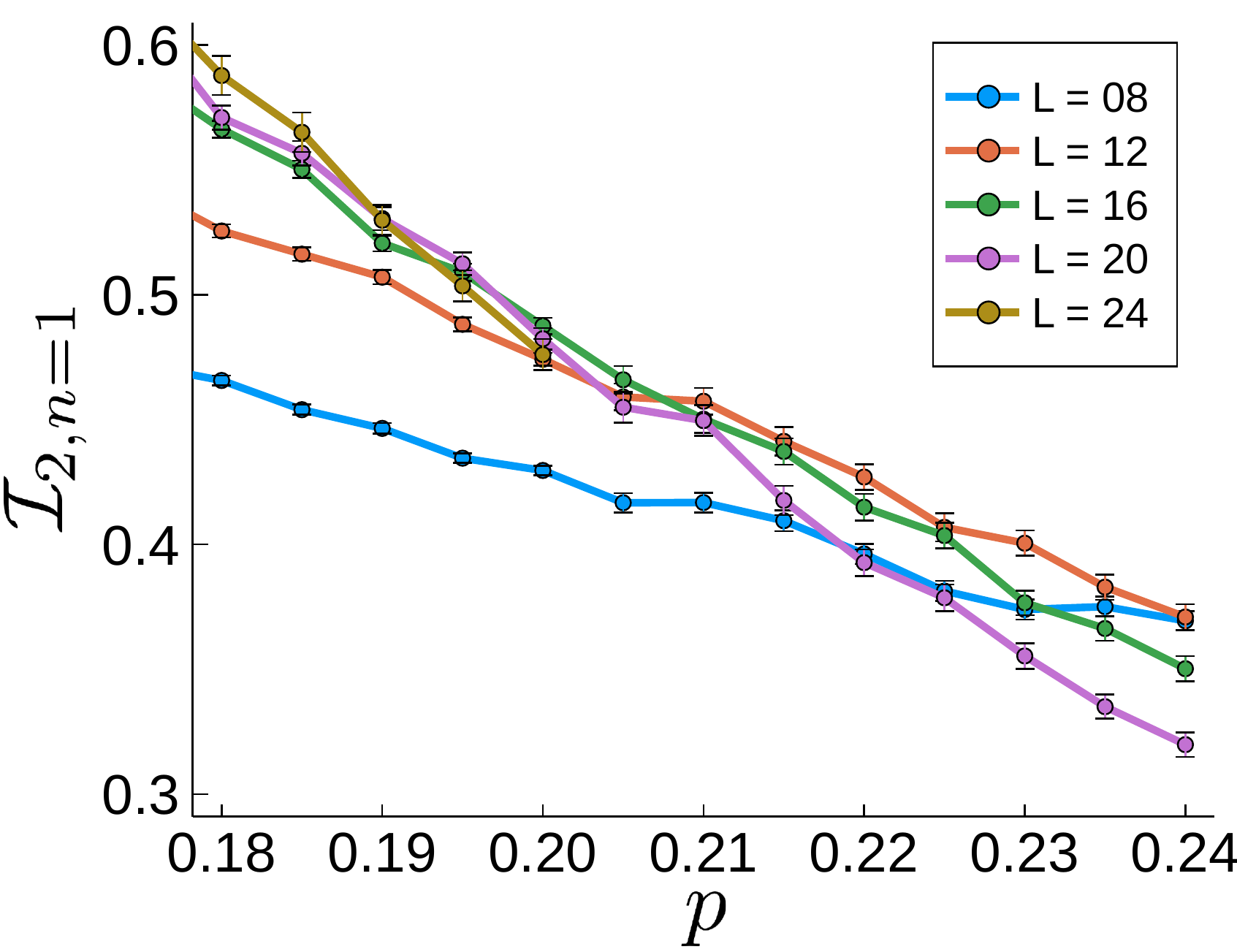}}\\
\subfloat[]{\includegraphics[width=.30\columnwidth]{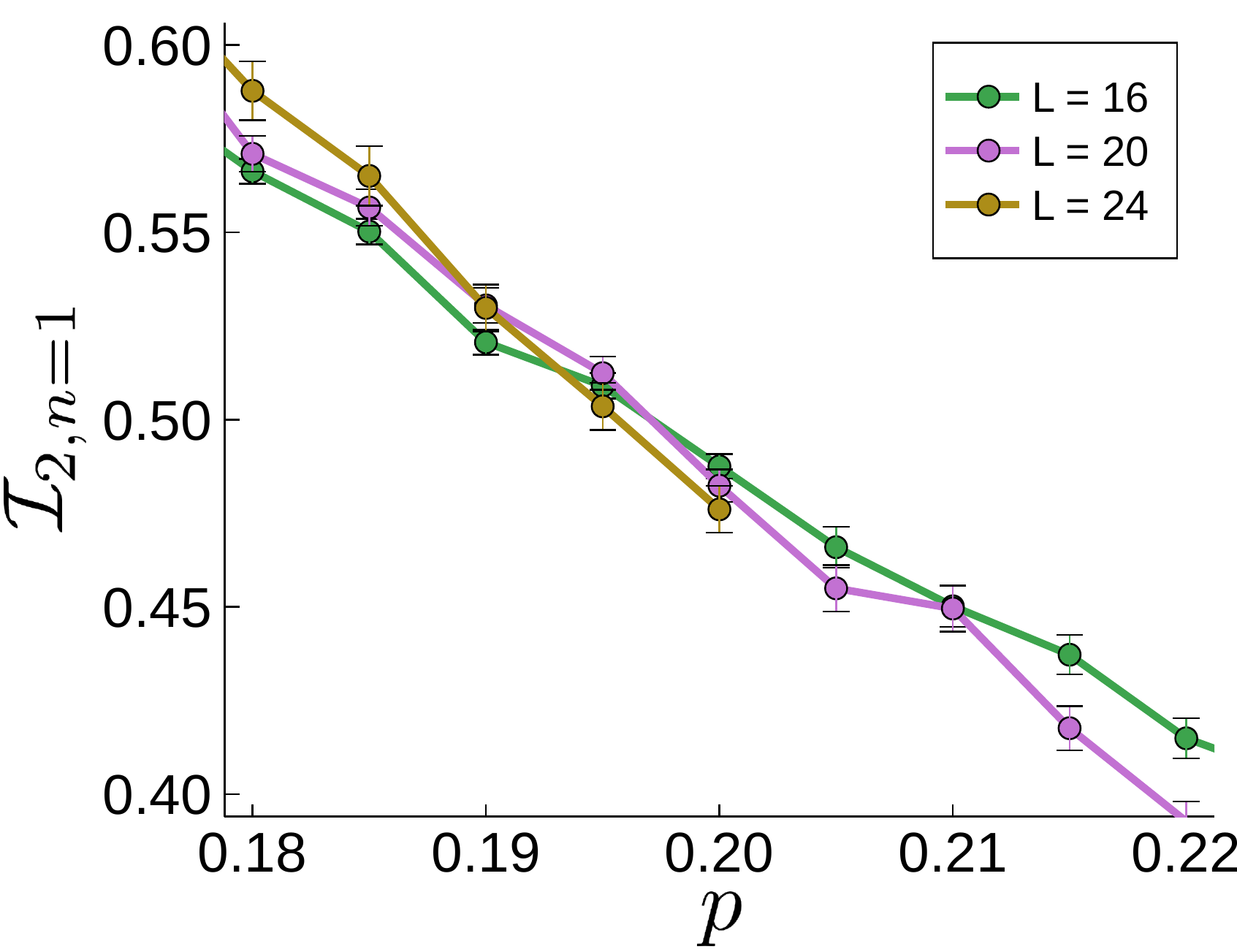}}
\subfloat[]{\includegraphics[width=.30\columnwidth]{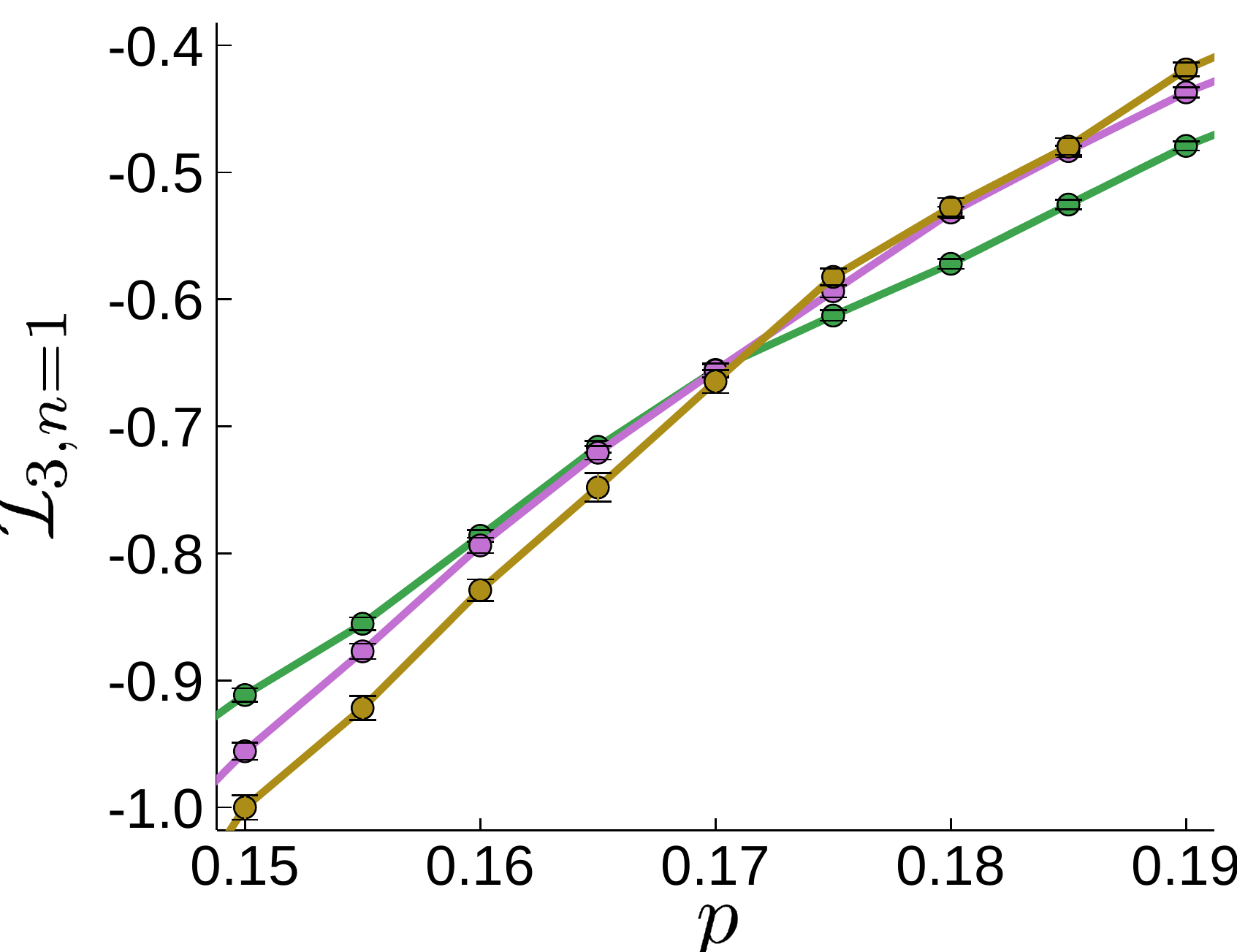}}
\caption{\emph{Mutual information crossing.} By studying the stabilizer circuit, (a), we find the bipartite mutual information between antipodal regions of size $L/4$ contains a crossing at $p_c$ that drifts heavily for small system sizes. The Haar circuit shows similar finite size drifts in the bipartite mutual information, (b) and (c), that are absent in the (d) tripartite mutual information.}
\label{fig:crossComp}
\end{figure}

The bipartite mutual information, $\mathcal{I}_{2,n=1}$, between antipodal regions of size $L/4$ contains a crossing at $p_c$ but suffers from stronger finite size drifts than $\mathcal{I}_{3,n=1}$ as shown in Fig. \ref{fig:crossComp}. This is consistent with the behavior seen in the stabilizer circuit when compared out to larger sizes up to $L=512$.

\section{Dynamical Exponent $z$}
\begin{figure}[htbp]
\centering
\subfloat[]{\includegraphics[width=.40\columnwidth]{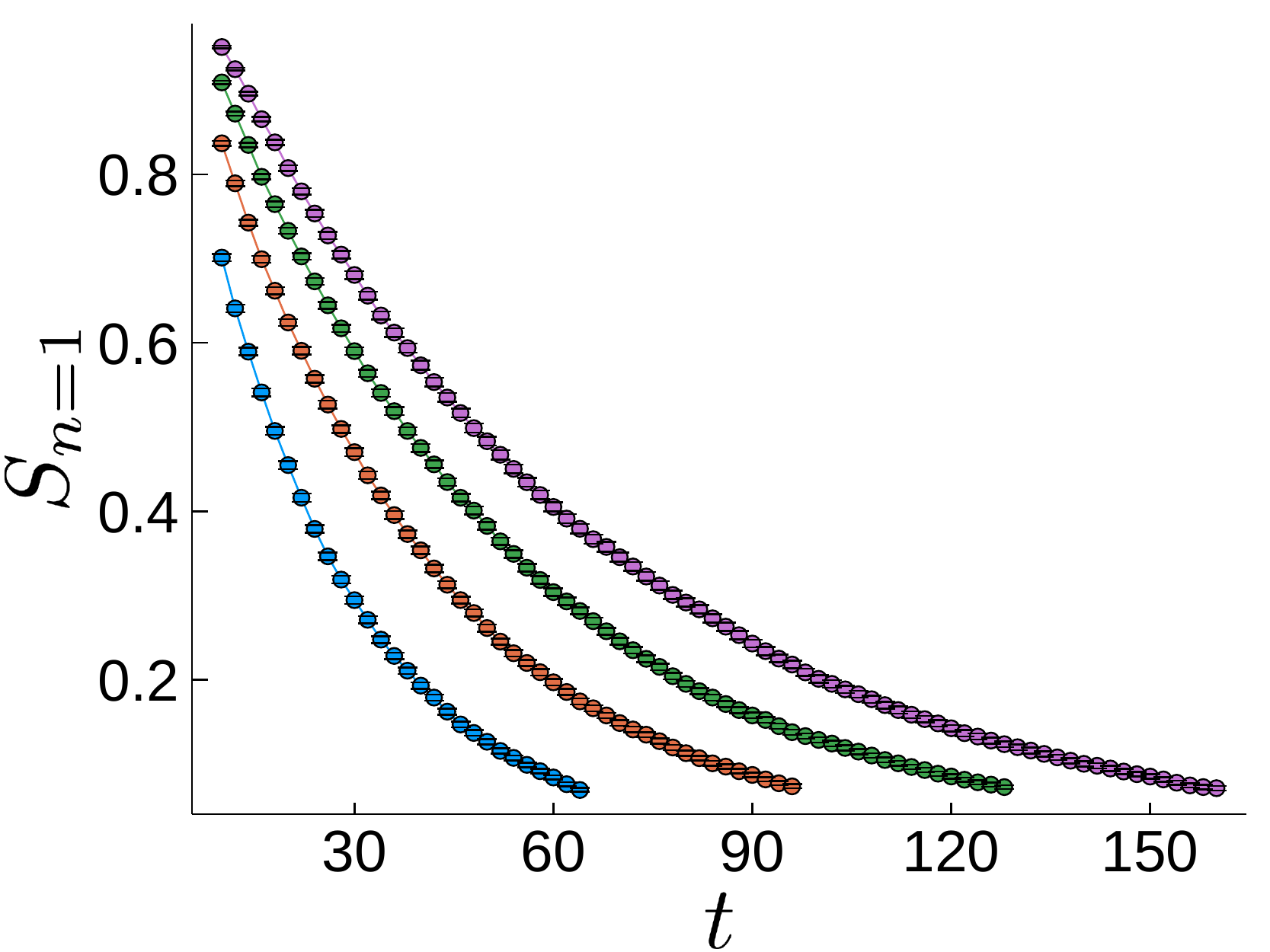}}
\subfloat[]{\includegraphics[width=.40\columnwidth]{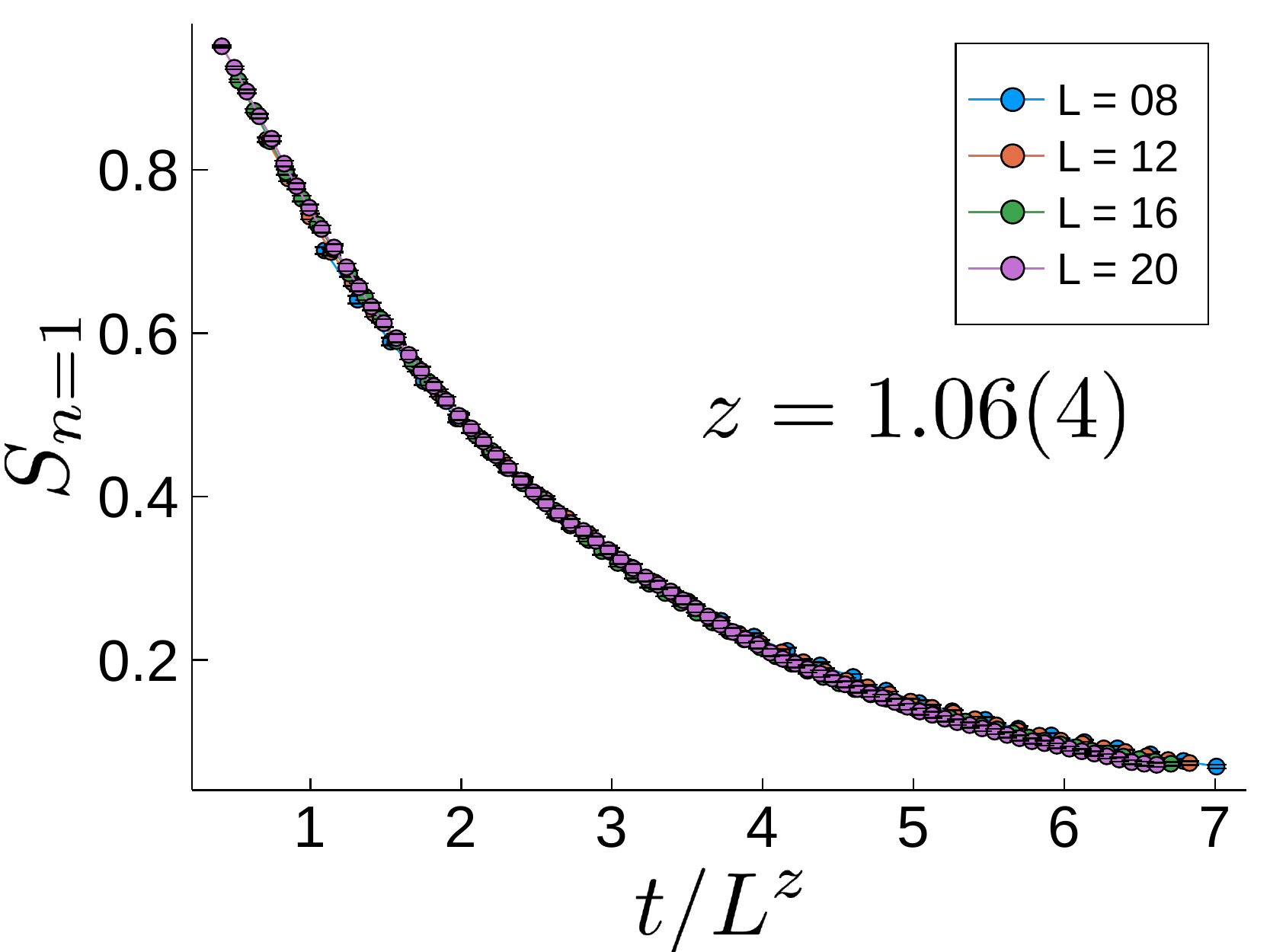}}
\caption{\emph{Dynamical exponent.} (a) The entanglement entropy decay as a function of time for an ancilla maximally entangled with a qubit in the circuit at $p = p_c$. (b) Through data collapse of the entanglement entropy we find the dynamical exponent $z = 1.06(4)$ which is close to the value of $z=1$ for conformal invariance.}
	\label{fig:z}
\end{figure}

To determine the dynamical exponent, $z$, we construct a circuit with periodic boundary conditions starting from an entangled Haar state such that an ancilla is maximally entangled with a qubit in the spin chain. We fix the measurement rate $p = p_c = 0.17$ and calculate the entanglement entropy decay of the ancilla as a function of time as shown in Fig. \ref{fig:z}. The entanglement entropy should scale as $S(t,L)\sim F(t/L^z)$ for an arbitrary scaling function $F$. From data collapse of system sizes $L = 12,16,20$, we find $z = 1.06(4)$ which is close to the value of $z=1$ for conformal invariance \cite{Ludwig2019,LCF2019}.

\section{Order Parameter}
\begin{figure}[htbp]
\centering
\subfloat[]{\includegraphics[width=.40\columnwidth]{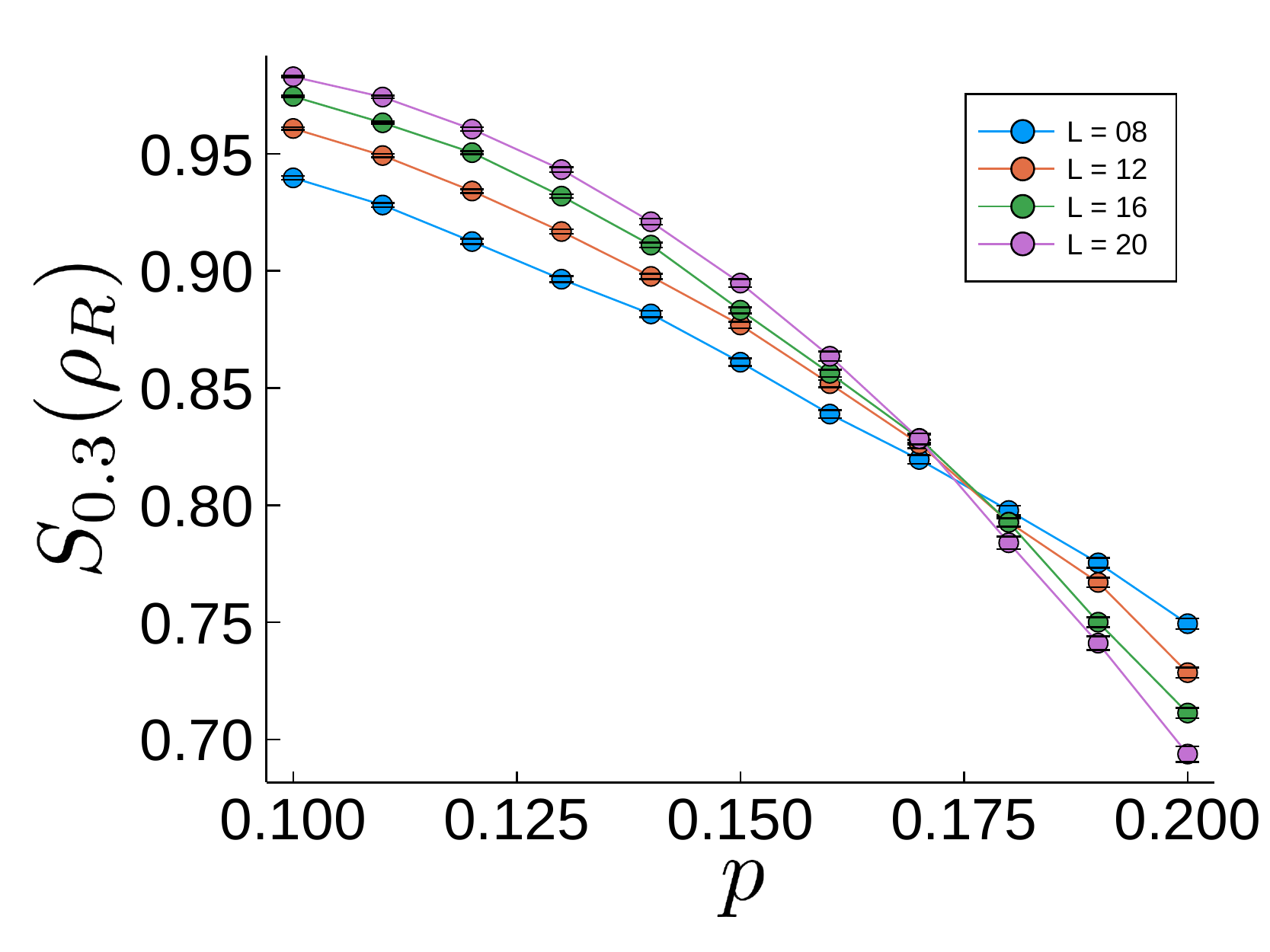}}
\subfloat[]{\includegraphics[width=.40\columnwidth]{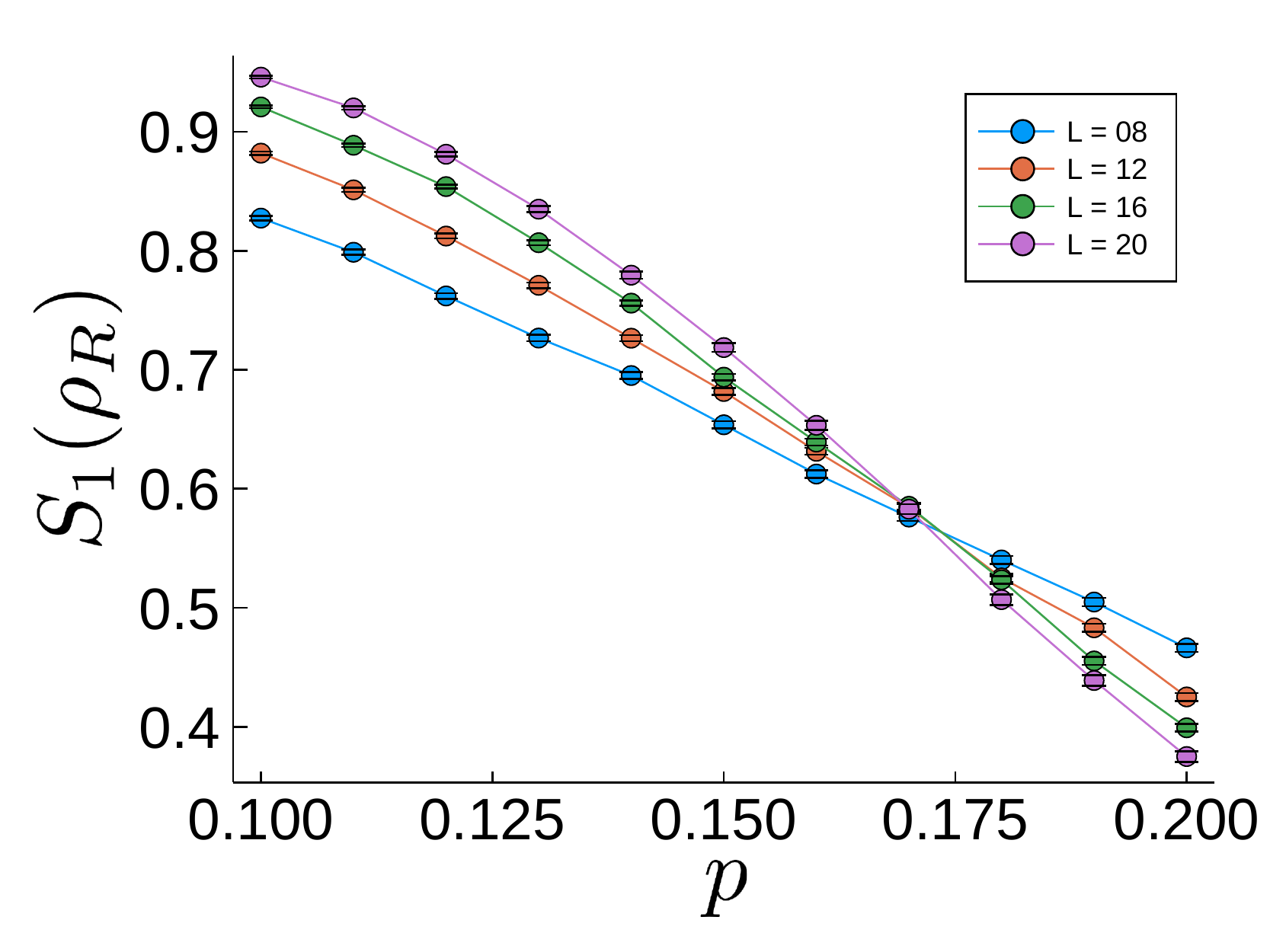}}
\caption{\emph{Order parameter.} The entanglement entropy of the reference qubit behaves as an order parameter for the transition. A crossing at similar $p_c$ is observed for the R\'enyi indices (a) $n = 0.3$ and (b) $n = 1$ indicating that all $n>0$ are described by the same transition.}
	\label{fig:op}
\end{figure}

An order parameter can be defined for the transition by introducing a reference qubit that is maximally entangled with a single site in the system and calculating its entanglement entropy. In Fig. \ref{fig:op} a crossing at similar values  of $p_c$ is observed for the R\'enyi indices (a) $n = 0.3$ and (b) $n = 1$ indicating that all $n>0$ are described by the same transition.

\section{Correlation Function}
The correlation function $C(t-t_0) = I_1(\tilde A,\tilde B) = S_1(\tilde A) + S_1(\tilde B)- S_1(\tilde A\cup \tilde B)$ is defined as the mutual information between two ancilla qubits entangled with the system at time $t_0$. The correlation function obeys the critical finite-size scaling form $C(t-t_0)\sim L^{-\eta}g((t-t_0)/L)$, where the exponent $\eta$ depends on the different boundary conditions. 

In the case of our numerical simulations we define the bulk exponent $\eta$ from the circuit with periodic boundary conditions starting from a product state. The circuit is run to a time $t_0 = 2L$ and the ancillas are maximally entangled with antipodal spins.  This corresponds to a two-point order parameter correlation function on a cylinder with complex coordinates for the two ancilla $(z_1',z_2')=(0,iL/2)$, which can be related to standard correlation functions in the complex plane through the conformal mapping $z = e^{2 \pi z'/L}$.  

\begin{figure}[htbp]
\centering
\includegraphics[width=0.75 \textwidth]{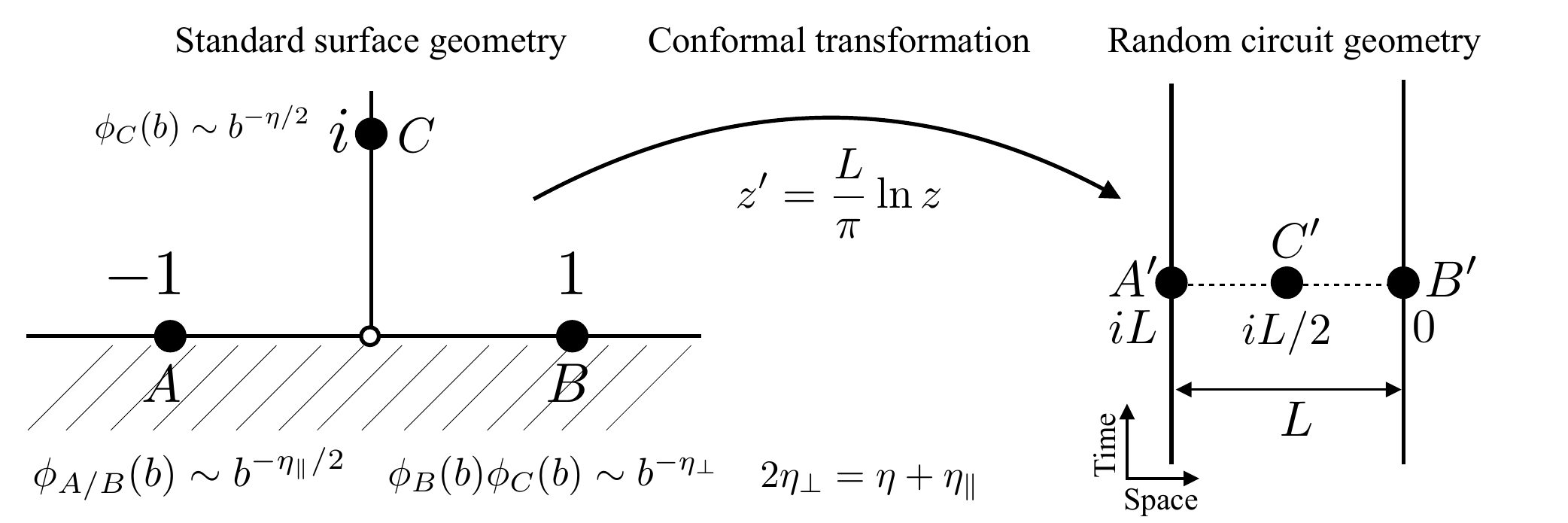}
\caption{\emph{Surface exponent definition.}  The standard surface geometry for defining surface exponents in a 2D system \cite{Binder83,Cardy84} can be related to the geometries we consider in the random circuit through the conformal transformation $z' = (L/\pi)\ln z$.  Here, $\phi_{A/B/C}(b)$ denotes the order parameter operator at scale $b$.}
\label{fig:conf}
\end{figure}

In Fig.~\ref{fig:conf}, we show how the surface exponents are related to different correlation functions in the random circuit geometry with open boundary conditions.  Here, $\phi_{A/B/C}(b)$ denotes the order parameter operator at scale $b$.  The scaling dimension in the bulk is $\eta/2$, the scaling dimension at the surface is $\eta_{\parallel}/2$, and the scaling dimension for two-point functions between the surface and bulk is given by
$\eta_{\perp}$ that satisfies the scaling relation $2 \eta_{\perp} = \eta + \eta_{\parallel}$ \cite{Binder83,Cardy84}.  In the random circuit geometry, the surface exponent $\eta_\parallel$ is given by a circuit with open boundary conditions starting from a product state.    The circuit is run to a time $t_0 = 2L$ and the ancillas are maximally entangled with the edge spins so that $(z_1',z_2')=(0,iL)$.  Another estimate of $\eta_\parallel$ is given by a circuit with periodic boundary conditions starting from a product state where at time $t_0 = 0$ an ancilla is maximally entangled with a spin in the system. The latter has non-universal early time dynamics that make it difficult extract $\eta_\parallel$. The surface exponent $\eta_\perp$ is given by a circuit with open boundary conditions starting from a product state. The circuit is run to a time $t_0 = 2L$ and the ancillas are maximally entangled with the edge and middle spins so that $(z_1',z_2')=(0,iL/2)$. 

\begin{figure}[htbp]
\centering
\subfloat[]{\includegraphics[width=.30\columnwidth]{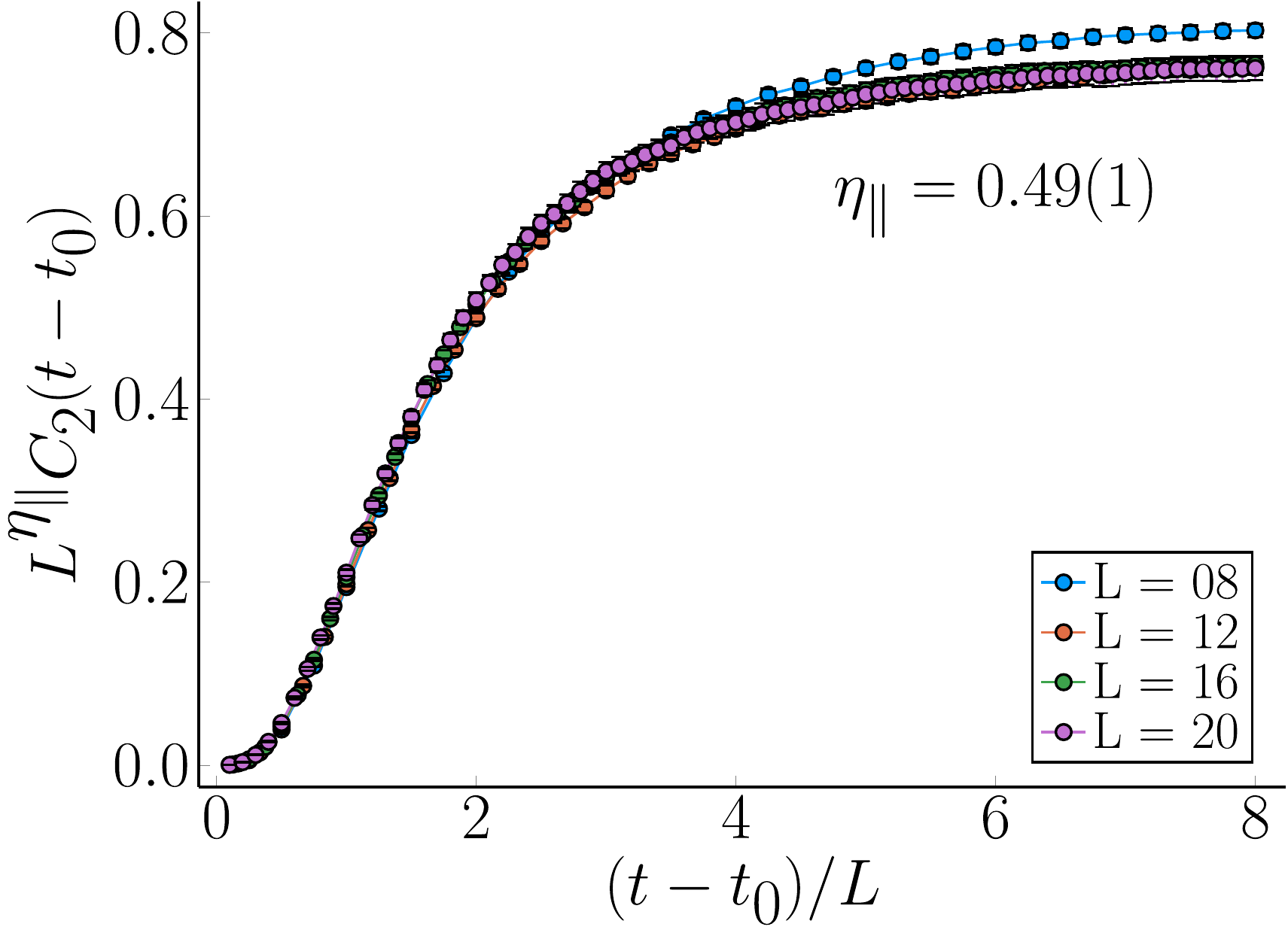}}
\subfloat[]{\includegraphics[width=.30\columnwidth]{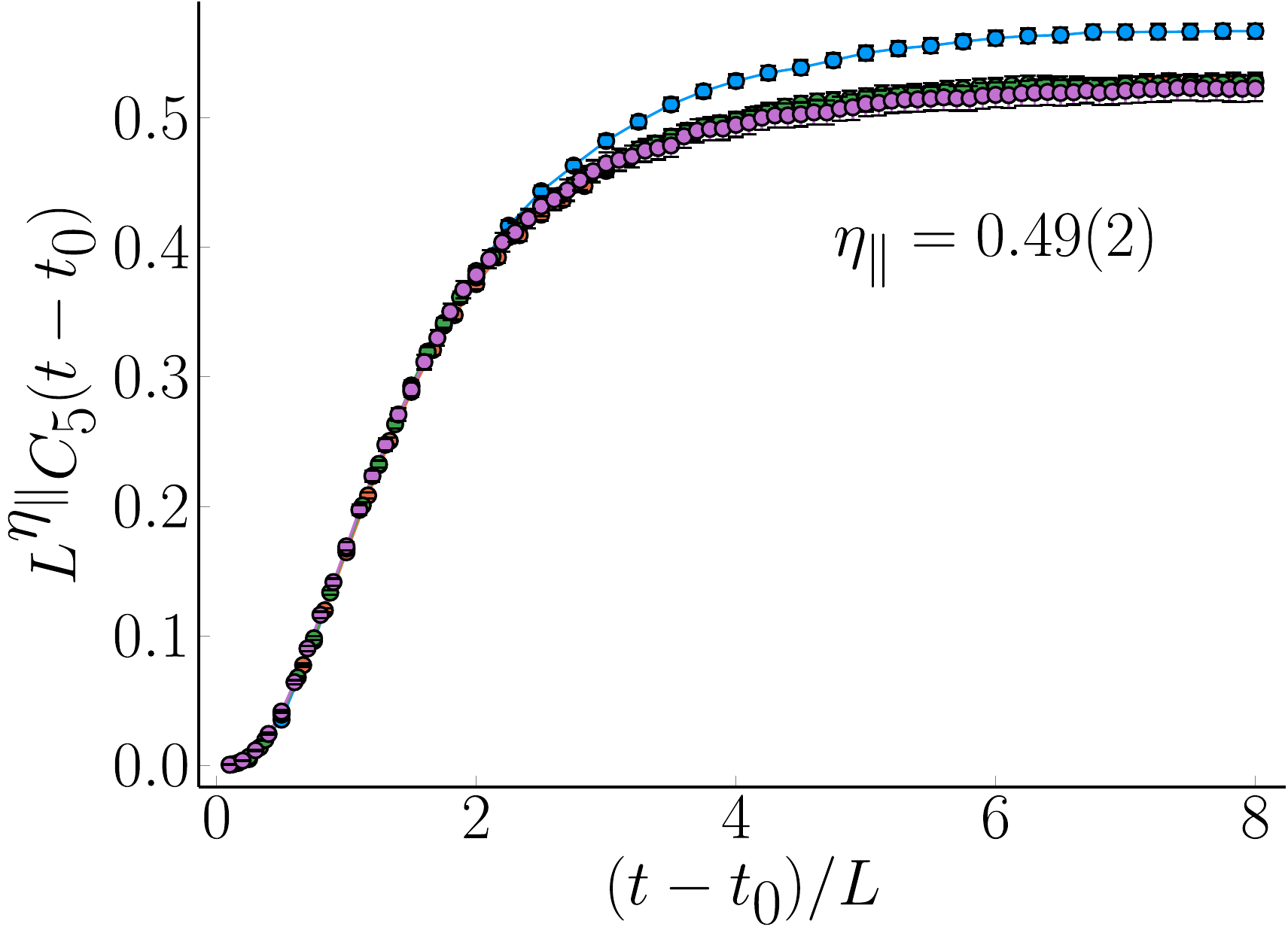}}
\subfloat[]{\includegraphics[width=.30\columnwidth]{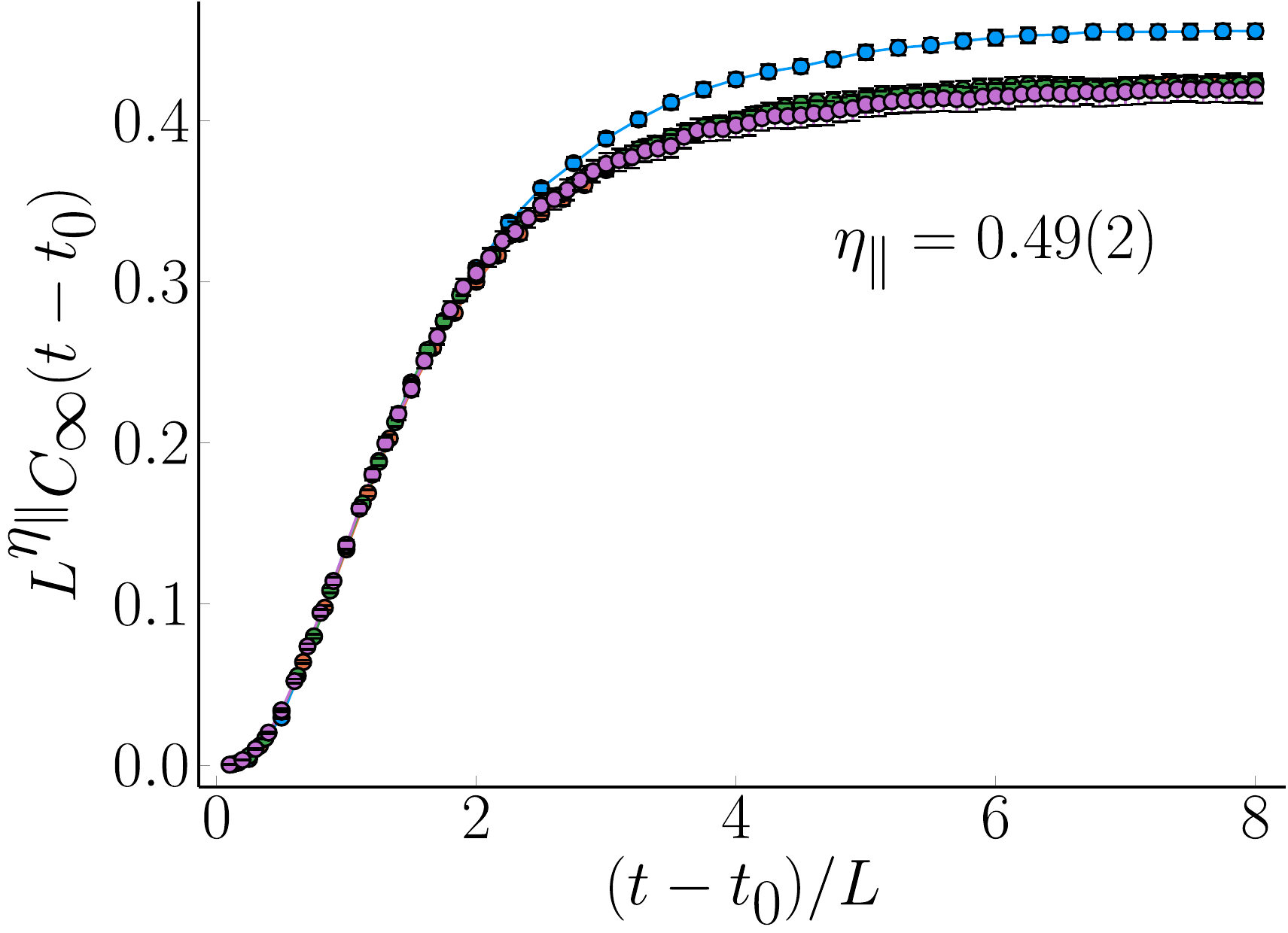}}
\caption{\emph{Surface exponent $\eta_\parallel$.} The surface $\eta_\parallel$ is given by a circuit with open boundary conditions starting from a product state. The circuit is run to a time $t_0 = 2L$ and the ancillas are maximally entangled with the edge spins.}
\label{fig:etaPara}
\end{figure}

\begin{figure}[htbp]
\centering
\subfloat[]{\includegraphics[width=.30\columnwidth]{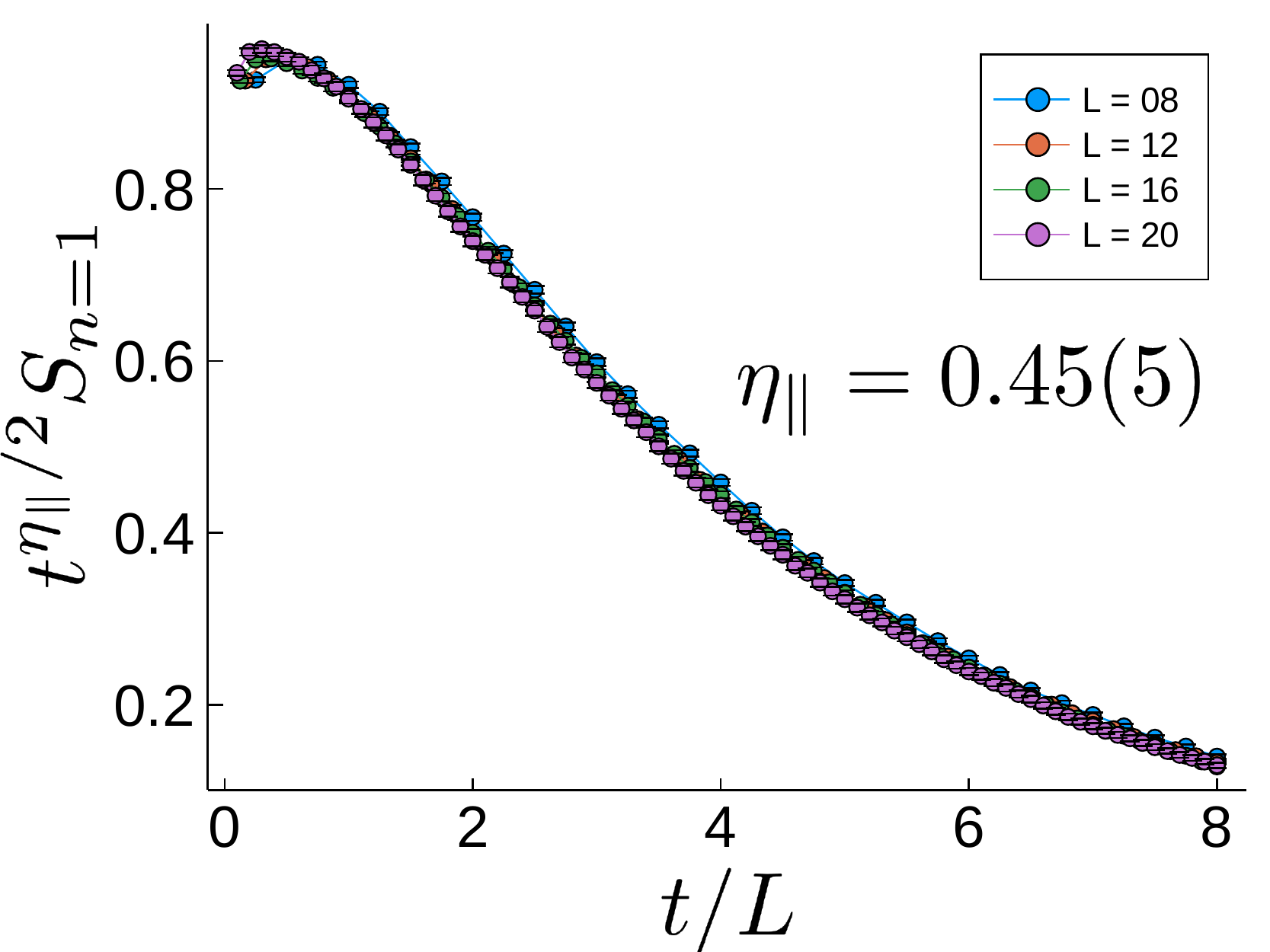}}
\subfloat[]{\includegraphics[width=.30\columnwidth]{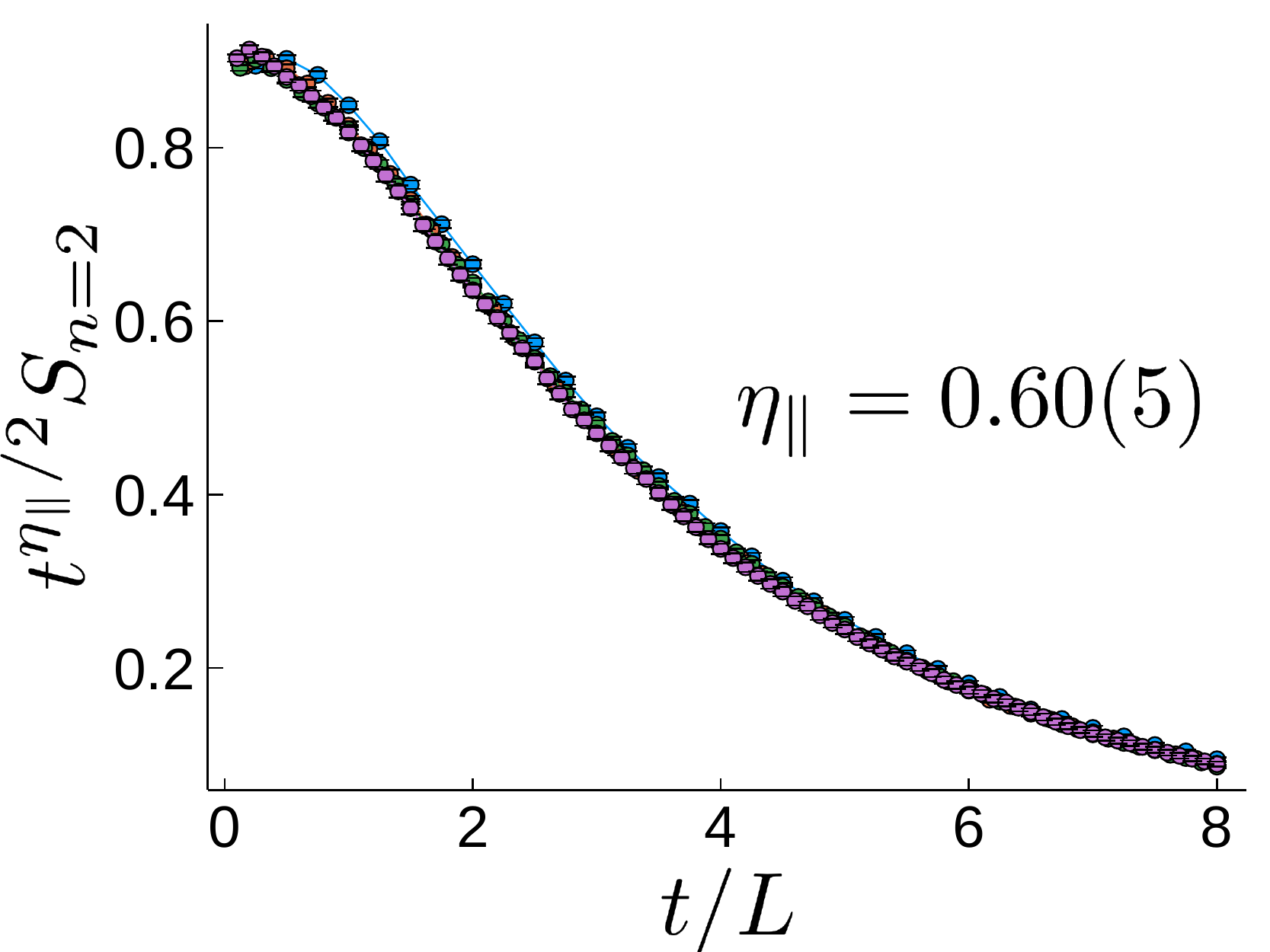}}
\subfloat[]{\includegraphics[width=.30\columnwidth]{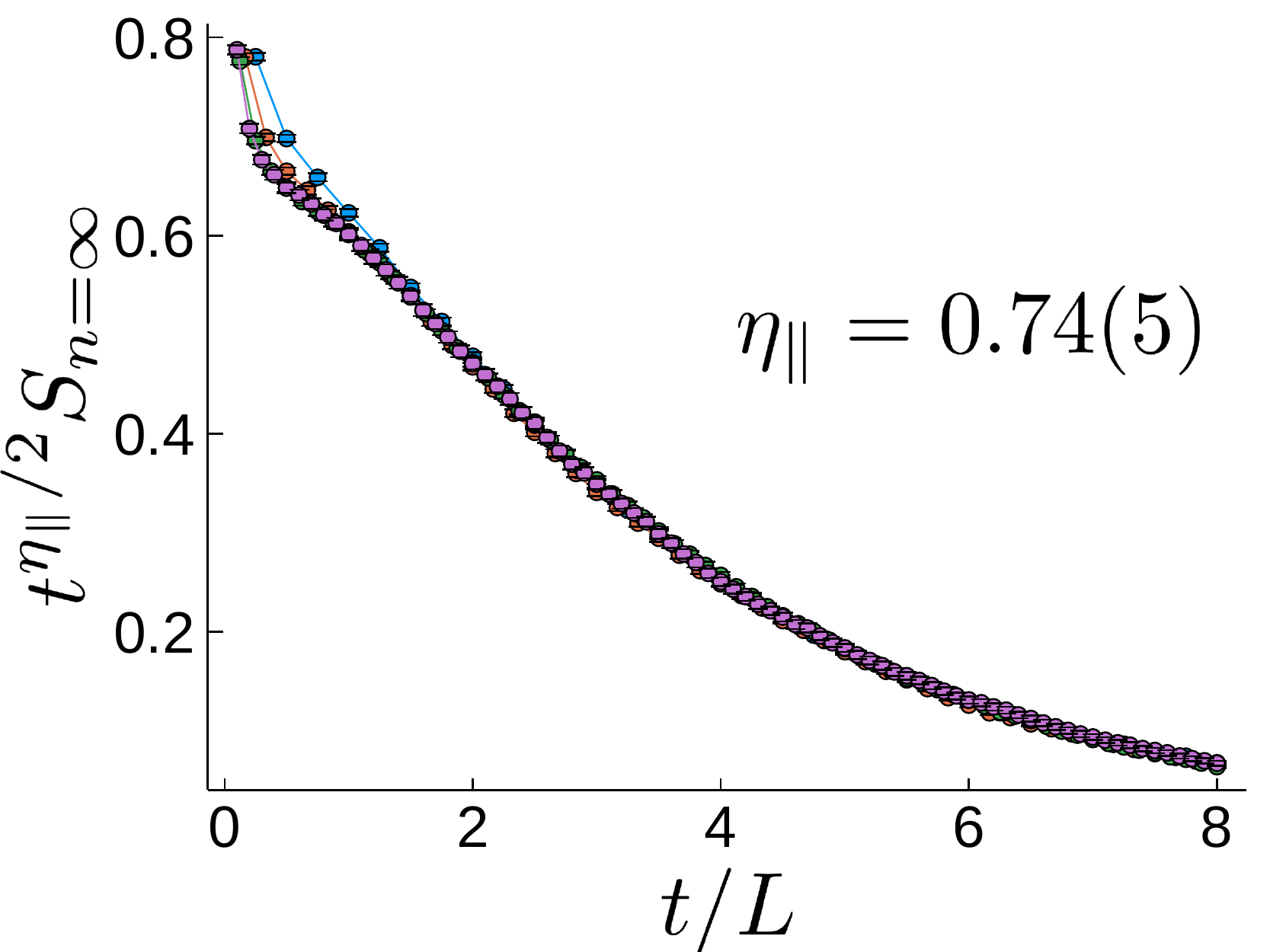}}\\
\subfloat[]{\includegraphics[width=.30\columnwidth]{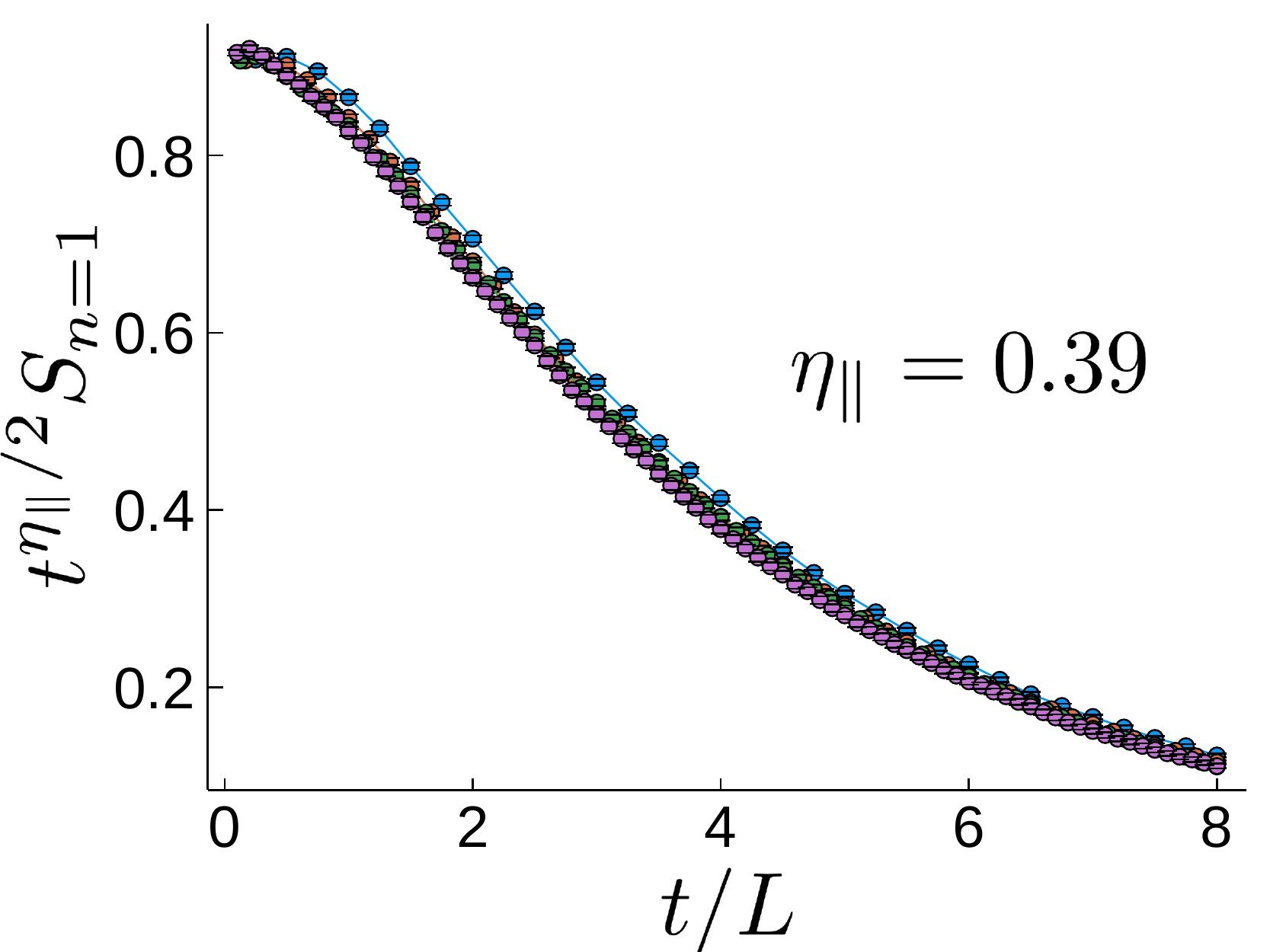}}
\subfloat[]{\includegraphics[width=.30\columnwidth]{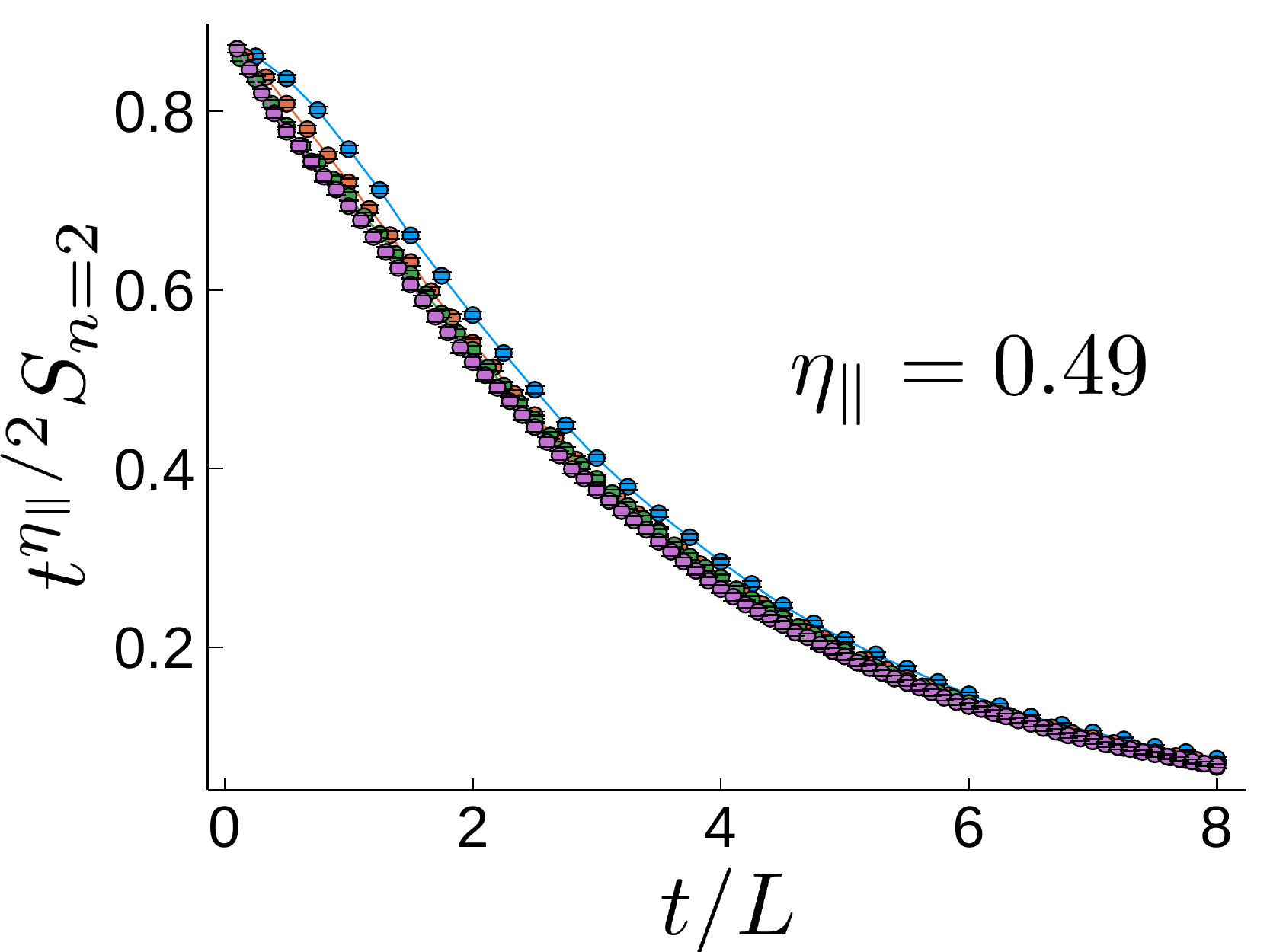}}
\subfloat[]{\includegraphics[width=.30\columnwidth]{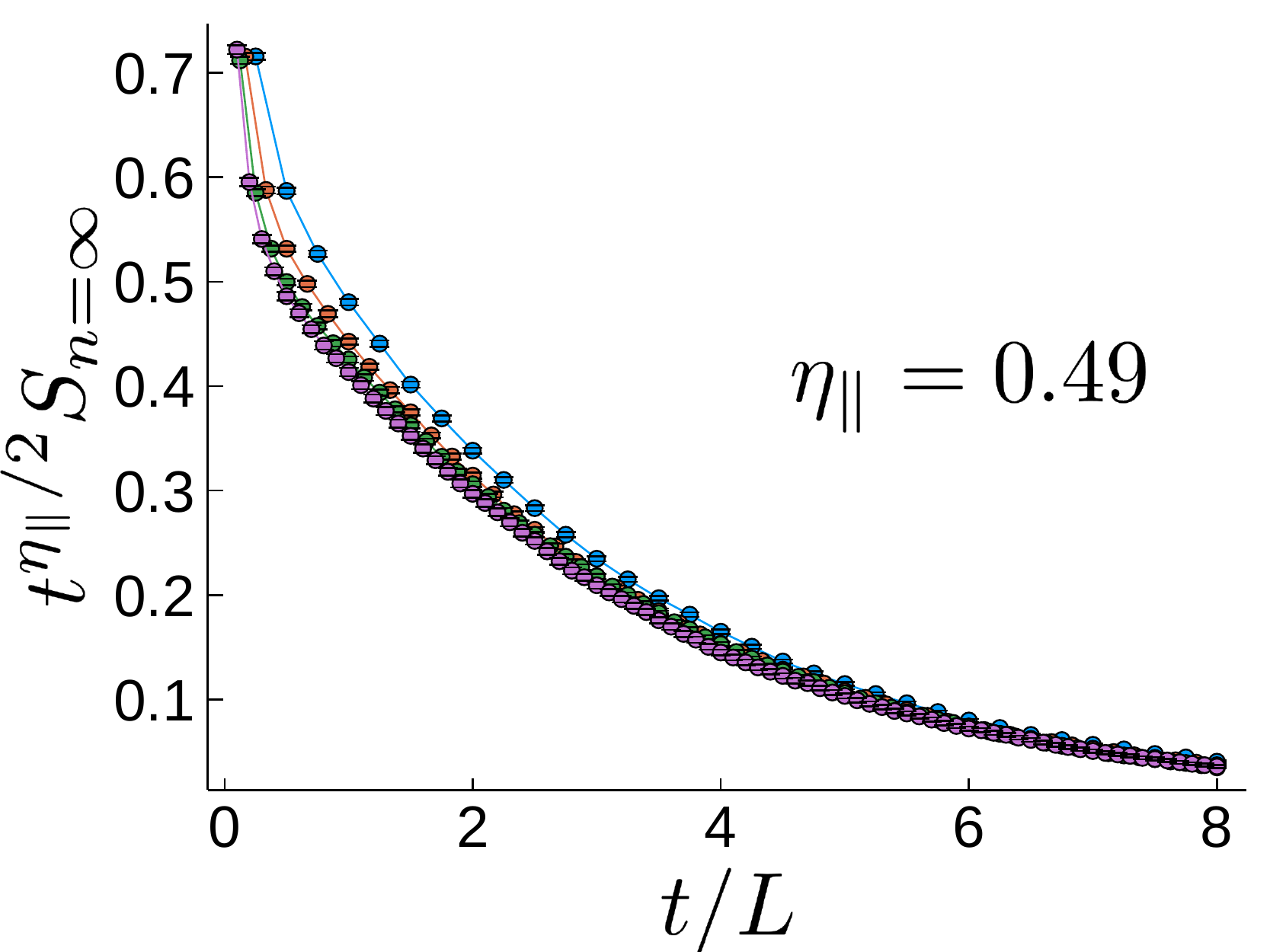}}
\caption{\emph{Surface exponent $\eta_\parallel$.} The surface $\eta_\parallel$ is also given by a circuit with periodic boundary conditions starting from a product state where at time $t_0 = 0$ an ancilla is placed in a maximally entangled state with a spin in the system. The collapse of the data for different R\'enyi indices $n = 1,2,$ and $\infty$ is shown in (a), (b), and (c), respectively. These results for $\eta_\parallel$ are different than the estimate used in the main text due to non-universal early time dynamics. However, at late times, the data collapses using the values of $\eta_\parallel$ from the main text, as shown in (d), (e), and (f).}
\label{fig:etaPara2}
\end{figure}

\begin{figure}[htbp]
\centering
\subfloat[]{\includegraphics[width=.30\columnwidth]{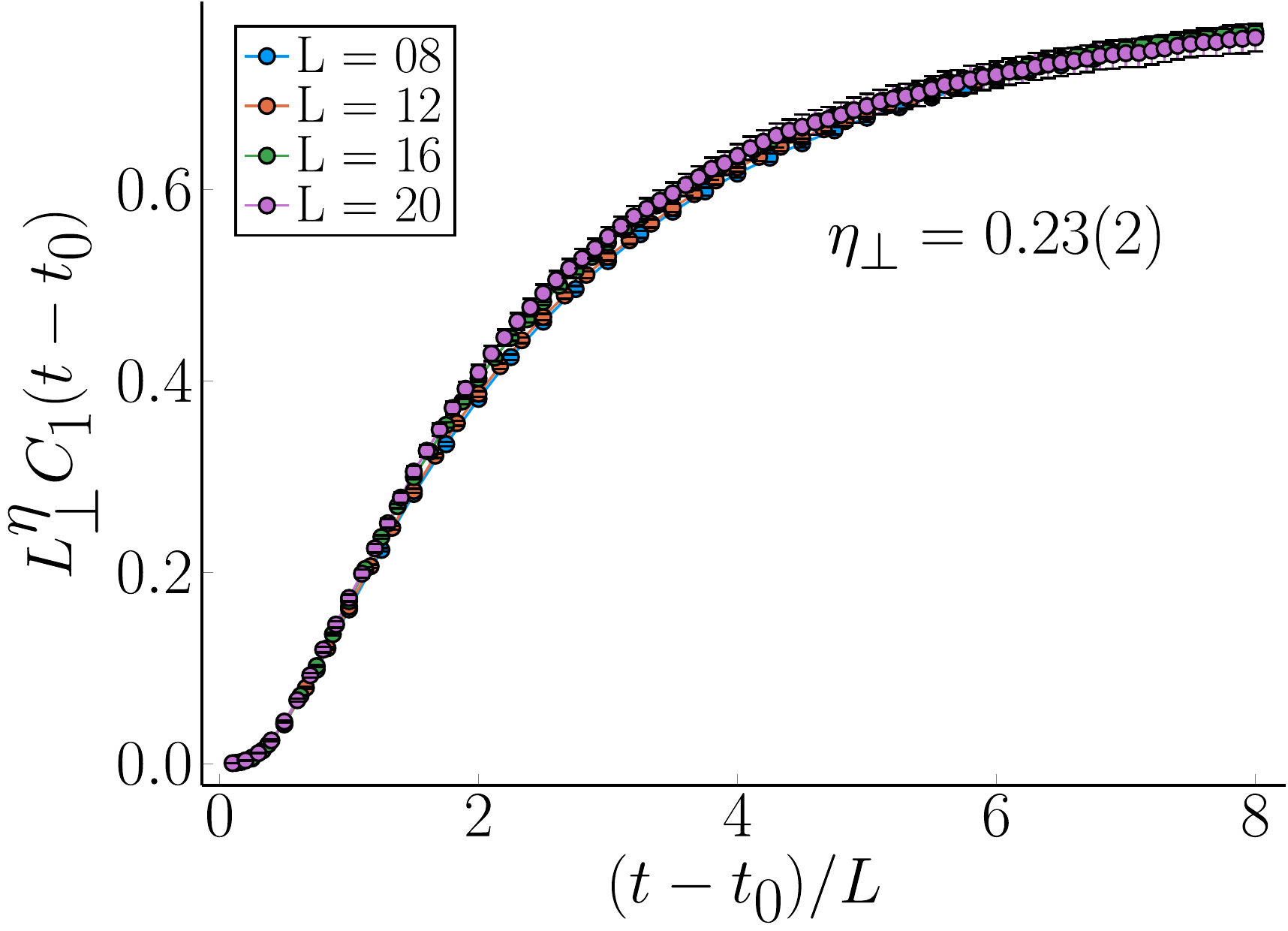}}
\subfloat[]{\includegraphics[width=.30\columnwidth]{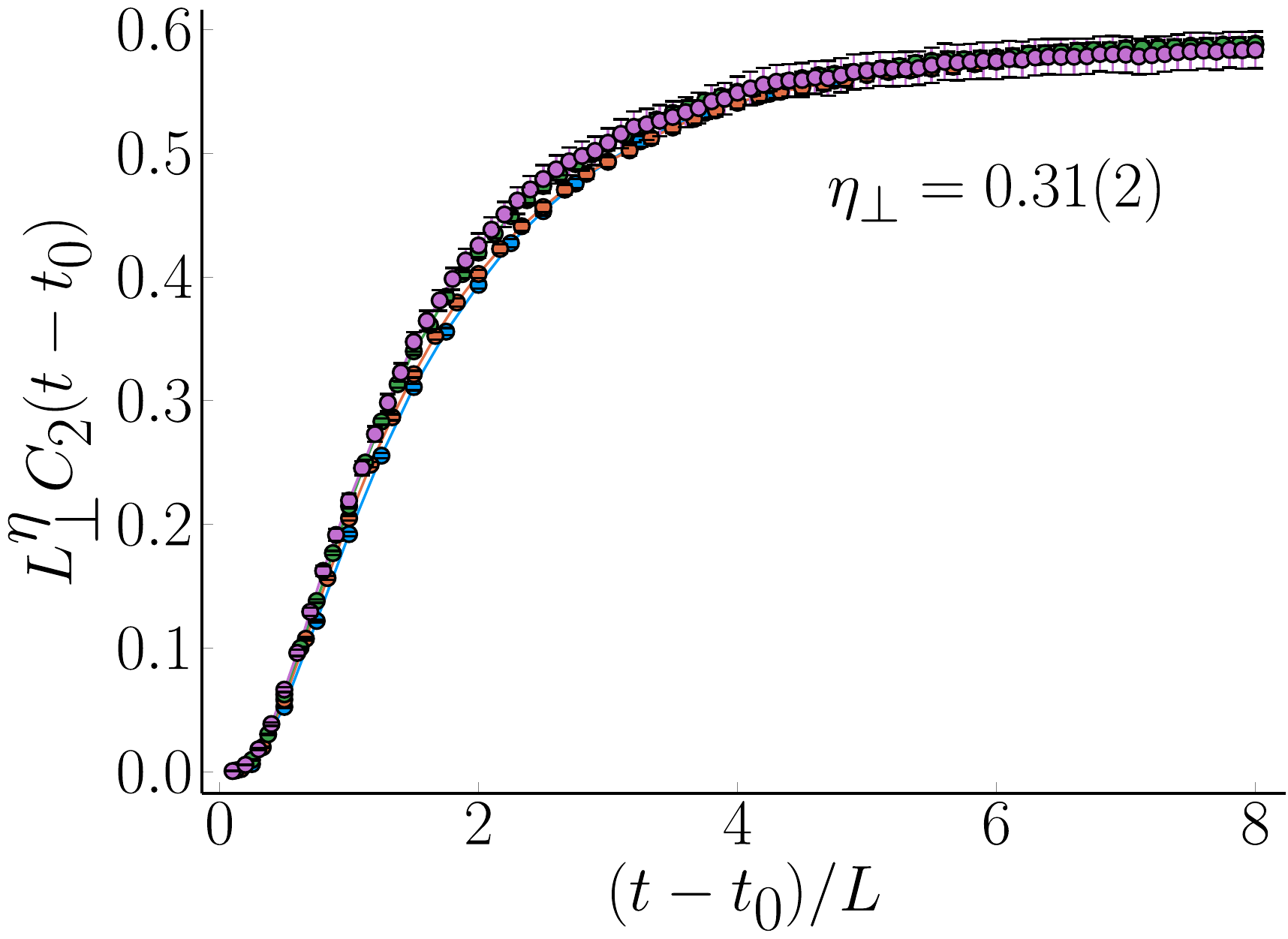}}
\subfloat[]{\includegraphics[width=.30\columnwidth]{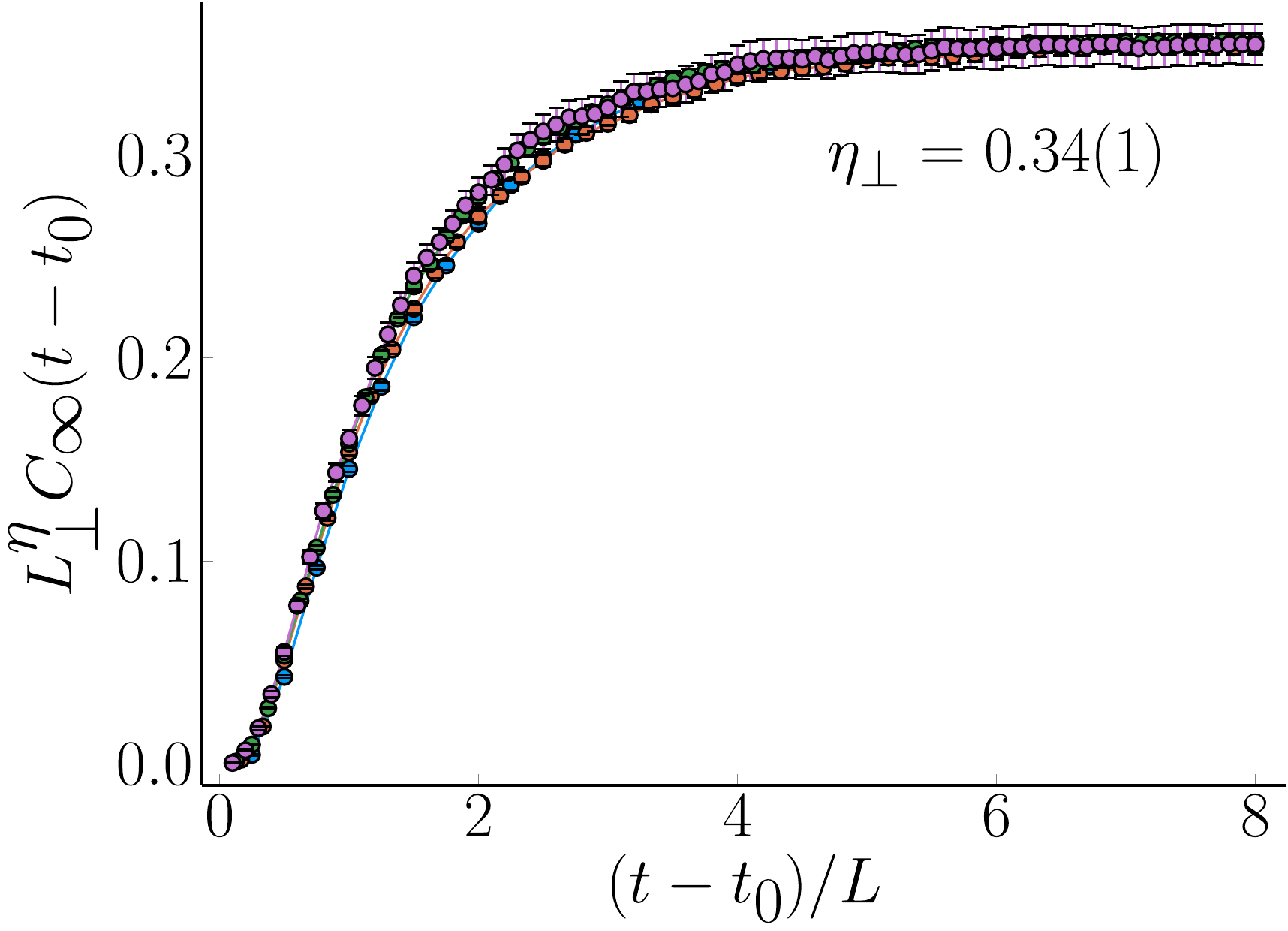}}
\caption{\emph{Surface exponent $\eta_\perp$.} The surface $\eta_\perp$ is given by a circuit with open boundary conditions starting from a product state. The circuit is run to a time $t_0 = 2L$ and the ancillas are maximally entangled with the edge and middle spins.}
\label{fig:etaPerp}
\end{figure}

The main text contains results for $\eta$ and $\eta_\parallel$ for the R\'enyi index $n = 1$. For an arbitrary value of the R\'enyi index we can define the correlation function as $C_n(t-t_0) = I_n(\tilde A,\tilde B) = S_n(\tilde A) + S_n(\tilde B)- S_n(\tilde A\cup \tilde B)$. In Figs. \ref{fig:eta} and \ref{fig:etaPara} the other R\'enyi's ($n = 2,5,\infty)$ are shown to have similar quality of data and collapse. In Fig. \ref{fig:etaPara2} we show results for the alternate estimate of $\eta_\parallel$ where we find values different than those reported in the main text, however, we find the values from the main text hold reasonably well at late times where the dynamics are expected to be universal. Fig. \ref{fig:etaPerp} also shows results for $\eta_\perp$ for $n = 1,2,\infty$.

\section{Order Parameter Dynamics}

\begin{figure}[htbp]
\begin{center}
\includegraphics[width = 0.75\textwidth]{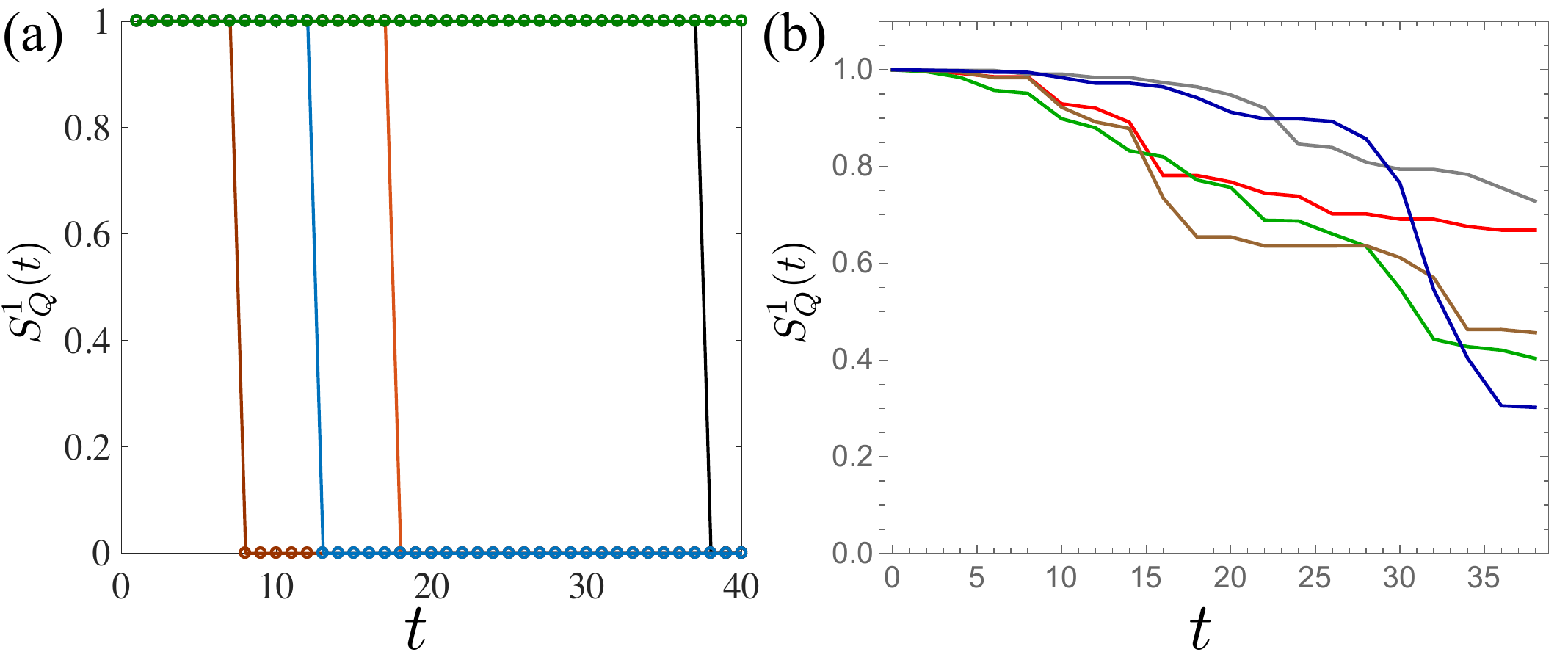}
\caption{\emph{Order parameter dynamics.} For fixed gates and measurement locations we average the entropy of a single reference qubit over measurement outcomes to determine the order parameter $S_Q^1(t)$ for this circuit \cite{Gullans2019b}. We took an initial state consisting of a pseudo-random stabilizer state/Haar random state between the reference and system in the case of Clifford/Haar gates. (a) Dynamics of $S_Q^1(t)$ for five different realizations of the stabilizer circuit at $p=p_c^{C}$.  The purification of the reference occurs in a single time step at a time that depends on the choice of circuit. (b) Dynamics of $S_Q^1(t)$ for five different realizations of the Haar random circuit at $p=p_c^H$.  In sharp contrast to the stabilizer circuits, the purification occurs gradually over many measurements.  We took $L=12$ in both panels and each curve in (b) is averaged over 10 000 trajectories.  No such averaging is required in (a) because $S_Q^1$ is independent of the measurement outcomes for stabilizer circuits. }
\label{fig:trajectories}
\end{center}
\end{figure}

In Fig.~\ref{fig:trajectories}, we show the dynamics of the order parameter $S_Q^1$ for the case of the Clifford and Haar gate models at their respective critical measurement rates for $L=12$.  Here, $S_Q^1$ is the entropy of the reference qubit averaged over measurement outcomes (trajectories) for a fixed realization of the circuit and measurement locations.  The initial state was chosen to be a pseudo-random stabilizer/Haar random state of the system and reference.  The pseudo-random stabilizer state was generated by running a depth $2L$ circuit without measurements on an initial product stabilizer state.  In the case of the stabilizer circuits, the purification of the reference qubit happens instantaneously in a single measurement.  The universal decay curve with time only arises after averaging over random choices of gates or measurement locations. For the Haar random circuit, in contrast, the randomness in the measurement outcomes is sufficient to lead to a nontrivial  decay for the order parameter.  This is an important qualitative distinction in the critical dynamics between these two models.  In the case of the stabilizer circuit, the order parameter dynamics are not self-averaging even in the thermodynamic limit, whereas the intrinsic randomness of the measurement outcomes is enough to lead to (at least partial) self-averaging behavior for the Haar model.  

Another interesting observation is that $S_Q^1$ is monotonically decreasing with time in both cases. Using the relation between  $S_Q^1$ and the mutual information between the reference qubit and an effective environment associated with the measurement outcomes \cite{Gullans2019b,Gullans2019,ChoiBaoQiAltman2019}, such a monotonic decay implies that, for these circuits, the information about the state of the reference qubit flows irreversibly into the environment.  This trend is consistent with the fact that the unitary-projective measurement dynamics effectively samples the quantum trajectories of a system interacting with a perfect Markovian environment.

\section{Minimal-cut picture for mutual information}

\begin{figure}[htbp]
\begin{center}
\includegraphics[width = 0.55\textwidth]{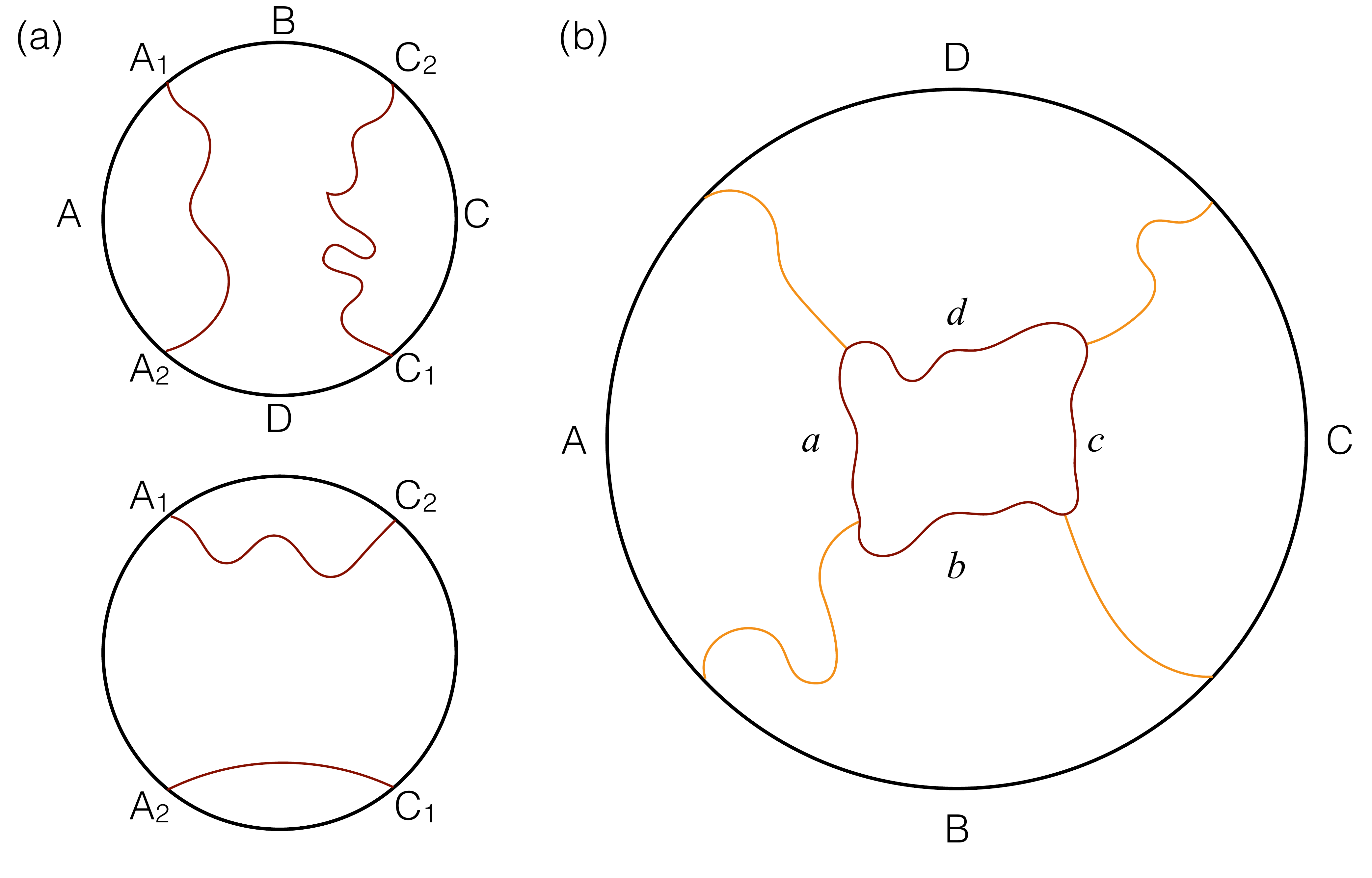}
\caption{\emph{Minimal cut.} Minimal cut diagrams for partitions of a system into four equal segments. (a)~Schematic minimal cut diagrams for the bipartite mutual information $\mathcal{I}_{2,0}(A:C)$. For the region $A (C)$, one requires the optimal path from $A_1$ to $A_2$ ($C_1$ to $C_2$), shown in the upper figure. However, for the region $A \cup C$ one only needs paths with endpoints at $A_1, A_2, C_1, C_2$; there are two of these, shown in the upper and lower figures. When the lower cut determines $S_0(A \cup C)$, the mutual information is nonzero. (b)~Diagram for computing the tripartite mutual information $\mathcal{I}_3$. It is convenient to separate the minimal cuts into ``boundary'' parts (denoted by light orange lines) and ``bulk'' parts (denoted by dark red lines). In $\mathcal{I}_3$ the boundary parts cancel but the bulk parts do not.}
\label{minimalcut}
\end{center}
\end{figure}

In this section we briefly discuss the behavior of the bipartite and tripartite mutual information, $\mathcal{I}_{2,n}$ and $\mathcal{I}_{3,n}$, within the minimal cut picture. For Haar-random circuits, the minimal cut picture is exact for $n = 0$, regardless of the local Hilbert-space dimension; within the minimal cut, the entanglement transition is a percolation transition. We will restrict our discussion to partitions of a system of size $L$, with periodic boundary conditions, into four equally sized pieces (each of length $L/4$) labeled $A,B,C,D$ as in Fig.~\ref{minimalcut}. Specifically, we will consider $\mathcal{I}_{2,0}(A:C)$ and $\mathcal{I}_{3,0}(A:B:C)$; for simplicity, in what follows we will suppress the indices and just call these quantities $\mathcal{I}_2$ and $\mathcal{I}_3$. Although our discussion is quantitative only for this specific choice of observable, we have found that the generic and percolation transitions are qualitatively similar, so our analysis offers useful guidance for the general case. 

\begin{figure}[htbp]
\centering
\includegraphics[width=.40\columnwidth]{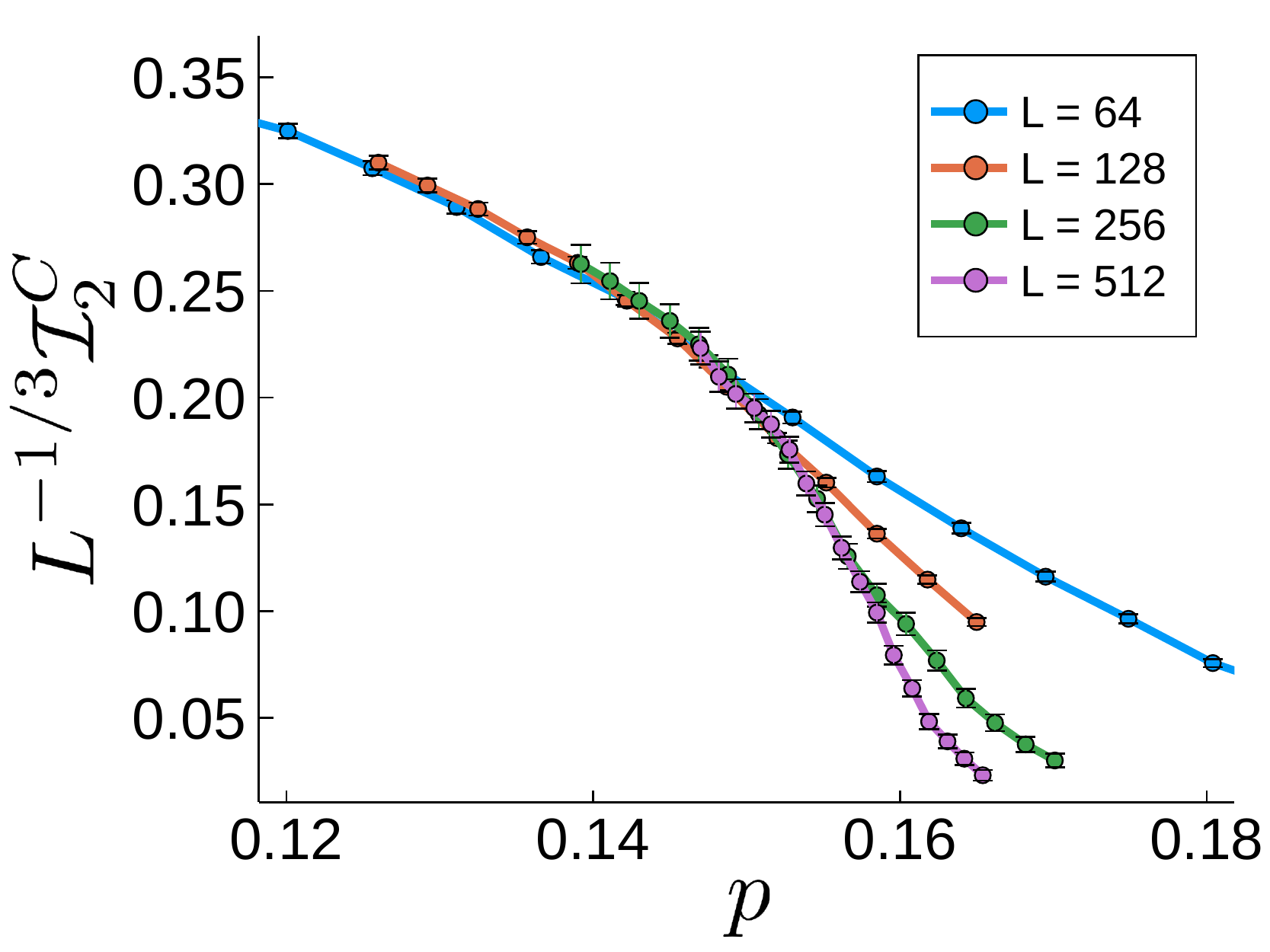}
\caption{\emph{Mutual information scaling.} The mutual information in the volume-law phase for the stabilizer circuits scales as $L^{1/3}$ as expected from the minimal cut picture.}
	\label{fig:MISc}
\end{figure}

\begin{figure}[htbp]
\centering
\subfloat[]{\includegraphics[width=.30\columnwidth]{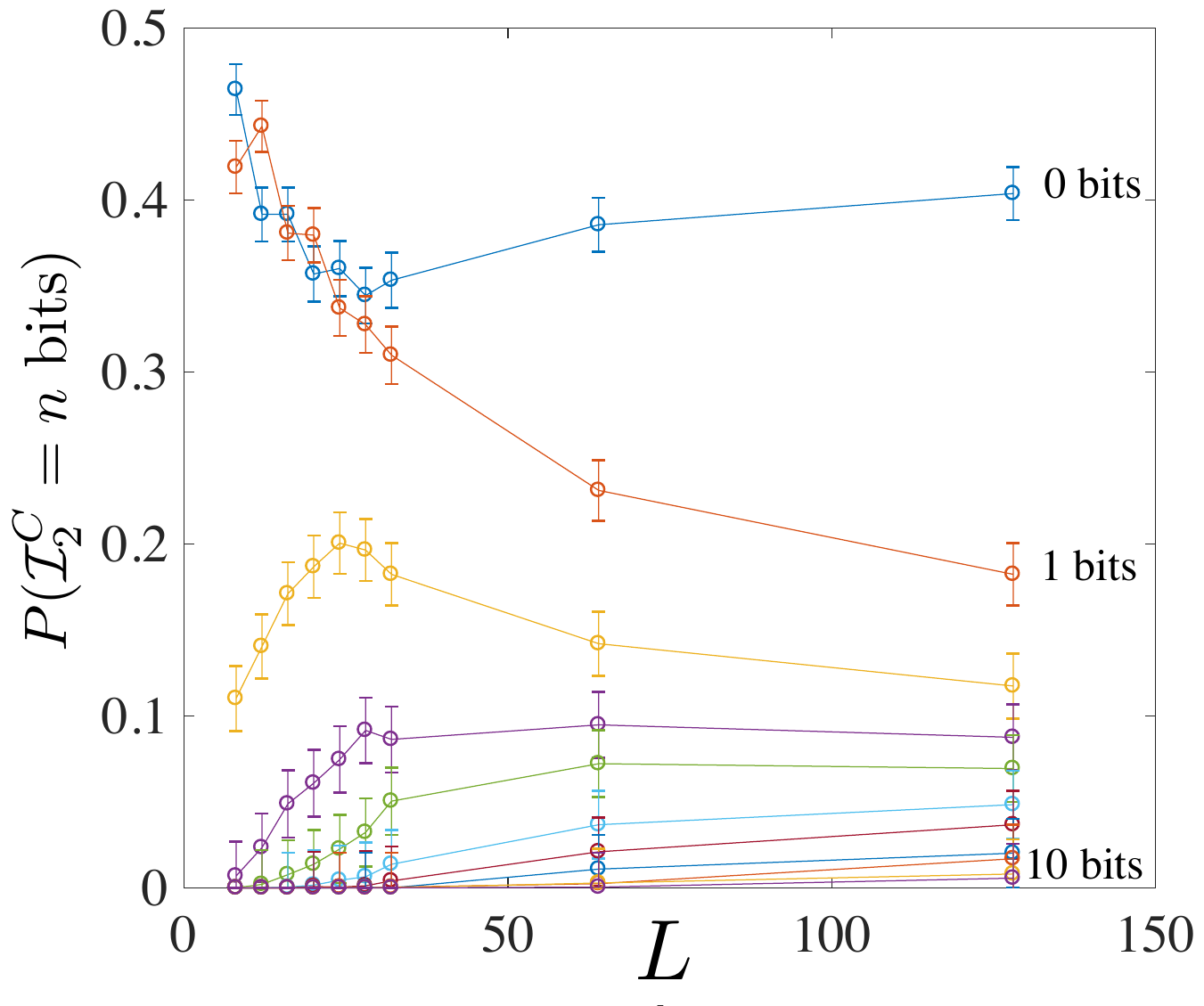}}
\subfloat[]{\includegraphics[width=.30\columnwidth]{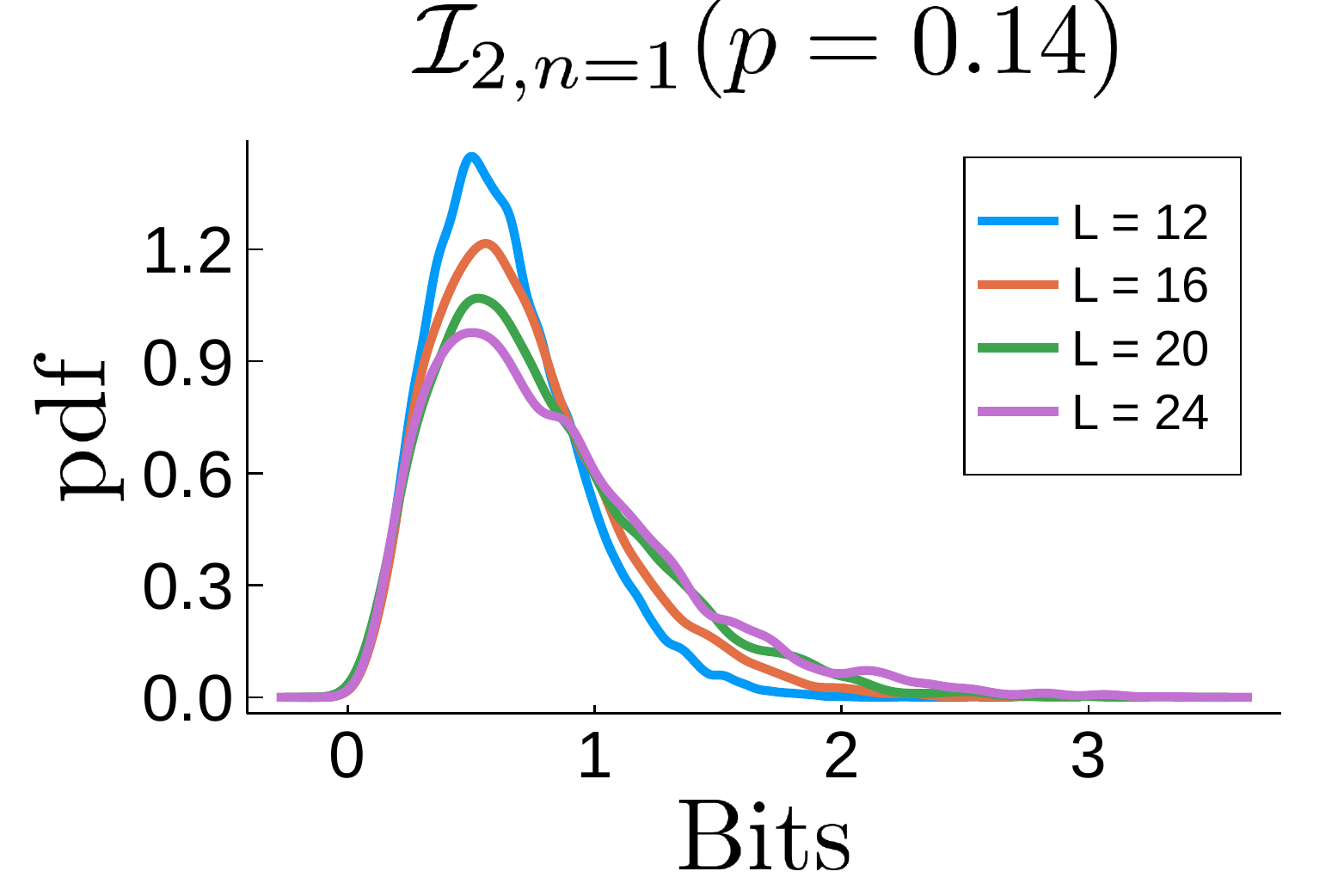}}
\subfloat[]{\includegraphics[width=.30\columnwidth]{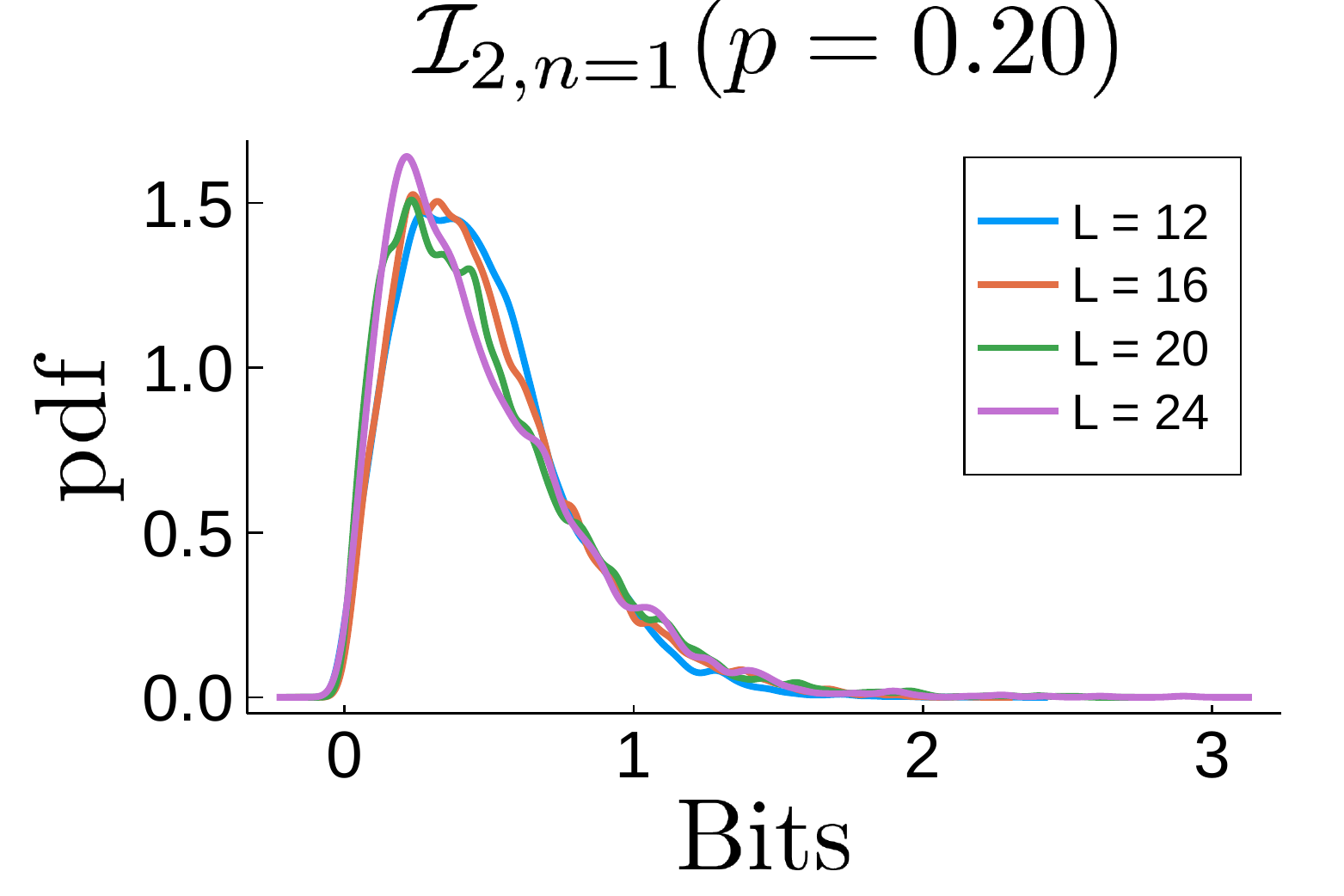}}
\caption{\emph{Mutual information probability distributions.} (a) Stabilizer circuits at $p = 0.08$, i.e., deep in the volume-law phase. Although $\mathcal{I}_2$ grows with system size, $L$, the probability of finding zero mutual information appears to approach a constant, as the minimal cut picture would predict. (b),(c): analogous data for Haar circuits. Unlike the stabilizer case, here $\mathcal{I}_2$ is continuously distributed.}
	\label{fig:hist}
\end{figure}

We begin with $\mathcal{I}_2 = S(A) + S(C) - S(A \cup C)$; we assume the circuit has been run out to times considerably longer than system size. The entanglement of $A$ is given by the minimal number of bonds that must be crossed by a polymer beginning and ending at endpoints of $A$. For the disjoint region $A \cup C$, one has two ways to connect the polymers (Fig.~\ref{minimalcut}a), $A_1 A_2, C_1 C_2$ and $A_2 C_1, C_2 A_1$; the minimal cut is the shorter of these. In the absence of measurements, the two arrangements cut the same number of bonds, so $\mathcal{I}_2 = 0$. Measurements locally cut bonds; thus, for nonzero measurement probability, in a given realization of the circuit, the two minimal-cut arrangements pass through different numbers of cut bonds. In half the realizations, the minimal cut for $A \cup C$ is the union of those for $A$ and $C$, and $\mathcal{I}_2 = 0$; in the other half, there is nonzero $\mathcal{I}_2$. Following a standard analysis of polymers in random media~\cite{kz} we find that the typical scale of fluctuations between the two minimal cuts, and therefore the mutual information in the volume law phase, scales as $\sim L^{1/3}$. This latter prediction is consistent with the data on Haar circuits (Fig.~\ref{fig:MISc}); moreover, the minimal-cut prediction that some nonvanishing fraction of the samples have zero mutual information in the thermodynamic limit is consistent with the data on stabilizer circuits in the volume law phase, as shown in Fig.~\ref{fig:hist} (by contrast, the distribution for Haar circuits is continuous, but at the achievable sizes we cannot conclude whether there is a distinct peak near zero mutual information). Thus the minimal cut picture appears to qualitatively describe many features of the mutual information beyond the $n = 0$ limit.

In the area law phase, there is a percolating cluster of bonds that are cut because of measurements. The polymer from each end of $A$ travels some distance into the circuit, finds the cluster of cut bonds (which we can regard as a large ``void''~\cite{Skinner2019}), and moves freely through it to join up with the polymer from the other end. The entanglement is set by the local geometry of getting from (say) $A_1$ to the void (the ``boundary'' sections in Fig.~\ref{minimalcut}b). In the area law phase there is no contribution from the ``bulk'' of the region and thus no distinction between the two ways of closing the minimal cut; the mutual information is thus zero, except in instances that are exponentially rare in the region size. 

The behavior at the critical point is intermediate. The polymer goes through a hierarchy of voids, each of which is larger than the previous one by some factor; to travel a distance $L$, therefore, requires crossing $\sim \log L$ bonds~\cite{Skinner2019}. To get from $A_1$ to $A_2$ (or $C_2$), the polymer first makes it out of $A_1$ into a void of size $\sim L$, crosses that void, and then goes through increasingly small voids until it reaches $A_2$ (or $C_2$). Crucially, the optimal path to the large void is the same regardless of which way one closes the minimal cut. Therefore, the differences between the two different minimal cuts come only from the largest few scales, and are thus $O(1)$; thus $\mathcal{I}_2$ at the critical point is expected to be constant. 

We now turn to $\mathcal{I}_3$, for which the analysis is similar but more involved. As noted before, we expect the local behavior near the cuts to be the same regardless of how the cuts will eventually line up; therefore we can schematically split up our entanglement cuts into ``bulk'' and ``boundary'' segments as in Fig.~\ref{minimalcut}b. 
In $\mathcal{I}_3$, the boundary segments cancel out, leading to the formula (expressed in the notation of that figure):

\begin{eqnarray}\label{i3mincut}
\mathcal{I}_3 & = & \min(a, b + c + d) + \min(b, a + c + d) + \min(c, a + b + d) \nonumber \\ && \qquad - \{ \min(a + b, c + d) + \min(a + c, b + d) + \min(a + d, b + c) \}.
\end{eqnarray}
In the volume law phase, $a,b,c,d$ are all comparable and volume-law in magnitude, so $\mathcal{I}_3 \approx -2a$ is negative and volume-law. In the area law phase, by contrast, all four numbers are essentially zero (since the polymer can move through a void), so $\mathcal{I}_3 = 0$ (since the boundary contributions cancel exactly). At the critical point, the ``bulk'' contribution is $O(1)$, as argued above, so $\mathcal{I}_3$ is also $O(1)$. 
%

One can see that this expression~\eqref{i3mincut} is never positive. Without loss of generality take $a \geq b \geq c \geq d$. The expression then simplifies to $\mathcal{I}_3 = \min(a, b + c + d)  - \min(a + d, b + c) - d$. There are two cases: (i)~$b + c \geq a + d$, in which case $\mathcal{I}_3 = a - (a + d) - d = - 2d \leq 0$; and (ii)~$a + d > b + c$, in which case $\mathcal{I}_3 = \min(a, b+c+d) - (b + c + d) \leq 0$. This establishes that $\mathcal{I}_3 \leq 0$ within the minimal cut picture; together with our previous argument that it should be $O(1)$ in magnitude, we conclude that $\mathcal{I}_3$ should be a negative number of order unity at the transition. This appears to be the case not just for percolation, as derived above, but also for Haar and stabilizer circuits. 

\bibliographystyle{apsrev4-1}
\bibliography{SupplementBib}

\clearpage

\section{Supplemental Figures}

\begin{figure}[htbp]
\centering
\subfloat[]{\includegraphics[width=.30\columnwidth]{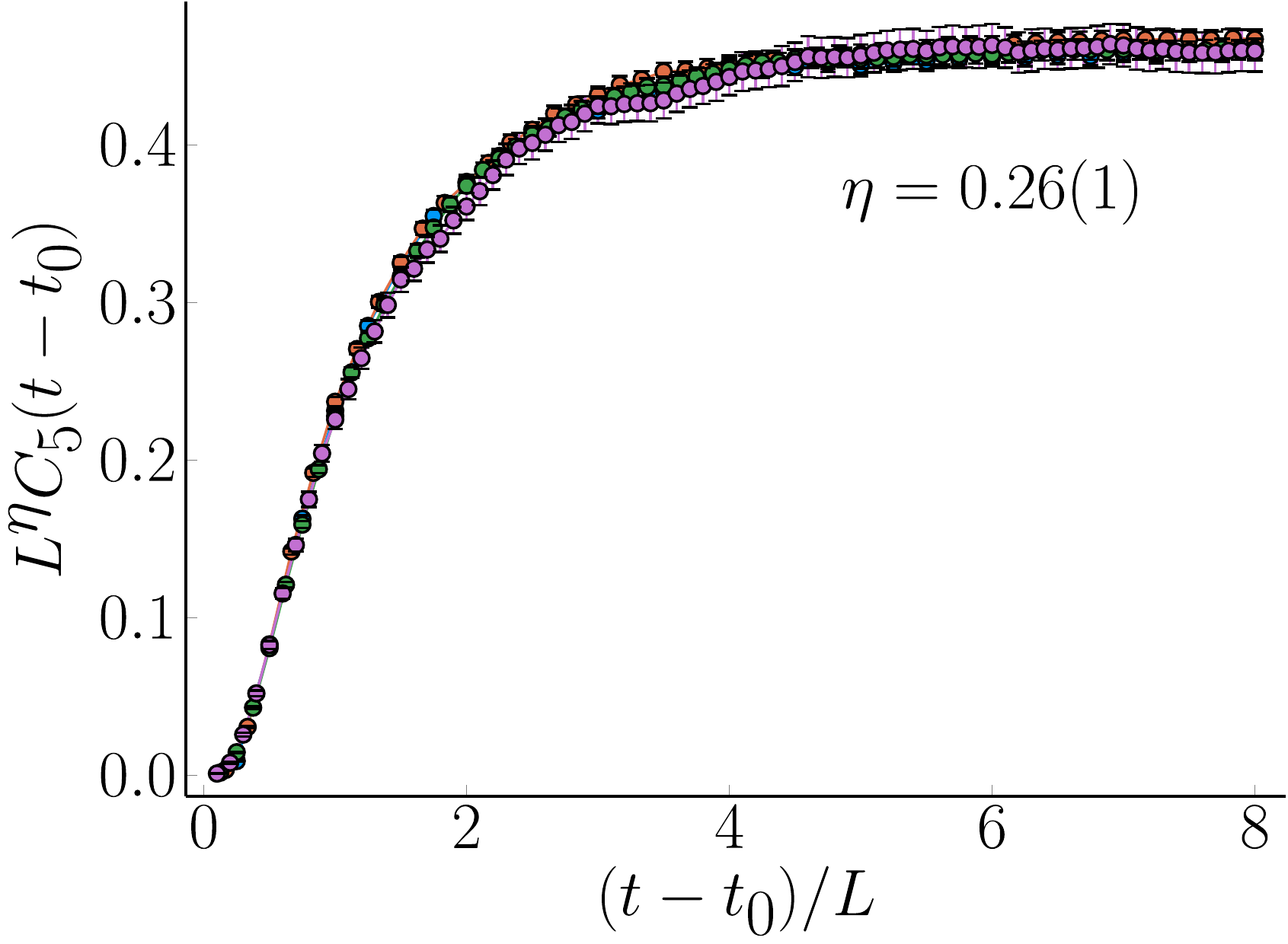}}
\subfloat[]{\includegraphics[width=.30\columnwidth]{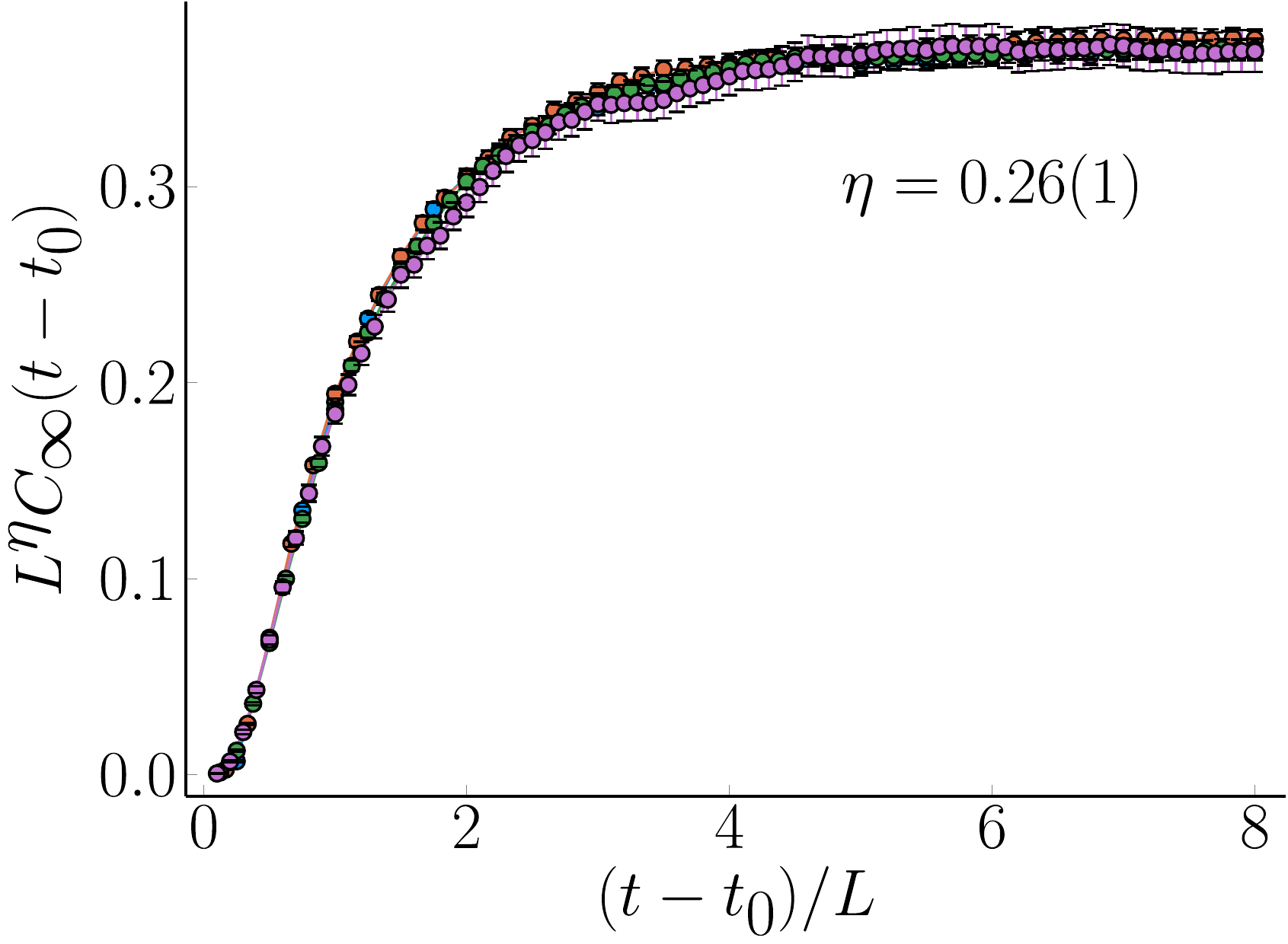}}
\caption{\emph{Bulk exponent $\eta$.} The bulk $\eta$ is given by a circuit with periodic boundary conditions starting from a product state. The circuit is run to a time $t_0 = 2L$ and the ancillas are maximally entangled with antipodal spins.}
\label{fig:eta}
\end{figure}

\end{document}